\documentclass{iopjournal}
\usepackage{ragged2e}
\justifying
\usepackage[utf8]{inputenc}
\usepackage{lmodern}
\usepackage{microtype}
\usepackage{amsmath,amssymb,amsfonts,mathtools,bbm}
\usepackage{bbold}
\usepackage{xcolor}
\usepackage{url}
\usepackage{hyperref}
\usepackage{graphicx}
\usepackage{enumitem}
\usepackage{booktabs}
\usepackage{makecell}
\usepackage{subfigure}
\usepackage{verbatim}
\usepackage{multirow}
\usepackage[normalem]{ulem}
\usepackage{booktabs}
\usepackage{array}
\newcommand{\V}{\mathcal{V}}

\begin{document}

\articletype{Review} 

\title{Exploring Exotic Spin-Dependent Interactions Beyond the Standard Model: Theoretical Foundations and Experimental Investigations}

\author{L. Y. Wu$^{1,2}$ and H. Yan$^{1,*}$}

\affil{$^1$Institute of Fundamental Physics and Quantum Technology, and School of Physical Science and Technology, Ningbo University, Ningbo, Zhejiang 315211, China}

\affil{$^2$Key Laboratory of Nuclear Physics and Ion-beam Application (MOE), Institute of Modern Physics, Fudan University, Shanghai 200433, China}

\affil{$^*$Author to whom any correspondence should be addressed.}

\email{yanhaiyang@nbu.edu.cn}


\begin{abstract}
\justifying
New interactions mediated by novel particles propose solutions to several important questions in modern physics. 
Axions serve as examples of such particles; they are lightweight and interact weakly with ordinary matter. 
This category of particles, including those similar to axions---termed Axion-Like Particles (ALPs)---arises from diverse theoretical frameworks, such as the Peccei-Quinn mechanism addressing the strong CP problem, string theory, and spontaneous supersymmetry breaking. 
Given their light mass and weak coupling, ALPs are also possible candidates for cold dark matter.
Introducing these new interactions mediated by novel particles not only tackles several challenges in modern physics but also raises a crucial question: Are there undiscovered interactions beyond the Standard Model? 
Many of the interactions predicted by these theories are spin-dependent, which is the primary focus of this review. 
In this review, we first outline the theoretical foundations for investigating exotic spin-dependent interactions, highlighting their importance in various models beyond the Standard Model.
We examine the potential roles of new lightweight particles in mediating these interactions, which may enhance our understanding of dark matter.
Relevant formulas derived from theoretical models are included to support experimental investigations. 
Following this theoretical framework, we conduct a detailed review of recent experimental efforts to detect these exotic interactions.
A systematic review of current constraints on these interactions is presented, along with an assessment of various detection approaches.
\end{abstract}

\section{Introduction}\label{intr}
Spin is an intrinsic property of fundamental particles. Wigner's classification of the irreducible representations of the Poincar\'{e} group serves as the foundation of particle physics~\cite{ramond_field_theory}. 
Fundamental particles are categorized based on these irreducible representations, which are defined by their mass and spin~\cite{weinberg_qft}. 
Consequently, spin is as fundamental to a particle's nature as its mass. 
Spin-dependent interactions are both rich and profound, particularly in their connection to axions.

Fundamental physics faces several profound unsolved problems, with the strong CP problem and the nature of dark matter (DM) being particularly fascinating. 
The proposal of the axion, a hypothetical particle that could simultaneously address both issues, has attracted widespread research interest. 
Experimentally, searching for exotic spin-dependent interactions is one strategy for confirming or ruling out the existence of the axion. In terms of chronological development, the axion emerged from a mechanism proposed to address the strong CP problem. 
In 1977, Peccei and Quinn introduced the PQ mechanism to address the strong CP problem in quantum chromodynamics (QCD)~\cite{Peccei1977PRD, Peccei1977PRL}. 
They proposed a new  U(1) global symmetry that spontaneously breaks, offering an elegant and promising solution to the strong CP problem. 
Almost simultaneously, Wilczek and Weinberg identified that this mechanism could lead to the emergence of new light pseudoscalar particles, now known as axions~\cite{Wilczek1978PRL, Weinberg1978PRL}. 
Due to their light masses and weak interactions with ordinary matter, axions are also considered potential candidates for cold DM~\cite{Preskill1983PLB, Abbott1983PLB, Dine1983PLB, Marsh2016PR,chadha-day2022SA}. 
In cosmological observations, the axion model provides a compelling explanation for experimental phenomena such as the cosmic microwave background and the large-scale structure of the universe~\cite{Hlozek2015PRD, Desjacques2018PRD, OHare2024arxiv}. In addition, in some theoretical models, axion-like fields are proposed as potential candidates for dynamical dark energy~\cite{Cicoli2012JCAP}. 
Unlike the cosmological constant, which is constant, these fields vary over time and space. The concept is that a slowly evolving field could mimic dark energy by driving the universe's accelerated expansion. 
The interaction strength $f_a$ of the axion---specifically the QCD axion that emerged from the strong CP problem---is related to its mass $m_a$ and varies according to different models such as KSVZ and DFSZ models~\cite{Kim1979PRL, Shifman1980NPB, Dine1983PLB}. 
However, various theories, such as supersymmetry extensions and string theory, predict the existence of many pseudoscalar particles that share properties similar to axions in terms of their couplings to standard-model particles~\cite{Svrcek2006JHEP, Arvanitaki2010}. 
In these cases, the particle masses and interaction strengths are not necessarily related.
Therefore, axions are generally considered in a broader context beyond the QCD axion.
In the experiment, it is also sensible to search for axions across the entire parameter space extended by both $f_a$ and $m_a$. 
The category of pseudoscalar particles was later expanded to include vector particles, such as the Z$^\prime$ boson and paraphoton~\cite{Fayet1980aPLB, Fayet1980PLB, Langacker2009RMP, Holdom1986PLB, Dobrescu2005PRL}. 
In this work, we collectively refer to all such new, light, and weakly interacting particles---whether scalar, pseudo-scalar, vector, or axial-vector---as axion-like particles (ALPs).

\begin{figure}[htbp]
  \centering
   \includegraphics[width=0.9\linewidth]{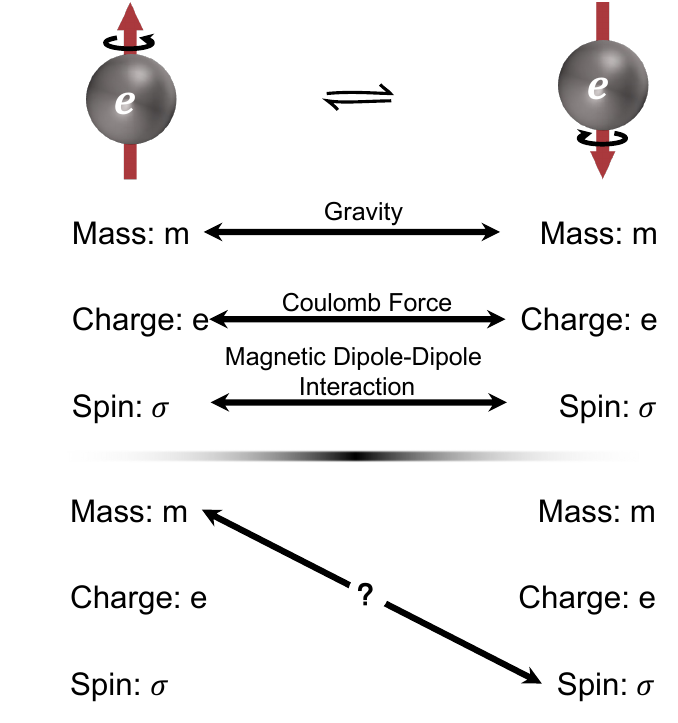}
  \caption{\justifying  Schematic representations of particle interactions. The upper part of the figure illustrates conventional interactions between two fundamental particles, where like properties are coupled: gravity couples mass, the Coulomb force couples electric charge, and the magnetic dipole-dipole interaction couples spin. The lower part presents a hypothetical scenario in which a new interaction couples different fundamental properties, specifically the mass of one particle to the spin of another, {\color{black}such as the $V_{\rm SP}$ interaction (\ref{sp1_g})}. Such cross-property couplings, if they exist, would represent novel forms of fundamental interactions beyond the SM.}
  \label{fig1}
\end{figure}

These new particles not only offer solutions to several significant problems in modern physics but also pose a critical question: Are there undiscovered interactions beyond the Standard Model (SM)?
While spin participates in known interactions via magnetic or electric moments, no fundamental interaction has yet been discovered that is directly generated by spin, unlike mass or charge.
As an intrinsic property of particles, it is natural to expect spin to play a more fundamental role in potential new interactions.
Spin-dependent interactions arise naturally within the framework of the Standard-Model Extension (SME)~\cite{Kostelecky2001PRD}.
The possibility that new interactions might be spin-dependent is particularly intriguing.
In addition to motivations like addressing the dark matter puzzle or the strong CP problem, there may be broader reasons to explore this idea.
As illustrated in Fig.~\ref{fig1}, conventional interactions couple like properties: gravity couples masses, the Coulomb force couples electric charges, and the magnetic dipole-dipole interaction couples spins.
This naturally raises the question: could there exist new interactions that couple different intrinsic properties?
For example, as suggested in the lower part of Fig.~\ref{fig1}, could a novel interaction couple a particle’s mass to another particle’s spin?
Given that mixing between different degrees of freedom is not uncommon in modern physics, it is both reasonable and essential to conduct precision experiments to test for the existence of such unconventional interactions.

In 1984, Moody and Wilczek first proposed that new macroscopic spin-dependent interactions between fermions could be mediated by axions~\cite{Moody1984PRD}.
Subsequently, Dobrescu and Mocioiu~\cite{Dobrescu2006JHEP} extended this idea to a more general framework that includes both spin-0 and spin-1 boson exchange.
Moreover, independent of the details of the underlying fundamental theory, and assuming only rotational invariance, they classified all possible long-range interactions between two fermions into 16 distinct types.
Each of these interactions involves a Yukawa-like potential or its derivatives; accordingly, we refer to them as Yukawa-type interactions.

Various non-Yukawa-type interactions have been theoretically proposed, including those in which spin couples directly to a classical background field.
A prominent example is the axion field.
In addition to mediating spin-dependent interactions as a propagator, ALPs can also couple directly to spin as a classical field that fills space.
If ALPs are assumed to be dark matter candidates, then sufficiently light ALPs would have a high number density, enabling them to be treated as a classical background field.
This situation is analogous to the large photon occupation number in a coherent electromagnetic wave, which justifies its classical-field description.
Torsion-induced spin-gravity interaction provides another example of spin coupling to a classical field.
In extensions of general relativity (GR), spin is proposed to generate spacetime torsion-analogous to the frame-dragging effect caused by the angular momentum of macroscopic bodies~\cite{Hehl1976RMP}.
Similarly, violations of CPT and Lorentz symmetries can give rise to background fields that couple to spin~\cite{haridass1977GRG, Leitner1964PR, Kostelecky2001PRD, Kostelecky2011PRD}.
The electric dipole moment (EDM) of particles may also be interpreted as a coupling between spin and a classical electric field.
Additional examples of non-Yukawa-type interactions include those mediated by unparticles-a scale-invariant field-which can induce long-range spin-dependent forces characterized by non-integer power-law potentials~\cite{Georgi2007PRL, Liao2007PRL}.
Moreover, Ref.~\cite{Costantino2020JHEP} proposes non-Yukawa-type spin-dependent interactions induced by bilinear scalar couplings.

\begin{figure}[htbp]
  \centering
  \includegraphics[width=0.9\linewidth]{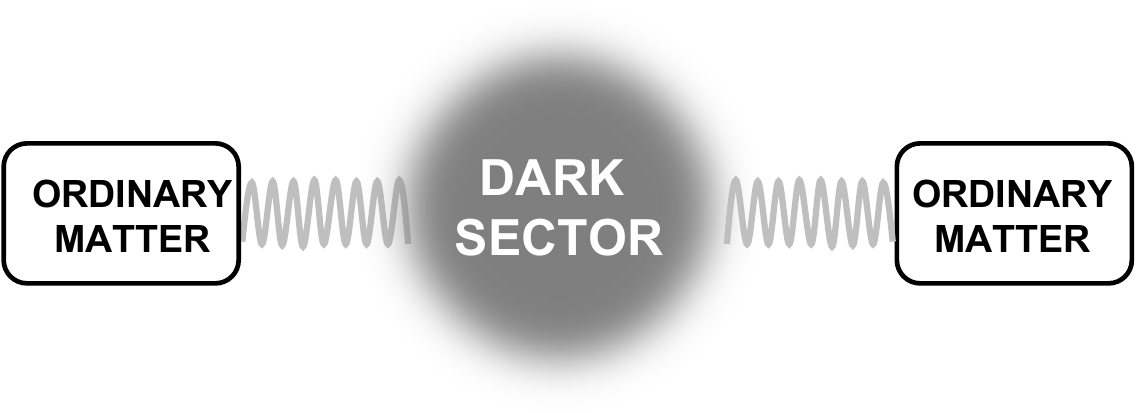}
  \caption{\justifying Detection strategy---Suppose ordinary matter is weakly coupled to the dark sector. In this case, extremely weak interactions may arise between ordinary objects. By performing precision measurements on ordinary matter, one can, in principle, infer the existence or absence of the dark sector.}
  \label{fig2}
\end{figure}

On the other hand, to detect is to interact. Any attempt to observe a subject necessarily involves some level of interaction. Without such interaction, observation becomes impossible. While we possess a comprehensive understanding of ordinary matter, our knowledge of the dark sector remains limited. 
Therefore, our strategy focuses on performing highly precise measurements in the ordinary sector, aiming to infer the presence of the dark sector through the interactions it mediates among ordinary-sector components, as illustrated in Fig.~\ref{fig2}.

Due to the light masses and weak couplings of these particles, precision measurement techniques provide a viable approach for detecting them, particularly through their predominantly spin-dependent interactions. 
The mass range of these particles spans many orders of magnitude, and, according to the de Broglie relation, their corresponding interaction ranges extend from atomic scales to astronomical distances. 
As a result, a variety of experimental techniques, each specialized in detecting particles of different masses, are required to complement one another in probing the full parameter space.

This article presents a comprehensive review of recent advances in the study of exotic spin-dependent interactions. 
{\color{black}Although significant progress in this field was reviewed in the comprehensive work by Safronova \textit{et al.}~\cite{Safronova2018RMP}, the rapid developments in recent years justify an updated overview. Notably, during the preparation of this manuscript, two related reviews addressing experimental searches for exotic spin-dependent interactions were published~\cite{Jiang2025ROPP, Cong2025RMP}. While these works provide valuable perspectives---Ref.~\cite{Jiang2025ROPP} focuses on spin-sensor-based approaches and Ref.~\cite{Cong2025RMP} surveys a broad spectrum of experimental strategies---the present review is deliberately positioned to make a distinct contribution. In particular, we provide a systematic synthesis of the underlying theoretical formalisms together with their concrete experimental implementations, with the explicit goal of serving as a practical roadmap for experimentalists. In addition, we include a dedicated discussion of noise suppression and systematic error mitigation, which constitute indispensable and widely shared techniques across diverse precision measurement platforms.}
Our focus is limited to investigations that explicitly involve spin-dependent phenomena. Nevertheless, some of the referenced theoretical models also predict spin-independent effects, which---though not central to this review---are included to a limited extent due to their relevance. In Sec.~\ref{theo}, we survey all known macroscopic spin-dependent interactions and discuss their theoretical foundations, including their transformation properties under discrete symmetry operations. We further categorize these interactions according to their experimental accessibility. In Sec.~\ref{expinv}, we review and classify the experimental approaches developed to detect such spin-dependent interactions. In Sec.~\ref{expcon}, we summarize the most up-to-date experimental constraints available. Finally, in Sec.~\ref{concl}, we offer our perspective on future directions in this active area of research.

While not all of our understandings, perspectives, or viewpoints may be correct, many are uniquely our own, shaped by years of dedicated work in this exciting field and reflected throughout this article.

\section{Theoretical Foundations}\label{theo}

In this section, we classify exotic spin-dependent interactions into two main categories: Yukawa-type and non-Yukawa-type interactions. Yukawa-type interactions have been more extensively studied, notably in the works of Moody and Wilczek, Dobrescu, Malta, and Fadeev, with an emphasis on Yukawa-like potentials~\cite{Moody1984PRD, Dobrescu2006JHEP, Fadeev2019PRA, Malta2016AHEP}. These interactions involve Yukawa-type potentials or their derivatives, arising from the finite mass of the mediator particle.
In contrast, non-Yukawa-type interactions include those in which spin couples directly to a classical background field. Representative examples include the axion field effect, background fields associated with Lorentz and CPT violation, torsion-induced spin-gravity couplings, and EDM interactions, which can be interpreted as spin coupling to an electric field. Other instances involve exotic interactions mediated by unparticles, as proposed by Georgi, Liao, and Wu~\cite{Georgi2007PRL, Georgi2007bPLB, Liao2007PRL, Wu2024JHEP}. Additionally, spin-dependent interactions induced by bilinear scalar couplings have been proposed and analyzed in detail by Costantino \textit{et al.}~\cite{Costantino2020JHEP}. In the theoretical sections, natural units ($\hbar = c = 1$) are used, whereas SI units are restored in the experimental discussions.

\subsection{Spin-Dependent Interactions Mediated by ALPs: The Yukawa Type}\label{theo_1}

\begin{figure}[htbp]
    \centering
    \includegraphics[width=0.8\linewidth]{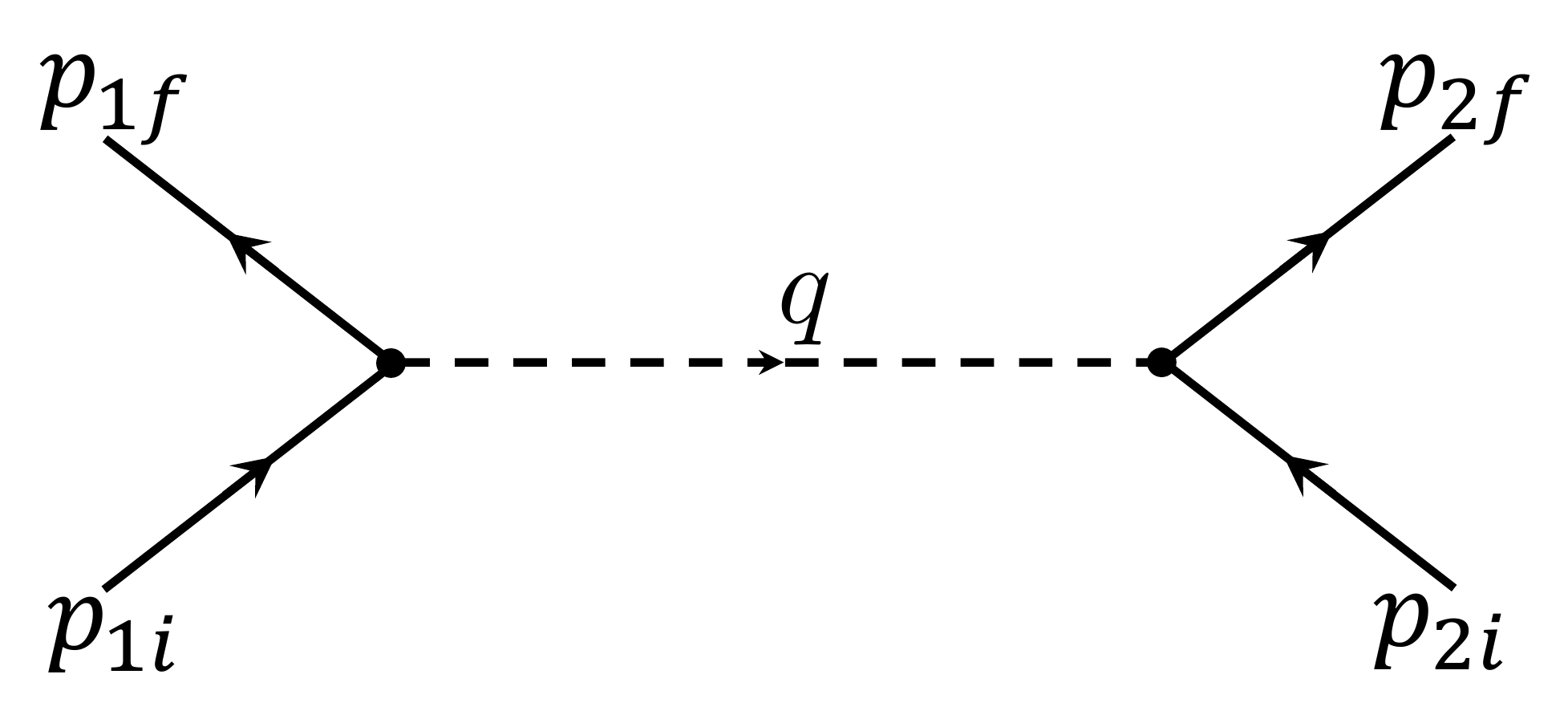}
    \caption{\justifying Representative tree-level Feynman diagram illustrating the interaction between two fermions via the exchange of a new mediator particle. The symbols $p_{1i}$ ($p_{1f}$) and $p_{2i}$ ($p_{2f}$) denote the initial (final) four-momenta of fermions 1 and 2, respectively, while $q$ represents the four-momentum transferred by the virtual mediator.}
    \label{fig3}
\end{figure}

We begin by introducing the Yukawa-type spin-dependent interactions mediated by ALPs. 
The general interactions between scalar $\phi$ and vector $X_\mu$ ALP fields and a fermion $\psi$ can be described by the following Lagrangian~\cite{Dobrescu2006JHEP}:
\begin{subequations}
    \begin{align}
    \begin{split}
    \label{eq1a}
        \mathcal{L}_{\rm SP}&=\phi\bar{\psi}(g_S +i\gamma_5g_P)\psi,
    \end{split}\\
    \begin{split}
    \label{eq1b}
        \mathcal{L}_{\rm VA}&=X_\mu\bar{\psi}\gamma^\mu(g_V+\gamma_5 g_A)\psi,
    \end{split}
\end{align}
\end{subequations}
where the dimensionless coupling constants $g_{S}$, $g_{P}$, $g_{V}$, and $g_{A}$ parameterize the strengths of the scalar (S), pseudoscalar (P), vector (V), and axial-vector (A) interactions, respectively, and $\gamma_\mu$ denote the Dirac matrices.

The tree-level diagram representing the interaction between two fermions via the exchange of ALPs, as described by the above Lagrangians, is shown in Fig.~\ref{fig3}.
In general, the scattering amplitude corresponding to this diagram can be expressed as:
\begin{equation}
    i\mathcal{M}=[-i\bar{u}(p_{1f})\Gamma_1u(p_{1i})]\mathcal{D}(q)[-i\bar{u}(p_{2f})\Gamma_2u(p_{2i})],
\end{equation}
where $u(p)$ ($\bar{u}(p)$) denotes the Dirac spinor, $\Gamma_i$ specifies the form of the interaction vertex, and $\mathcal{D}(q)$ represents the propagator of the ALP. Since the ALPs considered in this article are typically of low mass---corresponding to mesoscopic, macroscopic, or even astrophysical length scales---and are weakly coupled, a low-energy approximation is sufficient. The corresponding potential in position space is obtained from the scattering amplitude via the Fourier transform:
\begin{equation}
    V(r)=-\int \frac{d^3{\vec q}}{(2\pi)^3}\frac{\mathcal{M}}{4m_1m_2}e^{i\vec{q}\cdot\vec{r}},
\end{equation}
where $m_1$ ($m_2$) is the mass of particle 1 (2), $\vec{q}=\vec{p}_{2f}-\vec{p}_{2i}$ is the four-momentum transferred by the virtual mediator, and $\vec{r}=\vec{r}_2-\vec{r}_1$.
All possible types of interactions generated from Lagrangians (\ref{eq1a}) and (\ref{eq1b}) are discussed in detail in Refs.~\cite{Dobrescu2006JHEP, Malta2016AHEP, Fadeev2019PRA}.
In Ref.~\cite {Dobrescu2006JHEP}, these interactions are categorized into 16 types based on spin structures allowed by rotational invariance.
Later, Refs.~\cite{Malta2016AHEP, Fadeev2019PRA} classify interactions according to their coupling types, facilitating the interpretation of measurements for different theoretical models.

Although tremendous progress has been made and our understanding of these interactions has deepened, we believe that a more structured and systematic categorization would greatly benefit experimental efforts.
In this work, building upon previous classification schemes, we further organize and summarize these interactions in a more coherent and accessible manner.
Specifically, we classify the interactions according to two main criteria: the spin order, which denotes the number of spin operators involved, and the gradient order (i.e., the power of $\nabla$), which reflects the number of spatial derivatives acting on the original Yukawa-type potential.
On the one hand, the spin order influences experimental complexity: lower spin order corresponds to simpler experimental setups, since preparing and controlling polarized spins requires substantial effort and technical capability.
On the other hand, the gradient order affects the interaction's spatial behavior and detectability.
A higher power of $\nabla$ yields a stronger $1/r$ dependence, leading to faster decay of the interaction at macroscopic distances and thereby making it increasingly difficult to detect.
Moreover, for interactions sharing the same spin and gradient orders, we further order them by their velocity dependence---interactions with explicit velocity dependence generally introduce additional technical challenges and tend to be less constrained experimentally.
To support experimental searches, we have systematically organized all known spin-dependent interactions according to these criteria as follows:

\paragraph{Spin Order-0}
\begin{itemize}[left=0pt,itemsep=2pt,topsep=2pt]
    \item $\nabla^0~\text{Potentials}$\\
    \vspace{-0.5em}
    \begin{subequations}
    \begin{align}
    \begin{split}
        \label{ss0}
        V_{\rm SS}(r)&=\underbrace{-g^1_Sg^2_S}_{f_1}\underbrace{\frac{e^{-r/\lambda}}{4\pi r}}_{\V_1},
    \end{split}\\
    \begin{split}
        \label{vv0}
        V_{\rm VV}(r)&=\underbrace{g^1_Vg^2_V}_{f_1}\underbrace{\frac{e^{-r/\lambda}}{4\pi r}}_{\V_1},
    \end{split}
    \end{align}
    \end{subequations}
    \vspace{-0.5em}
\end{itemize}
\paragraph{Spin Order-1}
\begin{itemize}[left=0pt,itemsep=2pt,topsep=2pt]
    \item $\nabla^0~\text{Potentials}$\\
    \vspace{-0.5em}
    \begin{equation}
        \label{va1_v}
        V_{\rm VA}=\underbrace{g^1_Vg^2_A}_{f_{12+13}}\underbrace{\vec{\sigma}_2\cdot\vec{v}\frac{e^{-r/\lambda}}{4\pi r}}_{\V_{12+13}},
        \end{equation}
    \vspace{-0.5em}
\end{itemize}
\begin{itemize}[left=0pt,itemsep=2pt,topsep=2pt]
    \item $\nabla^1~\text{Potentials}$\\
    \vspace{-0.5em}
    \begin{subequations}
    \begin{align}
    \begin{split}
            \label{sp1_g}
            V_{\rm SP} &=\underbrace{\frac{g^1_Sg^2_P}{2}}_{f_{9+10}}\underbrace{\vec{\sigma}_2\cdot\nabla\frac{e^{-r/\lambda}}{4\pi m_2r}}_{\V_{9+10}},
        \end{split}\\
    \begin{split}
        \label{ss1_vg}
            V_{\rm SS} &=\underbrace{-g^1_Sg^2_S\frac{m_1}{4(m_1+m_2)}}_{f_{4+5}}\underbrace{\vec{\sigma}_2\cdot(\vec{v}\times\nabla)\frac{e^{-r/\lambda}}{4\pi m_2r}}_{\V_{4+5}},
        \end{split}\\
    \begin{split}
        \label{vv1_vg}
            V_{\rm VV} &=\underbrace{-g^1_Vg^2_V\frac{m_1+2m_2}{4(m_1+m_2)}}_{f_{4+5}}\underbrace{\vec{\sigma}_2\cdot(\vec{v}\times\nabla)\frac{e^{-r/\lambda}}{4\pi m_2r}}_{\V_{4+5}},
        \end{split}\\
    \begin{split}
        \label{aa1_vg}
            V_{\rm AA} &=\underbrace{-g^1_Ag^2_A\frac{m_2^2}{4m_1(m_1+m_2)}}_{f_{4+5}}\underbrace{\vec{\sigma}_2\cdot(\vec{v}\times\nabla)\frac{e^{-r/\lambda}}{4\pi m_2r}}_{\V_{4+5}}.
        \end{split}
    \end{align}
    \end{subequations}
    \vspace{-1.5em}
\end{itemize}
\paragraph{Spin Order-2}
\begin{itemize}[left=0pt,itemsep=2pt,topsep=2pt]
    \item $\nabla^0~\text{Potentials}$\\
    \vspace{-0.5em}
    \begin{subequations}
    \begin{align}
    \begin{split}
    \label{aa2_0}
        V_{\rm AA} &=\underbrace{-g^1_Ag^2_A}_{f_2}\underbrace{\vec{\sigma}_1\cdot\vec{\sigma}_2\frac{e^{-r/\lambda}}{4\pi r}}_{\V_2},
    \end{split}\\
    \begin{split}
    \label{aa2_vv}
        V_{\rm AA} &=-\underbrace{\frac{g^1_Ag^2_A}{2}}_{f_8}\underbrace{(\vec{\sigma}_1\cdot\vec{v})(\vec{\sigma}_2\cdot\vec{v})\frac{e^{-r/\lambda}}{4\pi r}}_{\V_8},
    \end{split}\\
    \end{align}
    \end{subequations}
    \vspace{-0.5em}
\end{itemize}
\begin{itemize}[left=0pt,itemsep=2pt,topsep=2pt]
    \item $\nabla^1~\text{Potentials}$\\
    \vspace{-0.5em}
    \begin{subequations}
    \begin{align}
    \begin{split}
    \label{va2_g}
        V_{\rm VA} &=\underbrace{-\frac{1}{2}g^1_Vg^2_A}_{f_{11}}\underbrace{(\vec{\sigma}_1\times \vec{\sigma}_2)\cdot\nabla\frac{e^{-r/\lambda}}{4\pi m_1r}}_{\V_{11}},
    \end{split}\\
    \begin{split}
        \label{va2_vvg}
        V_{\rm VA}&=\underbrace{\frac{m_1}{8(m_1+m_2)}g^1_Vg^2_A}_{f_{16}}\underbrace{\{
    [\vec{\sigma}_1\cdot(\vec{v}\times \nabla)](\vec{\sigma}_2\cdot \vec{v})+(\vec{\sigma}_1\cdot \vec{v})[\vec{\sigma}_2\cdot(\vec{v}\times \nabla)]
    \}\frac{e^{-r/\lambda}}{4\pi m_1r}}_{\V_{16}},
    \end{split}
    \end{align}
    \end{subequations}
    \vspace{-0.5em}
\end{itemize}
\begin{itemize}[left=0pt,itemsep=2pt,topsep=2pt]
    \item $\nabla^2~\text{Potentials}$\\
    \vspace{-0.5em}
    \begin{subequations}
    \begin{align}
    \begin{split}
    \label{pp2_gg}
        V_{\rm PP} &=\underbrace{\frac{g^1_Pg^2_P}{4}}_{f_3}\underbrace{(\vec{\sigma}_1\cdot\nabla)(\vec{\sigma}_2\cdot\nabla)\frac{e^{-r/\lambda}}{4\pi m_1m_2r}}_{\V_3},
    \end{split}\\
    \begin{split}
    \label{vv2_gg}
        V_{\rm VV} &=
        \frac{1}{4m_1m_2}\nabla^2\Big[\underbrace{g^1_Vg^2_V}_{f_2}\underbrace{(\vec{\sigma}_1\cdot\vec{\sigma}_2)\frac{e^{-r/\lambda}}{4\pi r}}_{\V_2}\Big]\\
        &-\underbrace{\frac{g^1_Vg^2_V}{4}}_{f_3}\underbrace{(\vec{\sigma}_1\cdot\nabla)(\vec{\sigma}_2\cdot\nabla)\frac{e^{-r/\lambda}}{4\pi m_1m_2r}}_{\V_3},
    \end{split}\\
    \begin{split}
    \label{aa2_gg}
        V_{\rm AA} &=\underbrace{\Big[\lambda^2m_1m_2g^1_Ag^2_A-\frac{m_1^2+m^2_2}{8 m_1m_2}g^1_Ag^2_A}_{f_3}]\underbrace{(\vec{\sigma}_1\cdot\nabla)(\vec{\sigma}_2\cdot\nabla)\frac{e^{-r/\lambda}}{4\pi m_1m_2 r}}_{\V_3},
    \end{split}\\
    \begin{split}
    \label{sp2_vgg}
        V_{\rm SP} &=\underbrace{g^1_Sg^2_P\frac{m_2}{8(m_1+m_2)}}_{f_{15}}\underbrace{\{
    [\vec{\sigma}_1\cdot(\vec{v}\times\nabla)](\vec{\sigma}_2\cdot\nabla)+(\vec{\sigma}_1\cdot\nabla)[\vec{\sigma}_2\cdot(\vec{v}\times\nabla)]
    \}\frac{e^{-r/\lambda}}{4\pi m_1m_2 r}}_{\V_{15}},
    \end{split}
    \end{align}
    \end{subequations}
    \vspace{-0.5em}
    \end{itemize}
where $g^1_i$($g^2_i$) is the coupling constant for particle 1(2) at the $i$-type vertex, $\vec{\sigma}_{1}$ ($\vec{\sigma}_{2}$) is the spin operator for particle 1 (2), $\vec{v} = \vec{v}_2 - \vec{v}_1$ is the relative velocity between particle 2 and particle 1, and $\hat{r}$ is the unit vector in the direction of point from particle 1 to 2. The interaction range is given by $\lambda = 1/m$, where $m$ is the mass of the ALP, corresponding to its Compton wavelength.

We annotated the relevant $\V_i$ terms using underbraces in accordance with the notation of Ref.~\cite{Dobrescu2006JHEP}, allowing for a clear comparison of the coupling constant conversions across different potentials.
It is important to note that the coefficients of the velocity-dependent interactions in Eqs.~(\ref{ss0})-(\ref{sp2_vgg}) differ slightly from those presented in Ref.~\cite{Dobrescu2006JHEP}.
These discrepancies arise from additional factors such as $m_1/(m_1 + m_2)$ or $m_1^2/(m_1 + m_2)^2$, which reflect different definitions of the velocity $\vec{v}$.
Specifically, Ref.~\cite{Dobrescu2006JHEP} defines $\vec{v}$ as the velocity of the lighter particle in the center-of-mass frame, whereas in our treatment, it represents the average relative velocity between the two particles before and after the interaction.
Moreover, the coupling factor $\lambda^2 m_1 m_2 g^1_A g^2_A$ in interaction~(\ref{aa2_gg}) arises from the longitudinal mode of the massive {\color{black}$Z^\prime$} boson.
This contribution was not included in Ref.~\cite{Dobrescu2006JHEP} but has since been analyzed in later studies~\cite{Malta2016AHEP, Fadeev2019PRA}.

A summary of the classification of these interactions---including their discrete symmetries and other relevant properties---is provided in Table~\ref{tab1}.
We reiterate that our focus is on interactions at mesoscopic, macroscopic, or even astrophysical scales, corresponding to low-mass or ultra-low-mass ALPs.
As a result, the $\V_{14}$ term from Ref.~\cite{Dobrescu2006JHEP} does not contribute at the current level of approximation and is therefore omitted.
{\color{black}In addition, the $\mathcal{V}_{6+7}$ interaction~\cite{Dobrescu2006JHEP}, which arises from axial-vector and tensor couplings and introduces additional free parameters beyond $g_S$, $g_P$, $g_V$, and $g_A$, is omitted from this review for simplicity and conciseness, given the limited experimental exploration to date.}
At atomic scales, however, classical quantities such as the velocity $\vec{v}$ and distance $r$ must be treated as quantum operators.
Additionally, for identical particles, the interaction potentials must exhibit definite symmetry under particle index exchange.
These considerations are discussed in detail in Refs.~\cite{Fadeev2019PRA, Fadeev2018PhD, Ficek2017PRA}.
A complete list---including tensor-type interactions between ALPs and fermions, along with their derivations---is provided in Appendix~\ref{appendix:a}.

\begin{table*}[htbp]
    \centering
    \caption{\justifying Categorization of Eqs.~(\ref{ss0})--(\ref{sp2_vgg}) follows the normalization scheme proposed by Dobrescu and Mocioiu in Ref.~\cite{Dobrescu2006JHEP}. The table summarizes selected interactions along with their key properties. Each row indicates the transformation behavior under charge conjugation (C), parity (P), and time reversal (T)---denoted as even ($+$) or odd ($-$)---as well as whether the interaction is spin-dependent (Spin Dep.), the number of spins involved (Spin Order), and the number of gradient operators in the interaction ($\nabla$ Power). Additional columns specify the velocity-dependent (Vel. Dep.), the distance-scaling behavior ({\color{black}Scaling with} $r$), the relevant coupling constants, and the corresponding Lagrangian terms. A ``$+$'' in the ``Spin Dep.'' and ``Vel. Dep.'' columns indicate the presence of the respective feature, while a ``$-$'' indicates its absence. The ``{\color{black}Scaling with} $r$'' column refers to interactions that scale as $r^n$, and the ``$\nabla$ Power'' column indicates the number of spatial derivatives acting on the Yukawa-type potential.}
\label{tab1}
\resizebox{\linewidth}{!}{
\begin{tabular}{ccccccccccc}
\toprule
$V$&~C&~P&~T&~Spin Dep.&~Spin Order &~$\nabla$ power &~Vel. Dep. &~{\color{black}Scaling with $r$}&Coupling(s) &Lagrangian Term(s) \\
 \midrule
$\V_{1}$&+&+&+&$-$&0&$0$&$-$&$-1$&$g_{S}g_{S}$,$g_{V}g_{V}$&$\mathcal{L}_{\rm SP}$,$\mathcal{L}_{\rm VA}$  \\
$\V_{2}$&+&+&+&$+$&2&$0$&$-$&$-1$&$g_{A}g_{A}$&$\mathcal{L}_{\rm VA}$  \\

$\V_{3}$&+&+&+&$+$&2&$2$&$-$&$-3$&$g_{V}g_{V}$,$g_{A}g_{A}$,$g_{P}g_{P}$&$\mathcal{L}_{\rm SP}$,$\mathcal{L}_{\rm VA}$  \\
$\V_{4,5}$&+&+&+&$+$&1&$1$&$+$&$-2$&$g_{V}g_{V}$,$g_{A}g_{A}$,$g_{S}g_{S}$&$\mathcal{L}_{\rm SP}$,$\mathcal{L}_{\rm VA}$  \\
$\V_{8}$&+&+&+&$+$&2&$0$&$+$&$-1$&$g_{A}g_{A}$&$\mathcal{L}_{\rm VA}$  \\
$\V_{9,10}$&+&-&-&$+$&1&$1$&$-$&$-2$&$g_{S}g_{P}$&$\mathcal{L}_{\rm SP}$ \\
$\V_{11}$&-&-&+&$+$&2&$1$&$-$&$-2$&$g_{V}g_{A}$& $\mathcal{L}_{\rm VA}$  \\
$\V_{12,13}$&-&-&+&$+$&1&$0$&$+$&$-1$&$g_{V}g_{A}$&$\mathcal{L}_{\rm VA}$  \\
$\V_{15}$   &+&-&-&$+$&2&$2$& $+$&$-3$&$g_{S}g_{P}$& $\mathcal{L}_{\rm SP}$ \\
$\V_{16}$   &-&-&+&$+$&2&$1$&$+$&$-2$&$g_{V}g_{A}$& $\mathcal{L}_{\rm VA}$  \\
 \bottomrule
\end{tabular}
}
\end{table*}

\subsection{Non-Yukawa Spin-Dependent Interactions}\label{theo_2}

In addition to the well-studied Yukawa-type interactions, several classes of theoretically proposed spin-dependent interactions fall outside this framework. These non-Yukawa-type interactions arise from fundamentally different mechanisms and often reflect novel extensions to established physical theories. In this section, we categorize and summarize three major types of non-Yukawa spin-dependent interactions. The first class involves spin coupling directly to classical background fields, such as axion or torsion fields, which may arise from dark matter, extensions of general relativity, or violations of Lorentz and CPT symmetries. The second class comprises interactions mediated by scale-invariant fields known as unparticles, which generate unconventional, non-integer power-law potentials. 
The third class includes interactions induced by bilinear scalar couplings at the quantum level, as recently proposed in theoretical studies. These categories encompass many of the leading alternatives to Yukawa-type interactions and are of growing interest in both theoretical and experimental research.

\subsubsection{Interactions in the Form of Spin Coupling to a Classical Field}\label{theo_3}
\quad\\                      
\noindent\textit{\S~1.1~Axion Field}\\
Axions produced in the early universe may manifest as a classical field~\cite{Preskill1983PLB, Abbott1983PLB, Dine1983PLB}.
This axion field spreads throughout space and, for example, can interact with nucleons via the coupling~\cite{Dror2023PRL, graham2013PRD}:
\begin{equation}
\label{alpfield}
\mathcal{L}=\textcolor{black}{\mathcal{L}_{\partial\phi}}
+
\mathcal{L}_{\rm EDM},
\end{equation}
where
\begin{equation}
\textcolor{black}{\mathcal{L}_{\partial\phi}}=
\textcolor{black}{g_{\partial\phi NN}}\partial_\mu \phi
\bar{\psi}_N\gamma^\mu\gamma_5\psi_N ,
\end{equation}
and
\begin{equation}
\mathcal{L}_{\rm EDM}=-\frac{i}{2}g_{\rm EDM}\phi
\bar{\psi}_N\sigma^{\mu\nu}\gamma_5\psi_N F_{\mu\nu}.
\end{equation}
\textcolor{black}{Here, $\psi_{N}$ denotes the nucleon field, $F_{\mu\nu}$ is the electromagnetic field tensor, and $\phi$ is the scalar ALP field. The two terms in Eq.~(\ref{alpfield}) represent two distinct effective interactions: $\mathcal{L}_{\partial\phi}$ describes the derivative coupling between the ALP field and the nucleon axial current, while $\mathcal{L}_{\rm EDM}$ describes the axion-induced EDM-type coupling. The corresponding coupling constants $g_{\partial\phi NN}$ and $g_{\rm EDM}$ have mass dimensions $-1$ and $-2$, respectively, in natural units, so that both terms have the correct mass dimension for a Lagrangian density.}
In the non-relativistic limit, this interaction leads to effective magnetic and electric dipole couplings described by:
\begin{equation}
\label{eq5}
V = -\textcolor{black}{g_{\partial\phi NN}}\nabla \phi \cdot \vec{\sigma}-g_{\rm EDM}\phi\vec{E}\cdot \vec{\sigma},
\end{equation}
where $\vec{\sigma}$ is the nucleon spin operator and $\vec{E}$ is the electric field experienced by the nucleon.
This interaction can be interpreted as an axion-induced effective magnetic field,
\begin{equation}
\vec{B}^\prime=-\frac{2}{\gamma}(\textcolor{black}{g_{\partial\phi NN}}\nabla\phi+g_{\rm EDM}\phi\vec{E})
\end{equation}
acting on the nucleon spin, where $\gamma$ is the nucleon's gyromagnetic ratio.
The second term in Eq.~(\ref{eq5}) can also be understood as an axion-induced nuclear EDM interacting with an applied electric field $\vec{E}$.
Experimental searches for ALPs can be categorized into those targeting the tiny effective field~\cite{smorra2019N, Wu2019PRL, bloch2020JHEP, Kimball2020, Jiang2021bNP, Bloch2022SA, Lee2023PRX, Bloch2023NC, Dror2023PRL, Fierlinger2024PRL, zhang2025arXiv} and those aimed at detecting an oscillating nucleon EDM~\cite{Graham2011PRD, Abel2017PRX, Chu2019PRD, diluzio2024JHEP}.

One popular production mechanism of the ALP field is the so-called misalignment mechanism that \textcolor{black}{occurred} in the early Universe~\cite{Preskill1983PLB, Abbott1983PLB, Dine1983PLB}.
For the ALP masses below $1~\mathrm{eV}$ and when they comprise a significant fraction of the dark matter, the occupation number must be extremely large, allowing quantum fluctuations to be ignored.
Therefore, the ALP field can be viewed as an oscillating classical field given by
\begin{equation}
\phi(\vec{r},t)=a_0\cos(\vec{k}\cdot\vec{r}+\omega t+\varphi),
\end{equation}
where $a_0$ is the amplitude, $\vec{k}=m_a\vec{v}$ and $\omega\approx m_a$ \textcolor{black}{are} the wave vector and Compton frequency, \textcolor{black}{respectively}, and $\varphi$ is the random initial phase uniformly distributed in $2\pi$.
The field amplitude $a_0$ can be estimated by the Galactic dark-matter energy density $\rho_a = m_a^2a_0^2c^2/2\approx 0.3~{\rm GeV}/{\rm cm^3}$~\cite{Riccardo2010JCAP}.
Then, the spatial components of the derivative coupling of the axion field with nuclear spins simplify as follows:
\begin{equation}
\label{Hint}
V= \textcolor{black}{g_{\partial\phi NN}} \sqrt{2\rho_a} \sin(\vec{k}\cdot\vec{r}+\omega t+\varphi) \vec{v} \cdot \vec{\sigma},
\end{equation}
where $\vec{v}$ represents the expected average axion wind velocity, which in the galactic reference frame is $|\vec{v}|\sim 10^{-3}$~\cite{Centers2021NC, Kimball2020}.
Note that the phase $\varphi$ has occasionally been omitted in some of the previous literature~\cite{Jiang2021bNP, Kimball2020, graham2013PRD}, while it could be critically important when the axion field frequency under study is extremely low.\\

\noindent\textit{\S~1.2~Interactions Induced by Lorentz Violation and CPT Violation}\\

 Lorentz invariance states that the laws of physics are identical for all observers, irrespective of their relative motion or orientation in spacetime.
 This principle serves as a foundation for both the SM of particle physics and GR.
 A violation of Lorentz invariance would imply that spacetime possesses preferred directions, speeds, or reference frames, fundamentally challenging our understanding of physics.
 Moreover, the CPT theorem, a cornerstone of quantum field theory, asserts that all local, Lorentz-invariant quantum field theories must preserve CPT symmetry~\cite{Luders1954, Pauli1988}.
 Consequently, any observed violation of CPT symmetry would signal physics beyond the SM, questioning the foundational principles of quantum field theory and relativity.

The link between Lorentz violation and CPT violation is deeply connected to the CPT theorem. The theorem establishes that CPT symmetry---a combination of C, P, and T---is guaranteed in any local Lorentz-invariant theory.
If Lorentz invariance is violated, the conditions required for the CPT theorem break down, allowing for the possibility of CPT violation.
However, the relationship is not strictly bidirectional: while CPT violation generally implies Lorentz violation, Lorentz violation need not imply CPT violation~\cite{Greenberg2002PRL}.

Kosteleck\'{y} and collaborators developed a formalism encompassing all possible Lorentz-violating terms~\cite{Colladay1998PRD, Kostelecky2001PRD}. It bridges theoretical predictions with experimental observations. {\color{black} This framework was further expanded to incorporate the gravitational sector~\cite{Kostelecky2004PRD}, an area that has since been extensively investigated through various theoretical and experimental studies~\cite{Kostelecky2011RMP, Bailey2013PRD, Xiao2020JPG, Ivanov2021PLB, Ye2022PRD}.} One notable example from this framework is a background field coupling to the spin of a particle, expressed as 
\begin{equation}
V = -\vec{b} \cdot \vec{\sigma},
\end{equation}
where \( \vec{b} \) is a background field that is odd under CPT.

The Standard Model Extension (SME) provides a systematic framework for incorporating Lorentz- and CPT-violating terms into the particle-physics Lagrangian.
Spin-dependent terms are particularly interesting because they directly influence the dynamics of fermions, which can be probed experimentally.
According to the SME, there can be CPT-odd terms that violate both CPT and Lorentz invariance~\cite{Kostelecky1999PRD}:
\begin{equation}
\mathcal{L} = - b_\mu \bar{\psi} \gamma^5 \gamma^\mu \psi + \frac{1}{2} i g_{\mu\nu\lambda} \bar{\psi} \sigma^{\mu\nu} \gamma^5 \partial^\lambda \psi,
\end{equation}
where $b_\mu$ represents the background axial vector field, and $g_{\mu\nu\lambda}$ coefficients describe a tensor coupling that modifies the spin dynamics of fermions.

Additionally, there can be CPT-even terms that are Lorentz-violating but CPT-invariant~\cite{Kostelecky1999PRD}:
\begin{equation}
\mathcal{L} =  - \frac{1}{2} H_{\mu\nu} \bar{\psi} \sigma^{\mu\nu} \psi +id_{\mu\nu} \bar{\psi} \gamma^5 \gamma^\mu \partial^\nu \psi,
\end{equation}
where $H_{\mu\nu}$ coefficients represent a {\color{black} tensor} background field that couples to the spin-tensor current. The $d_{\mu\nu}$ are coefficients that introduce a spin-dependent coupling to the derivatives of the fermion field. 
These terms provide a systematic framework for exploring Lorentz and CPT symmetry violations, with distinct experimental signatures arising from each type of term.

In the non-relativistic limit, these terms reduce to effective potentials involving spin, momentum, and external fields. 
The leading-order low-energy approximations for the terms involving spin above are~\cite{Kostelecky1999PRD}:
\begin{equation}
\begin{aligned}
{V}_b &= - \vec{b} \cdot \vec{\sigma},\\
{V}_d &=  m\vec{d} \cdot \vec{\sigma},\\
{V}_g &= -m\vec{g}\cdot\vec{\sigma},\\
{V}_H &= \vec{H} \cdot \vec{\sigma},
\end{aligned}
\end{equation}
where $m$ is the mass of the fermion, $\vec{\sigma}$ is its spin, and $\vec{b}$, $\vec{d}$, $\vec{g}$, and $\vec{H}$ are spacelike components of these Lorentz-violating fields, taking the form $\vec{b}=(b_1, b_2, b_3)$, $\vec{d}=(d_{10}, d_{20}, d_{30})$, $\vec{g}=(g_{230}, g_{310}, g_{120})$, and $\vec{H}=(H_{23}, H_{31}, H_{12})$, respectively.\\

\noindent\textit{\S~1.3~Torsion and Spin-gravity (Spin-mass) Interactions}\\

Einstein's theory of GR describes gravity as a manifestation of spacetime curvature induced by matter, which in turn dictates the motion of matter. 
GR’s success has motivated physicists to investigate spacetime geometry further, prompting interest in torsion---a geometric property distinct from curvature that also characterizes spacetime and its interaction with matter.
In standard GR, torsion vanishes, and gravity is fully described by curvature alone.
However, numerous theoretical extensions of GR introduce torsion, usually arising from spin sources~\cite{Lehnert2014PLB}.
Although torsion couples to spin as naturally as curvature couples to energy-momentum, observable spin sources capable of generating measurable torsion remain elusive~\cite{Hehl1976RMP}. Furthermore, the unknown and potentially short range of torsion makes detection exceedingly challenging~\cite{Lehnert2014PLB}. 
These factors have historically discouraged experimental searches for torsion effects.

One more intuitive illustration of such a connection is the spin-gravity interactions that have a general form of 
\begin{equation}
    V=\chi\vec{\sigma}\cdot\vec{g},
\end{equation}
where $\chi$ is the coupling constant, $\vec{\sigma}$ is the particle spin, and $\vec{g}$ is the gravitational acceleration. The spin-0 particle with scalar and pseudoscalar coupling to the matter also produces the spin-gravity interaction (see Eq.~(\ref{sp1_g})). Since the spin-gravity interaction violates the P and T symmetries, it is also a starting point to test the fundamental symmetries in the gravitational interaction~\cite{Leitner1964PR, Dass1976PRL}. Moreover, under the assumption of CPT invariance, the most general forms of spin-gravity interactions that violate P and T, as well as P and C symmetries, were proposed by Hari-Dass \textit{et al.}~\cite{Dass1976PRL, haridass1977GRG}:
\begin{equation}
    V=\frac{GM}{2r^2}\big[\alpha_1\vec{\sigma}\cdot\hat{r}+\alpha_2\vec{\sigma}\cdot\vec{v}+\alpha_3\vec{\sigma}\cdot(\vec{v}\times \hat{r})\big],
\end{equation}
where $G$ is Newton's gravitational constant, $M$ is the mass of the gravity source, $\hat{r}$ is the unit vector of the particle relative to the gravity source, and $\alpha_1$ ($\alpha_2$, $\alpha_3$) is a dimensionless constant.

\textcolor{black}{It is useful to compare the phenomenological coefficients in the above equation with preferred-frame spin couplings constrained in the Standard-Model Extension (SME). In the framework of the SME developed by Kosteleck\'{y} \textit{et al.}, various effective potentials involving particle spins have been formulated~\cite{Kostelecky1999PRD,Kostelecky2018PRD}. Lorentz- and CPT-violating spin couplings for a fermion species $w$ are commonly expressed in terms of effective nonrelativistic coefficients such as $\tilde{b}^{w}_{J}$, $\tilde{d}^{w}_{J}$, and related tilde combinations, where $J=X,Y,Z$ denotes axes in the Sun-centered celestial-equatorial frame~\cite{Kostelecky1999PRD,Kostelecky2004PRD,Kostelecky2011RMP}. These coefficients generate anomalous spin-precession signals that have been precisely tested in a wide range of clock-comparison experiments, including atomic clocks and comagnetometers, which currently provide some of the most stringent sensitivities to Lorentz- and CPT-violating effects.}

\textcolor{black}{The three structures in the Hari-Dass interaction can be viewed phenomenologically as different preferred-direction or preferred-frame spin couplings: the $\alpha_1$ term couples the spin to the source direction $\hat{r}$, the $\alpha_2$ term couples the spin to the relative velocity $\vec{v}$, and the $\alpha_3$ term couples the spin to the transverse direction $\vec{v}\times\hat{r}$. Given the structural similarity between spin-gravity interactions and SME-induced preferred-frame effects, the corresponding experimental signatures may be difficult to distinguish solely from spin-precession data, since both can lead to sidereal or annual modulations as the laboratory frame rotates and is boosted relative to the Sun-centered frame. Therefore, clock-comparison and comagnetometer experiments can also be used to constrain or benchmark torsion and spin-gravity interactions with high precision. The mapping is not generally one-to-one, because the SME coefficients are defined in a general effective-field-theory framework, whereas $\alpha_1$, $\alpha_2$, and $\alpha_3$ parameterize specific gravitationally motivated spin-mass couplings.}

\textcolor{black}{Comagnetometers provide some of the most stringent constraints on such preferred-frame spin couplings. For example, $^{3}$He/$^{129}$Xe comagnetometers have constrained the equatorial component of the neutron spin coupling to the level of $\tilde{b}^{n}_{\perp}<8.4\times10^{-34}~{\rm GeV}$~\cite{Allmendinger2014PRL}. Alkali-metal/noble-gas comagnetometers, such as K--$^{3}$He and K--Rb--$^{21}$Ne systems, provide complementary constraints on neutron-sector vector and tensor SME coefficients, including boost-dependent and quadrupolar combinations involving $\tilde{d}^{n}_{J}$ and related coefficients~\cite{Brown2010PRL, Smiciklas2011PRL}. These limits serve as important benchmark sensitivities for interpreting spin-gravity and spin-mass searches in terms of preferred-frame spin interactions, although a precise conversion to bounds on $\alpha_1$, $\alpha_2$, and $\alpha_3$ requires specifying the source geometry, relative velocity, and the underlying model.}

The spin-gravity interaction also provides a possible origin of the violation of the Einstein equivalence principle (EEP). For example, if the EEP is violated, particles with different spin orientations would experience different forces in free fall~\cite{Duan2016PRL}.

In a gravitational field, general relativity predicts two key effects: the Lense-Thirring effect (frame-dragging) and the de Sitter effect (geodetic precession), collectively referred to as gravitomagnetic effects~\cite{Schiff1960PRL}. These effects cause additional precession of a classical spin system---such as a gyroscope---orbiting a massive body like the Earth, as confirmed by the Gravity Probe B experiment~\cite{Everitt2011PRL}.
Analogously to the EP, one would expect a similar precessional effect for a particle with intrinsic spin in a gravitational field. For a spin bound to an Earth-based orbit, its precession under these two effects can be described by~\cite{Fadeev2021PRD}:
\begin{equation}
\label{gs}
V=g\bigg\{\frac{1}{2}\frac{G}{r^3}[3(\vec{J}\cdot\hat{r})\hat{r}-\vec{J}]+\frac{3}{4}\frac{GM}{r^2}(\hat{r}\times \vec{v})\bigg\}\cdot\vec{\sigma},
\end{equation}
where $\hat{r}$ is the unit vector pointing from the particle to the center of the Earth, $M$ is the Earth's mass, $\vec{J}$ denotes the Earth's angular momentum, and $\vec{v}$ is the velocity of the particle.
The gyro-gravitational ratio $g$ characterizes the coupling between intrinsic spin and gravity; if the EEP holds, $g = 1$.
The first term in Eq.~(\ref{gs}) appears in multiple theoretical frameworks and is commonly referred to as the gravitomagnetic effect~\cite{Adler2012PRD, Audretsch1981PRD, Kostelecky2011PRD, Ye2022PRD}.
A possible experimental approach using a ferromagnetic gyroscope has been proposed in Ref.~\cite{Fadeev2021PRD} to detect the general-relativistic precession of intrinsic spin.\\

\noindent\textit{\S~1.4~The EDMs}\\

In his early work on the Dirac equation for electrons~\cite{dirac1928}, Dirac noted the presence of an imaginary interaction term:
\begin{equation} 
\sim ie\vec{\sigma}\cdot\vec{E} 
\end{equation}
where $e$ is the electron charge, and $\vec{E}$ is the electric field. This term appears alongside the $\vec{\sigma}\cdot\vec{B}$ term, which describes the coupling of the electron's magnetic moment to the magnetic field. The latter is considered one of the triumphs of Dirac's theory, as it correctly predicts the electron's gyromagnetic ratio, a longstanding puzzle at the time. Regarding the $\vec{\sigma}\cdot\vec{E}$ term, Dirac initially interpreted it as representing the EDM of the electron. However, due to its purely imaginary nature, he speculated that it might not have any physical significance.

The electron's EDM is closely related to spin-dependent interactions. On one hand, the interaction between the EDM and an external electric field can be expressed as~\cite{Purcell1950PR}:
\begin{equation} 
V_\text{EDM} = -d_e \vec{\sigma} \cdot \vec{E} \end{equation}
Where $d_e$ is the magnitude of the electron's EDM, in this form, the interaction manifests as a coupling between the spin $\vec{\sigma}$ and the electric field $\vec{E}$. On the other hand, this interaction inherently breaks both P and T symmetries, similar to the scalar-pseudoscalar coupling. Furthermore, when considering the electric vertex for fundamental particles such as the muon and the neutron, these interactions can induce EDMs for these particles.

The electron EDM and neutron EDMs serve as highly sensitive probes for new physics beyond the SM. From the perspective of spin-dependent interactions, these searches can be interpreted as investigations into exotic couplings between spin and electric fields. Notably, SP couplings---whether of Yukawa or non-Yukawa type---can, in principle, induce intrinsic EDMs in fundamental particles as discussed in Ref.~\cite{Yan2019EPJC}. Therefore, precision measurements of particle EDMs provide powerful constraints on SP-type interactions.

\begin{table*}[htbp]
    \centering
    \caption{\justifying Categorization of the interactions mediated by unparticles follows the normalization scheme proposed by Dobrescu and Mocioiu in Ref.~\cite{Dobrescu2006JHEP}. The table presents selected interactions along with their key properties. Each row indicates the transformation behavior under charge conjugation (C), parity (P), and time reversal (T)---denoted as even ($+$) or odd ($-$)---as well as whether the interaction is spin-dependent (Spin Dep.), the number of spins involved (Spin Order), and the number of gradient operators in the interaction ($\nabla$ Power). Additional columns specify the velocity-dependent (Vel. Dep.), the distance-scaling behavior ({\color{black}Scaling with} $r$), the relevant coupling constants, and the corresponding Lagrangian terms. A ``$+$'' in the ``Spin Dep.'' and ``Vel. Dep.'' columns denote the presence of the corresponding feature, while a ``$-$'' indicates its absence. The ``{\color{black}Scaling with} $r$'' column refers to interactions scaling as $r^n$, and the ``$\nabla$ Power'' column quantifies the number of spatial derivatives applied to the original potential.}
    \label{tab2}
    \resizebox{\linewidth}{!}{
    \begin{tabular}{ccccccccccc}
  \toprule
 $V$&~C&~P&~T     &~Spin Dep.        &~Spin Order &~$\nabla$ power &~Vel. Dep. &~{\color{black}Scaling with $r$}&Coupling(s) &Lagrangian Term(s)    \\
 \midrule
$\V_{1}$&+&+&+&$-$&0&$0$&$-$&$1-2d_\mathcal{U}$&$C_{S}C_{S}$,$C_{V}C_{V}$&$\mathcal{L}_{\rm SP}$,$\mathcal{L}_{\rm VA}$  \\
$\V_{2}$&+&+&+&$+$&2&$0$&$-$&$1-2d_\mathcal{U}$&$C_{A}C_{A}$&$\mathcal{L}_{\rm VA}$  \\
$\V_{3}$&+&+&+&$+$&2&$2$&$-$&$-1-2d_\mathcal{U}$&$C_{V}C_{V}$,$C_{A}C_{A}$,$C_{P}C_{P}$&$\mathcal{L}_{\rm SP}$,$\mathcal{L}_{\rm VA}$  \\
$\V_{4,5}$&+&+&+&$+$&1&$1$&$+$&$-2d_\mathcal{U}$&$C_{V}C_{V}$,$C_{A}C_{A}$,$C_{S}C_{S}$&$\mathcal{L}_{\rm SP}$,$\mathcal{L}_{\rm VA}$  \\
$\V_{8}$&+&+&+&$+$&2&$0$&$+$&$1-2d_\mathcal{U}$&$C_{A}C_{A}$&$\mathcal{L}_{\rm VA}$  \\
$\V_{9,10}$&+&-&-&$+$&1&$1$&$-$&$-2d_\mathcal{U}$&$C_{S}C_{P}$&$\mathcal{L}_{\rm SP}$ \\
$\V_{11}$&-&-&+&$+$&2&$1$&$-$&$-2d_\mathcal{U}$&$C_{V}C_{A}$& $\mathcal{L}_{\rm VA}$  \\
$\V_{12,13}$&-&-&+&$+$&1&$0$&$+$&$1-2d_\mathcal{U}$&$C_{V}C_{A}$&$\mathcal{L}_{\rm VA}$  \\
$\V_{15}$   &+&-&-&$+$&2&$2$&$+$&$-1-2d_\mathcal{U}$&$C_{S}C_{P}$& $\mathcal{L}_{\rm SP}$ \\
$\V_{16}$   &-&-&+&$+$&2&$1$&$+$&$-2d_\mathcal{U}$&$C_{V}C_{A}$& $\mathcal{L}_{\rm VA}$  \\
 \bottomrule
\end{tabular}
}
\end{table*}
\subsubsection{Interactions Mediated by Unparticles}\label{theo_22}

Another theoretical motivation for searching for spin-dependent interactions arises from the concept of unparticles~\cite{Georgi2007PRL}, which incorporates scale invariance into the SM. In this framework, a hidden sector---such as a Banks-Zaks field with a non-trivial infrared (IR) fixed point~\cite{Banks1982NPB}---interacts with the SM via exchange of heavy mediator particles. Below some energy scale $\Lambda$, the hidden sector flows toward an IR fixed point, where its interaction with SM fields can be described within the framework of effective field theory (EFT) as $\mathcal{O}_\mathcal{U}\mathcal{O}_\mathcal{SM}$. Here, $\mathcal{O}_\mathcal{U}$ and $\mathcal{O}_\mathcal{SM}$  are operators associated with the unparticle and SM fields, respectively, and $\mathcal{O}_\mathcal{U}$ exhibits scale invariance.
Unlike SM fields, the excitations of unparticle fields cannot be described in terms of conventional particles, as they do not possess a definite mass and are not constrained by standard dispersion relations. This unusual feature prevents unparticles from being detected by traditional collider-based searches. If unparticles exist, their signatures would manifest either as missing energy or momentum in high-energy collisions, or via exotic long-range interactions between SM particles at low energies.
Various experiments have investigated such exotic interactions mediated by unparticles, including both spin-independent interactions~\cite{Goldberg2008PRL, Deshpande2008PLB, Bertolami2009PRD, Wondrak2016PLB, Poddar2022EPJC, Garcia2008JCAP} and spin-dependent interactions~\cite{Liao2007PRL, Wu2024JHEP}. Focusing on the coupling between unparticles and fermions, the leading terms in the effective field theory can be written as~\cite{Wu2024JHEP}:
\begin{equation}
\label{eq15}
\begin{aligned}
\mathcal{L}&=\underbrace{(C_S\bar{\psi}\psi+C_P\bar{\psi}i\gamma_5\psi)\Phi_\mathcal{U}}_{\mathcal{L}_{\rm SP}}\\  &+\underbrace{(C_V\bar{\psi}\gamma_\mu\psi+C_A\bar{\psi}\gamma_\mu\gamma_5\psi)X^\mu_\mathcal{U}}_{\mathcal{L}_{\rm VA}},
\end{aligned}
\end{equation}
where $\Phi_\mathcal{U}$ and $X^\mu_\mathcal{U}$ denote the scalar and vector unparticle fields, respectively, each with an energy dimension of $d_\mathcal{U}$. The parameters $C_{S, P, V, A}$ represent the corresponding coupling constants. These couplings can be expressed as $C_i = c_i \Lambda^{1-d_\mathcal{U}}$, where $c_i$ is a dimensionless coefficient and $\Lambda$ is the energy scale characterizing the effective interaction. A typical choice is $\Lambda\sim 1$ TeV, allowing $c_i$ to be constrained experimentally.
In the non-relativistic limit, the Lagrangian in Eq.~(\ref{eq15}) induces fermion-fermion interactions mediated by unparticles, as illustrated in the Feynman diagram in Fig.~\ref{fig3}. From these interactions, six distinct types of spin-dependent potentials---SS, SP, PP, VV, VA, and AA---can be derived. The corresponding coordinate-space potentials, expanded up to $\mathcal{O}(v^{2})$ and classified according to the same scheme introduced earlier, are presented below.

\paragraph{Spin Order-0}
\begin{itemize}[left=0pt,itemsep=2pt,topsep=2pt]
    \item $\nabla^0~\text{Potentials}$\\
    \vspace{-0.5em}
    \begin{subequations}
    \begin{align}
    \begin{split}
        \label{uss0}
        V_{\rm SS}(r)&=\underbrace{-C^1_SC^2_S\frac{A_{d_\mathcal{U}}}{\pi}\Gamma(2d_\mathcal{U}-2)}_{f_1}\underbrace{\frac{1}{4\pi r^{2d_\mathcal{U}-1}}}_{\V_1},
    \end{split}\\
    \begin{split}
        \label{uvv0}
        V_{\rm VV}(r)&=\underbrace{C^1_VC^2_V\frac{A_{d_\mathcal{U}}}{\pi}\Gamma(2d_\mathcal{U}-2)}_{f_1}\underbrace{\frac{1}{4\pi r^{2d_\mathcal{U}-1}}}_{\V_1},
    \end{split}
    \end{align}
    \end{subequations}
    \vspace{-0.5em}
\end{itemize}
\paragraph{Spin Order-1}
\begin{itemize}[left=0pt,itemsep=2pt,topsep=2pt]
    \item $\nabla^0~\text{Potentials}$\\
    \vspace{-0.5em}
    \begin{equation}
    \label{uva1_v}
        V_{\rm VA}(r)=\underbrace{C^1_VC_A^2\frac{A_{d_\mathcal{U}}}{\pi}\Gamma(2d_\mathcal{U}-2)}_{f_{12+13}}\underbrace{\frac{\vec\sigma_2\cdot\vec v}{4\pi r^{2d_\mathcal{U}-1}}}_{\V_{12+13}},
    \end{equation}
    \vspace{-0.5em}
\end{itemize}
\begin{itemize}[left=0pt,itemsep=2pt,topsep=2pt]
    \item $\nabla^1~\text{Potentials}$\\
    \vspace{-0.5em}
    \begin{subequations}
    \begin{align}
    \begin{split}
    \label{usp1_g}
        V_{\rm SP}(r)&= \underbrace{C_S^1C_P^2\frac{A_{d_\mathcal{U}}}{2\pi}\Gamma(2d_\mathcal{U}-2)}_{f_{9+10}}\underbrace{\vec\sigma_2\cdot\nabla\frac{1}{ 4\pi m_2 r^{2d_\mathcal{U}-1}}}_{\V_{9+10}},
    \end{split}\\
    \begin{split}
    \label{uss1_vg}
         V_{\rm SS}(r)&= \underbrace{-C_S^1C_S^2\frac{A_{d_\mathcal{U}}}{\pi}\Gamma(2d_\mathcal{U}-2)\frac{m_1}{4(m_1+m_2)}}_{f_{4+5}}\underbrace{\vec\sigma_2\cdot( \vec v \times \nabla)\frac{1}{4\pi m_2r^{2d_\mathcal{U}-1}}}_{\V_{4+5}},
    \end{split}\\
    \begin{split}
    \label{uvv1_vg}
        V_{\rm VV}(r)&=\underbrace{-C_V^1C_V^2\frac{A_{d_\mathcal{U}}}{\pi}\Gamma(2d_\mathcal{U}-2)\frac{m_1+2m_2}{4(m_1+m_2)}}_{f_{4+5}}\underbrace{\vec\sigma_2\cdot(\vec{v}\times \nabla)\frac{1}{4\pi m_2r^{2d_\mathcal{U}-1}}}_{\V_{4+5}},
    \end{split}\\
    \begin{split}
    \label{uaa1_vg}
        V_{\rm AA}(r) &=\underbrace{-C_A^1C_A^2\frac{A_{d_\mathcal{U}}}{\pi}\Gamma(2d_\mathcal{U}-2)\frac{m^2_2}{4m_1(m_1+m_2)}}_{f_{4+5}}\underbrace{\vec\sigma_2\cdot(\vec v\times\nabla)\frac{1}{4\pi m_2 r^{2d_\mathcal{U}-1}}}_{\V_{4+5}},
    \end{split}
    \end{align}
    \end{subequations}
    \vspace{-0.5em}
\end{itemize}
\paragraph{Spin Order-2}
\begin{itemize}[left=0pt,itemsep=2pt,topsep=2pt]
    \item $\nabla^0~\text{Potentials}$\\
    \vspace{-0.5em}
    \begin{subequations}
    \begin{align}
    \begin{split}
        \label{uaa2_0}
            V_{\rm AA}(r)=\underbrace{- C_A^1C_A^2\frac{A_{d_\mathcal{U}}}{\pi}\Gamma(2d_\mathcal{U}-2)}_{f_2}\underbrace{\frac{\vec\sigma_1\cdot\vec\sigma_2}{4\pi r^{2d_\mathcal{U}-1}}}_{\V_2},
        \end{split}\\
    \begin{split}
        \label{uaa2_vv}
            V_{\rm AA}(r)&=-\underbrace{C_A^1C_A^2\frac{A_{d_\mathcal{U}}}{2\pi}\Gamma(2d_\mathcal{U}-2)}_{f_8}\underbrace{\frac{(\vec\sigma_1\cdot\vec v)(\vec\sigma_2\cdot\vec v)}{4\pi r^{2d_\mathcal{U}-1}}}_{\V_8}
        \end{split}
    \end{align}
    \end{subequations}
    \vspace{-0.5em}
\end{itemize}
\begin{itemize}[left=0pt,itemsep=2pt,topsep=2pt]
    \item $\nabla^1~\text{Potentials}$\\
    \vspace{-0.5em}
    \begin{subequations}
    \begin{align}
    \begin{split}
    \label{uva2_g}
        V_{\rm VA}(r)&=\underbrace{-C^1_VC^2_A\frac{A_{d_\mathcal{U}}}{2\pi}\Gamma(2d_\mathcal{U}-2)}_{f_{11}}\underbrace{(\vec\sigma_1\times \vec \sigma_2)\cdot \nabla\frac{1}{4\pi m_1r^{2d_\mathcal{U}-1}}}_{\V_{11}},
    \end{split}\\
    \begin{split}
    \label{uva2_vvg}
        V_{\rm VA}(r)&=\underbrace{\frac{m_1}{8(m_1+m_2)}C_V^1C^2_A\frac{A_{d_\mathcal{U}}}{\pi}\Gamma(2d_\mathcal{U}-2)}_{f_{16}}\\
        &\underbrace{\{[\vec{\sigma}_1\cdot(\vec{v}\times \nabla)](\vec{\sigma}_2\cdot\vec{v})+(\vec{\sigma}_1\cdot\vec{v})[\vec{\sigma}_2\cdot(\vec{v}\times \nabla)]\}\frac{1}{4\pi m_1r^{2d_\mathcal{U}-1}}}_{\V_{16}},
    \end{split}
    \end{align}
    \end{subequations}
    \vspace{-0.5em}
\end{itemize}
\begin{itemize}[left=0pt,itemsep=2pt,topsep=2pt]
    \item $\nabla^2~\text{Potentials}$\\
    \vspace{-0.5em}
    \begin{subequations}
    \begin{align}
    \begin{split}
    \label{upp2_gg}
        V_{\rm PP}(r) &=\underbrace{C_P^1C_P^2\frac{A_{d_\mathcal{U}}}{4\pi}\Gamma(2d_\mathcal{U}-2)}_{f_3}\underbrace{(\vec\sigma_1\cdot\nabla)(\vec\sigma_2\cdot\nabla)\frac{1}{4\pi m_1m_2 r^{2d_\mathcal{U}-1}}}_{\V_3},
    \end{split}\\
    \begin{split}
    \label{uvv2_gg}
        V_{\rm VV}(r)&=\frac{1}{4m_1m_2}\nabla^2\Big[\underbrace{C_V^1C_V^2\frac{A_{d_\mathcal{U}}}{\pi}\Gamma(2d_\mathcal{U}-2)}_{f_2}\underbrace{\vec\sigma_1\cdot\vec\sigma_2\frac{1}{4\pi r^{2d_\mathcal{U}-1}}}_{\V_2}\Big]\\
        &-\underbrace{C_V^1C_V^2\frac{A_{d_\mathcal{U}}}{4\pi}\Gamma(2d_\mathcal{U}-2)}_{f_3}\underbrace{(\vec\sigma_1\cdot\nabla)(\vec\sigma_2\cdot\nabla)\frac{1}{4\pi m_1m_2r^{2d_\mathcal{U}-1}}}_{\V_3},
    \end{split}\\
    \begin{split}
    \label{uaa2_gg}
        V_{\rm AA}(r)&=-\underbrace{C_A^1C_A^2\frac{A_{d_\mathcal{U}}}{\pi}\Gamma(2d_\mathcal{U}-2)\frac{m_1^2+m_2^2}{8m_1m_2}}_{f_3}\underbrace{(\vec\sigma_1\cdot\nabla)(\vec\sigma_2\cdot\nabla)\frac{1}{4\pi m_1m_2r^{2d_\mathcal{U}-1}}}_{\V_3},
    \end{split}\\
    \begin{split}
    \label{usp2_vgg}    
    V_{\rm SP}(r)&=\underbrace{C_S^1C_P^2\frac{A_{d_\mathcal{U}}}{8\pi}\Gamma(2d_\mathcal{U}-2)\frac{m_2}{m_1+m_2}}_{f_{15}}\\
    &\underbrace{\{[\vec{\sigma}_1\cdot(\vec{v}\times\nabla)](\vec{\sigma}_2\cdot\nabla)+(\vec{\sigma}_1\cdot\nabla)[\vec{\sigma}_2\cdot(\vec{v}\times\nabla)]\}\frac{1}{4\pi m_1m_2r^{2d_\mathcal{U}-1}}}_{\V_{15}}.
    \end{split}
    \end{align}
    \end{subequations}
    \vspace{-1 em}
\end{itemize}

Here, $A_{d_\mathcal{U}} = \frac{16\pi^{5/2}}{(2\pi)^{2d_\mathcal{U}}} \frac{\Gamma(d_\mathcal{U} + 1/2)}{\Gamma(d_\mathcal{U} - 1)\Gamma(2d_\mathcal{U})}$, where $\Gamma(x)$ denotes the Gamma function, and $C^1_i$ ($C^2_i$) represents the coupling constant for particle 1 (2) at vertex type $i$. We highlight the spin-dependent structures using underbraces beneath each term and label the coupling coefficients and potentials according to the notation used in Ref.~\cite{Dobrescu2006JHEP}.
We categorize the unparticle-induced potentials using the same scheme as that applied to the Yukawa-type interactions, as summarized in Table~\ref{tab2}.

As with interactions mediated by ALPs, these unparticle-mediated potentials must be symmetrized for identical particles, and both distance and momentum should be treated as quantum operators at the atomic scale. A detailed derivation of these potentials is provided in Ref.~\cite{Wu2024JHEP}.
Notably, these interactions exhibit a non-integer power-law dependence on the distance between the interacting particles. The energy dimension $d_\mathcal{U}$, which can take continuous values in the range $[1,2]$, significantly modifies the interaction behavior, especially at atomic scales~\cite{Wondrak2016PLB}.
From a theoretical perspective, since conformal invariance is a more general symmetry that extends scale invariance, conformal fields interacting with SM fields can also generate potentials similar to those induced by unparticles~\cite{Costantino2020JHEP, Grinstein2008PLB}.

\subsubsection{Interactions Induced by Bilinear Scalar Couplings}\label{theo_21}

\begin{figure}[t!]
    \centering
    \includegraphics[width=0.9\linewidth]{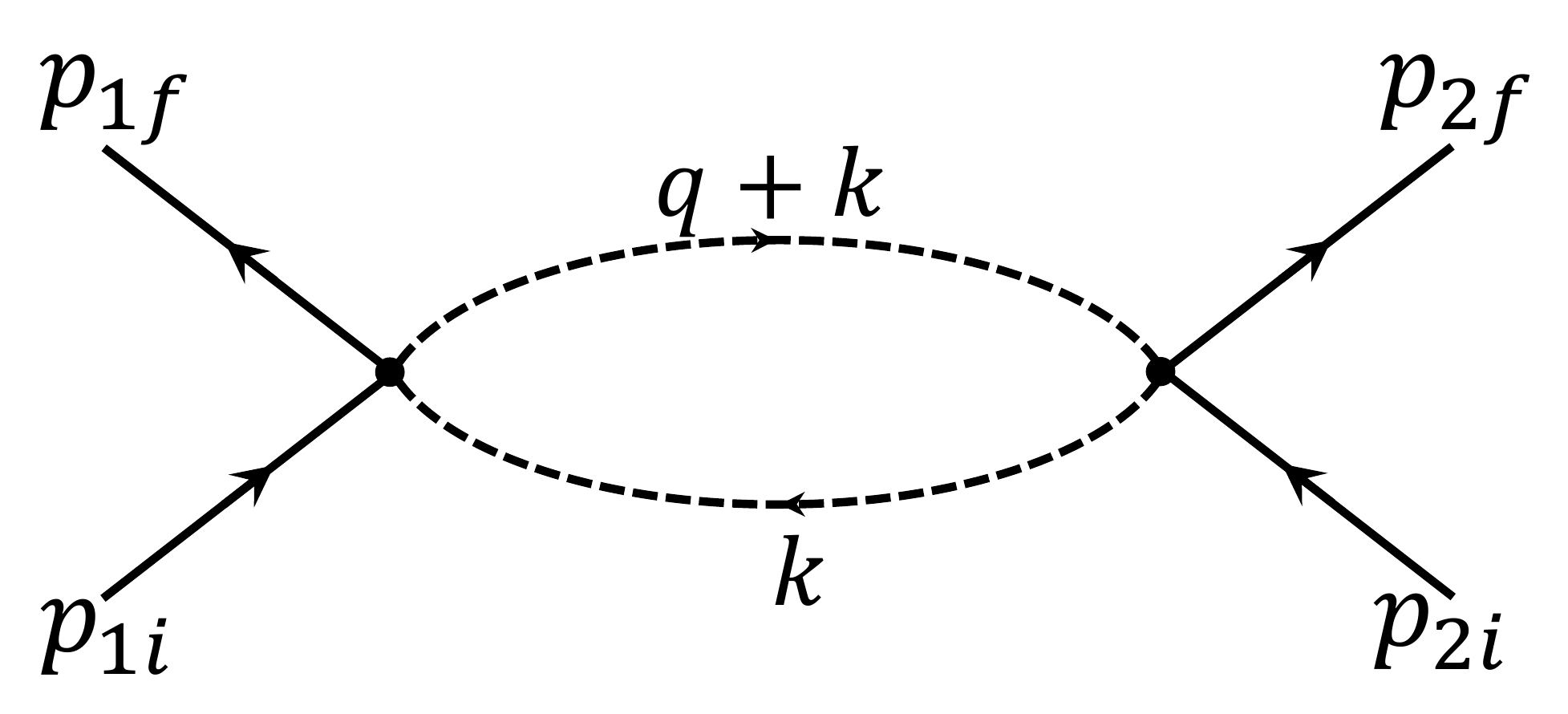}
    \caption{\justifying The loop-level Feynman diagram of two fermions interacting through exchanging two new particles. The $p_{1i}$ ($p_{1f}$) and $p_{2i}$ ($p_{2f}$) are the incoming (outgoing) four-momentum of particles 1 and 2, respectively. The transferred four-momentum carried by the virtual particle is denoted by $q$, while $k$ represents the loop integration variable.}
    \label{fig4}
\end{figure}

The Yukawa-like interactions described in Eqs.~(\ref{ss0})--(\ref{sp2_vgg}) arise from the exchange of a single bosonic particle, corresponding to a tree-level Feynman diagram.
However, under certain conditions, the contribution from such tree-level diagrams may be suppressed, or the linear couplings of these new particles to SM fields may be altogether absent~\cite{Olive2008PRD, Pospelov2013PRL, Fichet2018}.
In such cases, as proposed in Refs.~\cite{Aldaihan2017PRD, Fichet2018, Brax2018}, DM particles could couple to SM particles through bilinear interactions, inducing spin-independent forces that scale as $1/r^n$, where $n$ is a positive integer.
Further, the form of these bilinear couplings is extended to include the spin-dependent interactions in Ref.~\cite{Costantino2020JHEP}.
To stay on track, we will only discuss the spin-dependent types here. The bilinear coupling between the DM and the SM fermions has the form $\mathcal{O}_{SM}\mathcal{O}_{DM}$, where $\mathcal{O}_{SM}$ is the bilinear fermion operators with a form of $\bar{\psi}\Gamma \psi$, in which $\Gamma$ is any possible Lorentz structure, and $\mathcal{O}_{DM}$ is the bilinear of the DM particle.
As a result of the coupling, fermions could interact by exchanging two DM particles, as shown in Fig.~\ref{fig4}.
If two spin-0 DM particles are exchanged between the fermions, the corresponding Lagrangian is~\cite{Costantino2020JHEP}:
\begin{equation}
    \mathcal{L}=\frac{1}{\Lambda^3}\bar{\psi}\sigma^{\mu\nu}\psi\partial_\mu\phi\partial_\nu\phi^*,
\end{equation}
where $\phi$ is the spin-0 DM particle and $\Lambda$ is the scale at which this contact operator description breaks down.
In the non-relativistic limit, this Lagrangian will produce a spin-spin-dependent interaction taking the form of
{\small\begin{equation}
\label{vd0}
\begin{aligned}
    V  =\eta\frac{(\vec{\sigma}_1\cdot \nabla)(\vec{\sigma}_2\cdot \nabla)-(\vec{\sigma}_1\cdot\vec{\sigma}_2)\nabla^2}{32\pi^3\Lambda^6}\frac{K_2(2r/\lambda)}{\lambda^2r^3},
\end{aligned}
\end{equation}}
where $\lambda$ is the Compton wavelength of the DM particle, $r$ is the distance between the fermions, and $\eta$ is a discrete variable that equals 0 if the dark particle is self-conjugate and 1 otherwise. $K_2(2r/\lambda)$ denotes the modified Bessel function of the second kind.
Because this interaction is more strongly dependent on separation, it diverges much more sharply at small distances than the previously mentioned Yukawa interactions.
Here, we consider only one case of exchanging two spin-0 dark matter particles to illustrate the property of the non-Yukawa spin-dependent interactions.
A thorough study on the two DM particle exchange-induced exotic spin-dependent interactions can be found in Ref.~\cite{Costantino2020JHEP}.
The non-Yukawa-like interactions like Eq.~(\ref{vd0}) complement the interaction forms mediated by DM particles.

\section{Experimental Investigations}\label{expinv}

In this section, we outline the general principles underlying experimental setups for detecting exotic spin-dependent interactions. These detection methods are broadly categorized into torque-based and force-based approaches. Each category can be implemented using either resonance or off-resonance techniques. We provide a comprehensive review of the experimental techniques we are aware of that use this classification scheme. For the convenience of experimentalists, SI units are reinstated from this point onward.

\subsection{General Principles}\label{expinv_1}

Experimental searches for spin-dependent interactions generally rely on two fundamental detection principles: measuring torques or detecting forces induced by exotic fields. In contrast to cavity-based techniques widely used in axion searches, torque- and force-based methods directly probe the interactions between spin-polarized probes and nearby sources of exotic fields. This section outlines the core strategies and principles underlying such approaches. We first classify experiments by whether they measure torque or force, and then further distinguish between resonance and off-resonance detection techniques. Key experimental considerations---such as magnetic shielding to suppress external magnetic interference, modulation schemes to enhance signal discrimination, and co-magnetometer configurations to reduce common-mode noise---are also addressed. As an example, we describe a slow-drift removal technique that helps suppress noise. 

These spin-dependent interactions, which we compiled in the previous section, have a general form of 
\begin{equation}
    H_I = \alpha\vec{\sigma}\cdot \vec{V},
\end{equation}
where $\alpha$ represents the strength of the interaction, $\vec{V}$ is the exotic field produced by the spin-dependent interactions, and $\vec{\sigma}$ denotes the spin.
Compared with the magnetic dipole interaction between a spin and a magnetic field, this interaction is analogous to that of a magnetic dipole moment in a magnetic field, described by the Hamiltonian
\begin{equation}
\label{B_p}
H_I=\alpha\vec{\sigma}\cdot\vec{V}=\frac{\hbar\gamma}{2}\vec{\sigma}\cdot\left(\alpha\frac{2}{\hbar\gamma}\vec{V}\right)=-\vec{\mu}\cdot\vec{B}^\prime,
\end{equation}
where $\vec{\mu}$ is the magnetic moment and $\gamma$ is the gyromagnetic ratio of the polarized particle.
Accordingly, the exotic field $\vec{V}$ can be interpreted as an effective magnetic field 
\begin{equation}
\vec{B}^\prime=-\alpha\frac{2}{{\hbar\gamma}}\vec{V}.
\end{equation}
This analogy enables the use of magnetometry-inspired detection schemes in the search for exotic spin-dependent interactions.

It is worth noting that although $\vec{B}^\prime$ mimics the behavior of a conventional magnetic field, it may exhibit different transformation properties under discrete symmetries. For instance, a real magnetic field---arising from VV coupling---is invariant under C, P, and T transformations. In contrast, an effective field $\vec{B}^\prime$ induced by SP coupling is odd under both P and T, while one arising from VA coupling is odd under both C and P. These distinct symmetry properties provide a valuable handle to discriminate signals from exotic interactions against conventional magnetic background noise, as demonstrated in Ref.~\cite{Wu2022PRL}.

\subsubsection{Torque-Based vs. Force-Based Detection}\label{expinv_11}

In torque-based measurements, one detects the rotation or spin precession induced by the effective field acting on a polarized spin ensemble. The resulting torque is given by
\begin{equation}
\vec{\tau} = \vec{\mu} \times \vec{B}^\prime = -\alpha \frac{2}{\hbar \gamma} \vec{\mu} \times \vec{V}.
\end{equation}
\textcolor{black}{Here, $\vec{V}$ denotes the spin-dependent interaction written in vector form, and $\alpha$ is a proportionality factor determined by the definition of the effective field. Torque-based measurements are therefore directly sensitive to the spin-dependent interaction itself, or equivalently to the effective magnetic field acting on polarized spins. In practice, the signal can appear as a mechanical torque, a spin-precession signal, or a frequency shift. Such methods benefit from large spin polarization, long spin-coherence time, and high-precision readout of spin dynamics.}

In force-based measurements, one detects the force induced by the effective field acting on the magnetic moment of the polarized spins:
\begin{equation}
\vec{F} = \nabla(\vec{\mu} \cdot \vec{B}^\prime)=-\alpha \frac{2}{\hbar\gamma}\nabla(\vec{\mu} \cdot\vec{V}).
\end{equation}
\textcolor{black}{Thus, force-based measurements are sensitive to the spatial gradient of the spin-dependent interaction. This makes them naturally suited to experiments in which the source mass or spin source is spatially modulated, or where the interaction varies rapidly over the source--probe separation. The force signal is typically read out using mechanical oscillators or other displacement-sensitive devices.}

\textcolor{black}{The quantitative difference between these two approaches can be illustrated using a velocity-dependent spin interaction with a Yukawa-type radial dependence. As a representative point-particle potential, we consider}
\begin{equation}
\textcolor{black}{
\vec{V}(r)=V_0\frac{e^{-r/\lambda}}{r}\vec{\sigma}\cdot\vec{v},
}
\end{equation}
\textcolor{black}{where $r$ is the source--probe separation, $\lambda$ is the interaction range, $\vec{v}$ is the relative velocity between the source and the polarized spin, and $\vec{\sigma}$ is the spin operator. For force-based detection, the signal is determined by the spatial derivative of this potential,}
\begin{equation}
\textcolor{black}{
\vec{F}=-\nabla V(r).
}
\end{equation}
\textcolor{black}{For a fixed spin orientation, relative velocity, and experimental geometry, the radial dependence gives the characteristic force scaling}
\begin{equation}
\textcolor{black}{
F(r)\sim
V_0
e^{-r/\lambda}
\left(\frac{1}{r^2}+\frac{1}{\lambda r}\right)|\vec{\sigma}\cdot\vec{v}|.
}
\end{equation}
\textcolor{black}{Equivalently, the force signal contains an additional inverse-length factor compared with the potential itself,}
\begin{equation}
\textcolor{black}{
F(r)\sim |V(r)|\left(\frac{1}{r}+\frac{1}{\lambda}\right).
}
\end{equation}
\textcolor{black}{This result shows explicitly that force-based detection introduces an additional distance suppression through the gradient operation. More generally, taking the spatial gradient increases the signal's inverse-distance power. For example, a potential term that scales as $1/r^2$ yields a force term that scales as $1/r^3$ upon differentiation. This additional distance dependence can make force-based signals weaker at macroscopic distances, especially for long-range or slowly varying interactions.}

\textcolor{black}{By contrast, a torque-based experiment responds directly to the spin-dependent interaction or the associated effective field. For an ensemble of $N_s$ coherently polarized spins, the torque signal scales approximately as}
\begin{equation}
\textcolor{black}{
\tau \sim N_s V(r),
}
\end{equation}
\textcolor{black}{where the experimental geometry is usually arranged to maximize the torque response. This expression shows that torque-based methods do not require taking a spatial gradient of the potential. They are therefore often favorable for long-range or slowly varying interactions, provided that a sufficiently large polarized spin ensemble and long coherence time can be achieved. Force-based methods, on the other hand, can be advantageous for short-range interactions or strongly modulated source geometries, where the potential gradient is large.}

Experimentally, the force-based technique involves mechanical oscillators. In contrast, torque-based techniques encompass the torsion pendulum, atomic magnetometers, nuclear magnetic resonance (NMR), atomic beam methods, and atomic or molecular spectroscopy, which will be discussed in detail in the following sections. \textcolor{black}{Thus, the optimal choice between torque- and force-based approaches depends on the interaction range, source--probe geometry, achievable spin polarization, relative velocity modulation, mechanical susceptibility, magnetic shielding, spin-coherence time, and dominant noise sources.}

\subsubsection{On-Resonance vs. Off-Resonance Measurement Methods}\label{expinv_12}

Once the torque or force to be measured is specified, a critical design decision is whether to operate the detection system on- or off-resonance. 
On-resonant and off-resonant measurement techniques each offer distinct advantages depending on the nature of the system, whether it involves spin dynamics, mechanical oscillators, or other resonant modes. 
On-resonant methods involve driving the system at or near its natural frequency, thereby maximizing its response to external perturbations. 
Although this does not inherently enhance the signal-to-noise ratio (SNR), it enables frequency-selective detection, facilitates phase-sensitive techniques such as lock-in amplification, and allows precise control over systematic effects. 
These features make on-resonant detection particularly effective for probing narrowband signals, enhancing sensitivity near resonance, and rejecting broadband noise.

In contrast, off-resonant approaches do not require matching the drive frequency to the system's natural frequency; instead, they measure the system’s response directly---such as displacement, polarization, or magnetization---often via optical, electrical, or magnetometric methods.
Off-resonant methods typically offer broader bandwidth, greater robustness against frequency drift, and reduced susceptibility to nonlinear effects or parasitic couplings associated with strong resonant excitation. 
These techniques are especially advantageous for detecting slowly varying or broadband signals, monitoring long-lived states, or searching for ultra-low-frequency phenomena.
Ultimately, the choice between on- and off-resonant methods reflects a trade-off among spectral selectivity, bandwidth, and systematic control versus operational flexibility.

To illustrate the distinction between on-resonance and off-resonance methods, we consider the harmonic oscillator as a prototypical model. Many physical measurement systems, including torsion pendulums, mechanical resonators, and NMR setups, can be effectively modeled as driven, damped harmonic oscillators.
The general equation of motion for such a system is given by
\begin{equation}\label{ho}
    \frac{d^2 q(t)}{dt^2} + \omega_0^2 q(t) + \Gamma \frac{dq(t)}{dt} = f(t),
\end{equation}
where \( q(t) \) is the generalized coordinate of the oscillator, \( \omega_0 \) is its natural (resonance) frequency, \( \Gamma \) is the damping factor, and \( f(t) \) is the external driving force. Transforming to the frequency domain, the response function is
\begin{equation}
\label{qomega}
    q(\omega) = \frac{f(\omega)}{\omega_0^2 - \omega^2 + i\frac{\omega \omega_0}{Q}},
\end{equation}
where \( f(\omega) \) is the Fourier transform of the driving force and \( Q = \omega_0 / \Gamma \) is the quality factor of the oscillator.

At resonance (\( \omega = \omega_0 \)), the magnitude of the response simplifies to
\begin{equation}
    |q(\omega_0)| = \frac{Q |f(\omega_0)|}{\omega_0^2} = \frac{|f(\omega_0)|}{\omega_0 \Gamma}.
\end{equation}
In contrast, off-resonant responses are significantly weaker. For low-frequency driving (\( \omega \ll \omega_0 \)), the ratio of on-resonant to off-resonant amplitude scales as
\begin{equation}
    \frac{|q(\omega_0)|}{|q(\omega)|} \approx Q.
\end{equation}
For high-frequency driving (\( \omega \gg \omega_0 \)), the suppression is even stronger:
\begin{equation}
    \frac{|q(\omega_0)|}{|q(\omega)|} \approx \frac{\omega^2}{\Gamma \omega_0} > Q.
\end{equation}
This sharp contrast in response illustrates how resonance enhances the system response within a narrow frequency band, whereas off-resonance operation trades amplitude gain for a broader operational range and reduced sensitivity to resonance conditions.

Resonance phenomena are highly effective in amplifying weak signals. In practical implementations, mechanical oscillators can achieve $Q$ on the order of $\sim 10^4$~\cite{Leslie2014PRD}. In NMR, the transverse spin relaxation time $T_2$, also called the coherence time and defined as $T_2 = 1/\Gamma$, is commonly used to characterize the performance of polarized spin systems. 
The precision of magnetic field measurements based on spin precession frequency is primarily governed by two key parameters: the SNR and the transverse relaxation time $T_2$. 
Statistically, the achievable magnetic field sensitivity $\delta B$ with respect to a reference field $B_0$ extracted from the precession frequency is constrained by the Cram\'{e}r-Rao lower bound (CRLB)~\cite{Gemmel2010EPJD, yan2014aCCP}:
\begin{equation}
\label{eq4}
\delta B \geq \frac{\sqrt{8}}{\gamma\sqrt{f_{\rm BW}}T_2^{3/2}{\rm SNR}},
\end{equation}
where $f_{\rm BW}$ is the bandwidth determined by the sampling rate, and $\gamma$ is the gyromagnetic ratio of the spin.
A $^3$He NMR magnetometer developed in Ref.~\cite{Gemmel2010EPJD}, featuring a transverse relaxation time of up to $T_2 \approx 60$ hours and an SNR of approximately $5 \times 10^3$, achieved a magnetic field sensitivity of $\delta B \approx 10^{-4}\ {\rm fT}$ for a $400\ {\rm nT}$ field after one day of integration. Notably, the sensitivity scales with $T_2^{3/2}$ rather than linearly with $T_2$, highlighting the particular advantage of extending the coherence time: even moderate increases in $T_2$ can lead to substantial improvements in sensitivity.

\subsubsection{Noise Reduction}\label{expinv_13}

A key parameter for evaluating the sensitivity of these methods to exotic interactions is the SNR. Enhancing the signal and reducing the noise both improve the SNR and tighten the corresponding constraints.
Since further improving sensor sensitivity can be technically demanding, reducing noise is often a more practical approach.
A number of experimental techniques have been developed to design measurement schemes with this goal in mind thoughtfully.\\

\noindent\textit{\S~3.1~Magnetic Shielding}\\

Since the effective field associated with exotic spin-dependent interactions behaves like a magnetic field, a primary experimental challenge is suppressing background magnetic fields and their associated noise. This strategy leverages the fact that magnetic fields obey Maxwell’s equations, whereas exotic spin-dependent interactions do not. Multiple layers of $\mu$-metal shielding are commonly employed~\cite{Wu2022PRL, Arvanitaki2014PRL, Kim2019NC}, typically reducing ambient magnetic fields to the $\sim$nT level, with $\sim$pT-level shielding demonstrated in recent ten-layer configurations~\cite{Xu2024JP}.

While $\mu$-metal shielding effectively attenuates external magnetic fields, Johnson noise arising from eddy currents in the shielding material can become a limiting factor for ultra-sensitive magnetometers, such as SERF magnetometers. This noise, however, can be further reduced by using ferrite shielding~\cite{Dang2010APL, Kimball2016PRD}.

Magnetic shielding is generally assumed to influence only standard electromagnetic fields, leaving exotic spin-dependent interactions unaffected. Under typical conditions, magnetic shielding has little impact on sensitivity to exotic spin-dependent interactions, especially with co-magnetometers~\cite{Kimball2016PRD}. However, electron-coupled exotic fields can induce secondary fields in soft ferro- or ferrimagnetic shields, which must be considered in data interpretation. 
This also suggests using flux concentrators to enhance such signals.
Superconducting shielding, which offers even greater magnetic noise attenuation, has been successfully used in neutron EDM (nEDM) experiments.
Several proposals have suggested applying superconducting shielding to exotic spin-dependent interactions~\cite{Arvanitaki2010, Yan2014EPJC}, and its application was recently reported in Ref.~\cite{Crescini2022PRD}.

When searching for exotic fields generated by a polarized source---such as in dipole-dipole-type spin-dependent interactions---the magnetic field produced by the spin source (e.g., hard rare-earth magnets) must be carefully controlled to prevent it from generating spurious signals that exceed the background noise level. An effective strategy for reducing magnetic field leakage is to confine the magnetic flux near the source. For instance, Ji \textit{et al.} encapsulated a SmCo$_5$ spin source with three layers of pure iron, successfully reducing the external magnetic field from approximately 1T to below $10\mathrm{\mu T}$~\cite{ji2017PRD, Ji2023PRL}.\\

\noindent\textit{\S~3.2~Modulation and Lock-in Based Detection Technique}\\

While shielding helps reduce background noise, modulation provides a means of discriminating between signal and noise in the time domain. In the frequency domain, noise is typically not uniformly distributed; for instance, low-frequency regions are often dominated by $1/f$ noise, and various discrete noise peaks may appear across the spectrum. By modulating the signal to a higher frequency and employing phase-sensitive detection and amplification, the effective noise bandwidth can be significantly reduced, thereby substantially enhancing the SNR. This approach not only narrows the bandwidth over which noise contributes but also shifts the signal away from low-frequency regions dominated by $1/f$ noise.

The specific modulation strategy depends on the experimental configuration. For unpolarized sources, modulation can be realized through translational~\cite{Kim2019NC} or rotational~\cite{Wu2022PRL} motion of the source mass. For polarized sources, modulation is often achieved by periodically reversing the spin orientation, either via optical pumping~\cite{Wang2023SA} or using adiabatic fast passage (AFP) NMR~\cite{Vasilakis2009PRL}. Compared to mechanical modulation, spin-based modulation has the advantage of minimizing vibrational and mechanical disturbances.

In cases where celestial bodies such as the Sun or the Moon act as sources---or for background fields arising from axion dark matter or from Lorentz and CPT violation---the Earth's rotation naturally induces sidereal modulation of the measured signals in the laboratory frame~\cite{Wu2019PRL, Heckel2006PRL, Heckel2008PRD, Wu2023PRL}. Additionally, alternating the spin orientation of the probe itself is another modulation technique employed to extract weak exotic signals~\cite{Brown2010PRL}.\\

\noindent\textit{\S~3.3~Co-magnetometer Technique}\\

The co-magnetometer technique has a significant advantage in its capability to cancel common background field noise. A co-magnetometer comprises two spin species with different gyromagnetic ratios, each precessing at distinct Larmor frequencies given by:
\begin{eqnarray*}
\omega_1=\gamma_1B_0,\\
\omega_2=\gamma_2B_0,
\end{eqnarray*}
where $B_0$ is the holding field oriented along, for example, the $\hat{z}$ direction. The key measurable in this setup is the ratio of the two frequencies:
\begin{equation}
\label{romege}
\mathcal{R} =\frac{\omega_1}{\omega_2}=\frac{\gamma_1B_0}{\gamma_2B_0}.
\end{equation}
This ratio effectively cancels out fluctuations of the magnetic field $B_0$, whereas non-magnetic effects such as inertial rotations or exotic spin-dependent interactions, which typically do not scale with the gyromagnetic ratio, remain detectable. Therefore, the inherent background-field cancellation makes co-magnetometers particularly well-suited for investigating spin-dependent interactions, where background magnetic noise is often the primary limitation.

This technique is also commonly referred to as clock comparison, as it effectively compares the accumulated phases of two distinct spin-based clocks. In addition to the straightforward frequency ratio $\mathcal{R} =\omega_1/\omega_2$~\cite{Chupp1988PRA, Feng2022PRL, Zhang2023PRL}, alternative forms such as $\mathcal{R}=(\omega_1-\omega_2)/(\omega_1+\omega_2)$~\cite{Kimball2017PRD} are employed to further suppress common-mode noise.\\

\noindent\textit{\S~3.4~Technique for Removing Slow Drifts}\\

Despite employing the precision measurement techniques discussed above, slow drifts often persist in experimental data. To reduce these residual slow drifts, the [+1, -3, +3,-1] weighting method provides an effective and robust solution.

A concise description of this weighting approach, as outlined in Ref.~\cite{Chu2013PRD}, is as follows.
Suppose the measurement system exhibits a quadratic drift over time.
The measured signal can then be modeled as:
\begin{equation}\label{drft}
B_{\text{exp}}(t)=at^2+bt+c\pm B^\prime,
\end{equation}
where $a$, $b$, and $c$ are constants characterizing the slow drift, $t$ represents time, and $+B^\prime$ (or $-B^\prime$) denotes the signal of interest, which can be periodically inverted during measurement. 
Conducting a series of alternating $+B^\prime$ and $-B^\prime$ measurements at regular intervals $\Delta T$, we obtain a sequence such as $B_{\text{exp}}(\Delta T)(+B^\prime)$, $B_{\text{exp}}(2\Delta T)(-B^\prime)$, $B_{\text{exp}}(3\Delta T)(+B^\prime)$, $B_{\text{exp}}(4\Delta T)(-B^\prime)$, and so forth.

Merely subtracting the $-B^\prime$ measurement from the $+B^\prime$ measurement eliminates only the constant drift term. However, applying the [+1,-3,+3,-1] weighting scheme effectively extracts the desired signal as follows:
\begin{eqnarray*}
 B^\prime=\frac{1}{8}[B_{\rm exp}(\Delta T)(+B^\prime)-3B_{\rm exp}(2\Delta T)(-B^\prime)\\
 +3B_{\rm exp}(3\Delta T)(+B^\prime)-B_{\rm exp}(4\Delta T)(-B^\prime)].
\end{eqnarray*}

This weighting method removes slow drifts up to quadratic order, which is typically sufficient for precision measurement applications.
This approach is particularly familiar in nuclear physics, where data acquisition periods can extend up to 100 days. The drift described in Eq.~(\ref{drft}) can also be interpreted as extremely low-frequency systematic noise. By periodically flipping the signal, one effectively modulates it to a frequency significantly higher than the drift-induced noise frequency, thereby improving noise reduction.
Indeed, Ref.~\cite{Snow2011PRC} demonstrates that employing a linear drift removal algorithm at a relatively low flipping frequency significantly reduces measurement uncertainty, achieving reductions of up to 10$\%$.\\

\subsection{Torque Based Detection}\label{expinv_3}

In this subsection, we review several experimental techniques for probing exotic spin-dependent interactions via torque measurements.
On-resonance methods encompass various NMR-based approaches---such as conventional NMR, zero-field NMR, NMR sideband search, NMR co-magnetometers, and spin-relaxation methods---as well as atomic spectroscopy.
Off-resonance methods include torsion pendulums, alkali-metal magnetometers, atomic co-magnetometers, and beam-based techniques employing neutron beams, muon beams, and $^3$He beams.\\

\subsubsection{On-Resonance Measurement Method}\label{expinv_32}
\quad\\
\noindent\textit{1.1~NMR Based Methods}\\

\noindent\textit{\S~1.1.1~Conventional NMR}\\

\begin{figure}
    \centering
    \includegraphics[width=\linewidth]{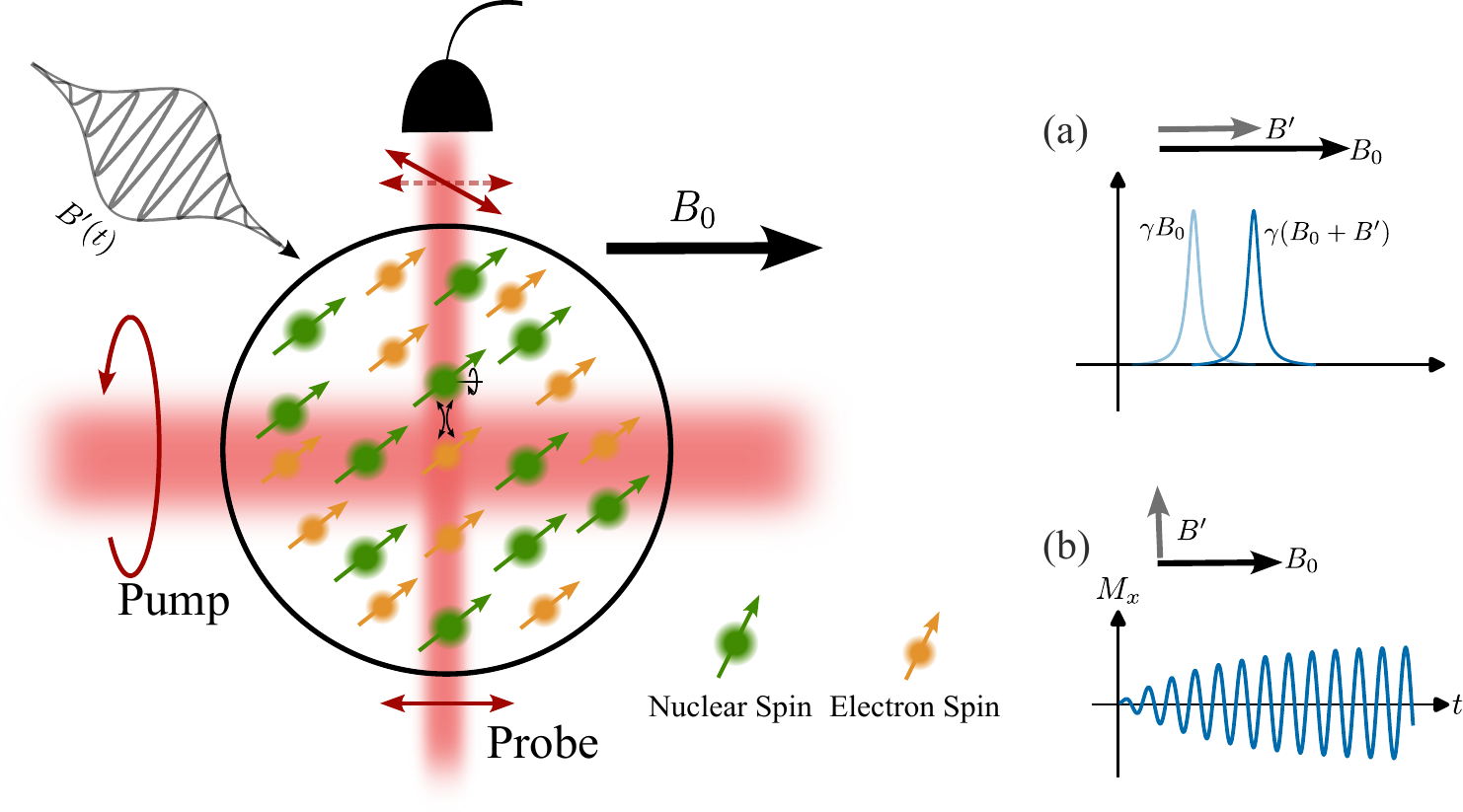}
    \caption{\justifying \color{black}A schematic illustration of the NMR-based detection method for exotic spin-dependent interactions. The effective magnetic field $\vec{B}'(t)$ generated by the exotic potential perturbs the Larmor precession of polarized nuclear spins in the holding field $\vec{B}_0$. Depending on its orientation and frequency, $\vec{B}'(t)$ can either shift the precession frequency or, under resonant conditions, induce transverse oscillations of the spins. The resulting precession signals are typically detected using an alkali-metal magnetometer (as illustrated) or other high-sensitivity sensors, such as SQUIDs.}
    \label{fig:nmr_detect}
\end{figure}

NMR is a high-precision technique for manipulating and measuring the spins of polarized nuclei, making it a natural platform for probing exotic spin-dependent interactions, as illustrated in Fig.~\ref{fig:nmr_detect}. By using resonance-based signal amplification, NMR provides a highly sensitive method for detecting new physics involving nuclear spins. Additionally, nuclear spins typically exhibit gyromagnetic ratios approximately three orders of magnitude smaller than those of electrons, which intrinsically enhances their isolation from background magnetic fields and associated noise.
The possibility of long-range spin-dependent interactions between protons, mediated by new light-mass particles, was first constrained by Ramsey in 1979 through measurements of the nuclear radiofrequency spectrum of H$_2$~\cite{Ramsey1979PA}. Today, a wide array of experimental efforts---both ongoing and proposed---utilize NMR-based techniques to search for exotic spin-dependent interactions and dark matter. Notable examples include the Cosmic Axion Spin Precession Experiment (CASPEr)~\cite{Garcon2018QST}, the Axion Resonant Interaction Detection Experiment (ARIADNE)~\cite{Arvanitaki2014PRL}, and spin-based amplifier schemes~\cite{Jiang2021bNP, Su2021SA, Wang2022PRL}.

Among the various experimental efforts, the ARIADNE experiment stands out for its innovative use of a resonantly driven source to amplify the system’s response. ARIADNE is specifically designed to probe ALP-mediated spin-dependent interactions in the ALP mass range of $10^{-6}$ to $10^{-3}$~eV~\cite{Arvanitaki2014PRL}. 
The experiment employs hyperpolarized $^3$He gas, contained within a quartz vessel, as a sensitive detector for the effective magnetic field generated by a nearby unpolarized or polarized source mass.
The source mass is modeled as a cylindrical rotor with a cross-section resembling a ``sprocket''.
When the cylinder rotates about its axis, it generates a time-varying effective magnetic field whose spectral components appear at harmonics of the rotation frequency. This geometry not only produces a spatially modulated interaction but also helps decouple the signal from ambient mechanical vibrations.
The rotation frequency is tuned to coincide with the Larmor frequency of the $^3$He nuclei under a static magnetic field, thereby enabling resonant amplification of the response.
Under the influence of this resonantly oscillating effective field, the polarized $^3$He develops a transverse magnetization that oscillates in time, allowing for sensitive detection of the interaction:
\begin{equation}
\label{mx}
M_x \propto  B^\prime T_2\cos(\omega_0t),
\end{equation}
where $B^\prime$ is the effective magnetic field, $T_2$ is the transverse spin relaxation time, and $\omega_0$ is the Larmor frequency.
$^3$He is an excellent probe due to its exceptionally long transverse relaxation time ($T_2 \approx 1000$~s in the liquid state) and the feasibility of achieving high nuclear polarization through optical pumping.
In addition to employing liquid-phase polarized $^3$He, the ARIADNE experiment plans to use superconducting magnetic shielding to further suppress background magnetic noise.
The resulting spin precession is detected using a SQUID magnetometer. 
Owing to the integration of these advanced techniques, ARIADNE's projected sensitivity to the $\V_{9+10}$ interaction is expected to exceed existing astrophysical constraints for interaction ranges around $\lambda\sim 10^{-2}$~m. 
Moreover, by scaling up the apparatus and operating in the liquid phase, the experiment could access parameter space relevant to the Peccei-Quinn axion.
The ARIADNE experiment is currently under active development~\cite{Lohmeyer2018Progress, Aggarwal2022PRR}. However, implementing these novel techniques also introduces considerable experimental complexity.

The CASPEr experiment is designed to search for axion fields through their coupling to nuclear spins. The collaboration is structured into two main experimental branches: CASPEr-Wind and CASPEr-Electric. CASPEr-Wind targets the coupling of the axion field to nucleons via an effective oscillating magnetic field (i.e., the first term in Eq.~(\ref{alpfield})). At the same time, CASPEr-Electric is sensitive to the axion-gluon coupling that induces an oscillating electric dipole moment in nuclei (i.e., the second term in Eq.~(\ref{alpfield})).
In both cases, the search is performed by sweeping the strength of an applied guiding magnetic field to scan for resonant enhancement at different frequencies, each corresponding to a particular ALP mass. The detector system employs a polarized liquid of $^{129}$Xe as the sensing medium, combined with various magnetometers, including atomic magnetometers, SQUIDs, and induction pickup coils.
The high density and long transverse coherence time ($T_2$) of liquid $^{129}$Xe enable significant amplification of the effective magnetic field, as described by Eq.~(\ref{mx}). Furthermore, using multiple magnetometer types enables coverage of a wide frequency spectrum, from near-zero to MHz. Recent developments and experimental results from the CASPEr collaboration are reviewed in Ref.~\cite{Kimball2020}.

Resonance amplification can be further enhanced by exploiting spin-exchange interactions between alkali-metal atoms and polarized noble-gas nuclei.
Due to both spin-exchange and intermolecular interactions between the alkali metal and noble gas atoms, an additional amplification factor $\kappa_0$ arises beyond what is predicted by Eq.~(\ref{mx}).
A gain factor of approximately $\sim 128$ has been reported for the $^{87}$Rb-$^{129}$Xe system~\cite{Jiang2021bNP}, achieving a magnetic field sensitivity of $18~\mathrm{fT/\sqrt{Hz}}$ at the $^{129}$Xe Larmor frequency ($\sim$10 Hz).
This enhancement is further improved to a factor of 5400 when the $^{129}$Xe is operated in the dark~\cite{Jiang2024PNAS}.
Compared to the $^{87}$Rb-$^{129}$Xe system, the K-$^3$He system offers even greater promise due to the significantly longer coherence time and larger gyromagnetic ratio of $^3$He spins.
The amplification factor for the K-$^3$He system is expected to reach $10^6$, and the standard quantum limit for such an amplification scheme is estimated to be $90~\mathrm{aT/\sqrt{Hz}}$~\cite{Jiang2024PNAS}.
Furthermore, the relatively accessible operational conditions of spin-based amplifiers have facilitated their deployment in experimental searches for spin-dependent interactions~\cite{Su2021SA, Wang2022PRL, Wang2023SA}.\\

\noindent\textit{\S~1.1.2~Zero-Field NMR}\\

Apart from traditional NMR experiments that require a nonzero external magnetic field, emerging zero-to-ultralow-field (ZULF) NMR techniques are increasingly being employed to search for exotic spin-dependent interactions, enabled by recent advances in magnetometer technology~\cite{Blanchard2020JMR, jiang2021FR}.
The J-coupling interaction between nuclear spins, described by $J\vec{I}_1\cdot\vec{I}_2$, provides a sensitive platform to probe dipole-dipole-type spin-dependent interactions.
J-coupling arises indirectly from hyperfine interactions between nuclei and surrounding electrons. 
In NMR spectroscopy, it typically appears as a spectral peak at $J$ Hz in a zero magnetic field, or as a peak splitting of $J$ Hz under small applied fields.
By comparing experimentally measured values of $J$ with high-precision theoretical predictions, one can place bounds on exotic spin-dependent interactions.
In Ref.~\cite{ledbetter2013}, a discrepancy between the measured J-coupling spectra of HD molecules and theoretical calculations was used to derive stringent constraints on such interactions at the atomic scale.
Conventional J-coupling measurements are often performed in strong magnetic fields to enhance nuclear polarization and improve SNR, but this comes at the cost of introducing unavoidable and strong background magnetic fields.
ZULF NMR, by contrast, enables precise J-coupling measurements in near-zero magnetic fields~\cite{Ledbetter2009JMR}, significantly reducing magnetic noise and systematic effects due to the background field.
When combined with increasingly accurate theoretical models~\cite{Puchalski2018PRL}, ZULF NMR offers a promising route toward establishing tighter bounds on exotic spin-dependent interactions. \\

\noindent\textit{\S~1.1.3~NMR Sideband Search}\\

In conventional NMR, resonance detection becomes less sensitive at low frequencies. In contrast, if an effective magnetic field is aligned with the main field, it can modulate the Larmor frequency or produce sidebands near the resonance frequency---without requiring resonance with the Larmor frequency---thus extending the detectable frequency range to very low values.
The sideband detection scheme works as follows: when a low-frequency effective magnetic field is aligned with the main field, it modulates the Larmor frequency, producing sidebands in the Fourier spectrum. These appear symmetrically around the Larmor peak, with offsets corresponding to the harmonic components of the effective field. The amplitude of each sideband is proportional to the strength of the effective magnetic field. To resolve sidebands from the precession signal, the frequency of the effective field must exceed the transverse relaxation rate.
This technique has been applied in the search for axion-nucleon couplings via ultralow-field NMR measurements, such as the detection of sidebands in the J-coupling spectroscopy of $^{13}$C-formic acid~\cite{Garcon2019SA}, and in the spin precession of $^{199}$Hg at approximately 10 Hz~\cite{Abel2023SP}.
These methods probe axion fields down to frequencies limited by spin relaxation, with sensitivity improving at lower frequencies.
In the context of exotic spin-dependent interactions, a time-varying effective field can also be generated by modulating the position of a mass source. 
A recent study searched for the $\V_{9+10}$ interaction by measuring sidebands in the free precession of $^{129}$Xe induced by a rotating germanium block~\cite{Shortino2024RSI}.
The projected sensitivity is expected to surpass existing constraints at an interaction range of $\lambda\sim 10^{-4}$~m.\\

\noindent\textit{\S~1.1.4~NMR  Co-magnetometers}\\

NMR co-magnetometers utilize nuclear species such as $^{199}$Hg, $^3$He, and $^{129}$Xe, as well as ultracold neutrons confined within a vessel. Although the gyromagnetic ratio of the nucleon is approximately three orders of magnitude smaller than that of the electron, its coupling to ambient magnetic fields is correspondingly weaker. As a result, the transverse relaxation time---typically scaling inversely with the square of the nuclear spin’s gyromagnetic ratio---is significantly longer than that of electron spins.
Consequently, the magnetic field sensitivity of nuclear spin-based magnetometers can be comparable to that of alkali-metal magnetometers, as indicated by Eq.~(\ref{deltaB}). Moreover, since exotic spin-dependent interactions couple to spin rather than to magnetic moment, nuclear spins---with their much smaller gyromagnetic ratios----offer superior energy resolution relative to electrons under the same level of magnetic noise.

NMR co-magnetometers generally operate in a free precession mode, where polarized nuclear spins undergo Larmor precession at a frequency set by the local magnetic field. The resulting precession signal is typically detected using external magnetometers placed near the sample cell, such as superconducting quantum interference devices (SQUIDs) or alkali-metal magnetometers. Alternatively, when nuclear spins are polarized via spin-exchange optical pumping (SEOP), the cohabiting alkali-metal vapor can serve as an internal magnetometer.
To improve robustness against magnetic field fluctuations, nuclear spin magnetometers often use co-magnetometer configurations that involve two atomic states or spin species with distinct gyromagnetic ratios. These configurations come in several forms. For instance, a single-species co-magnetometer based on the J-coupling spectrum of acetonitrile-2-$^{13}$C was demonstrated in Ref.~\cite{Wu2018PRL}. Dual-isotope systems include co-located ensembles of $^{199}$Hg-$^{201}$Hg~\cite{Lamoreaux1987PRL, Venema1992PRL} and $^{129}$Xe-$^{131}$Xe~\cite{Bulatowicz2013PRL, Feng2022PRL, Zhang2023PRL}. Noble gas pairs such as $^3$He-$^{129}$Xe have also been widely employed~\cite{Bear2000PRL, limes2018PRL, Allmendinger2014PRL}.
Most implementations operate in a clock-comparison mode, in which two distinct Larmor frequencies are monitored simultaneously. This configuration enables effective rejection of common-mode magnetic noise and drift, allowing for highly sensitive, long-term-stable measurements.

Co-magnetometers have been employed to search for various exotic spin-dependent interactions, including $\V_{9+10}$, spin-gravity couplings, and Lorentz-violating background fields. In the work of Venema \textit{et al.}~\cite{Venema1992PRL}, a $^{199}$Hg-$^{201}$Hg co-magnetometer was used to search for spin-gravity interactions, treating the Earth as an unpolarized source mass. The co-magnetometer was oriented such that the holding magnetic field aligned with the Earth’s rotation axis. In this configuration, the contribution of the Earth’s rotation $\Omega_\oplus$---known with high precision---can be accurately subtracted~\cite{Kimball2013AP}.
A similar approach was adopted by Zhang \textit{et al.}~\cite{Zhang2023PRL}, who employed an Rb-$^{129}$Xe-$^{131}$Xe co-magnetometer as a spin sensor, again using the Earth as the source. Bulatowicz \textit{et al.}~\cite{Bulatowicz2013PRL} also used an Rb-$^{129}$Xe-$^{131}$Xe co-magnetometer to search for $\V_{9+10}$ interactions, with a zirconia rod serving as the source. By translating the rod’s position, they distinguished frequency shifts induced by spin-dependent couplings from constant shifts associated with Earth’s rotation.
Similarly, in Ref.~\cite{Tullney2013PRL}, a cylindrical BGO crystal (Bi$_4$Ge$3$O${12}$) was periodically moved toward and away from a $^{129}$Xe-$^{131}$Xe co-magnetometer, and the resulting spin-dependent frequency shifts were detected using SQUID sensors configured in a gradiometric arrangement.
In searches for Lorentz- and CPT-violating background fields, the exotic field is assumed to remain fixed in a preferred reference frame. In such cases, sidereal modulation of the signal measured by Earth-based co-magnetometers provides a means to separate the exotic signal from effects tied to Earth’s rotation or any laboratory-fixed backgrounds~\cite{Lamoreaux1987PRL, Allmendinger2014PRL}.

Ultra-cold neutrons (UCNs), employed in the search for the neutron electric dipole moment (nEDM)~\cite{Baker2006PRL, Abel2020PRL} using the co-magnetometer principle, also serve as a powerful tool for constraining a wide range of spin-dependent new physics scenarios~\cite{Altarev2009PRL, Abel2017PRX, Afach2015PLB}. High-precision Ramsey interferometry is typically used to measure the tiny frequency shifts induced by EDMs or exotic spin-dependent interactions.
In the study by Afach \textit{et al.}~\cite{Afach2015PLB}, the $\V_{9+10}$ interaction was investigated using an nEDM apparatus by comparing the Larmor precession frequencies of UCNs and spin-polarized $^{199}$Hg atoms in a co-magnetometer configuration. 
In this setup, the chamber walls act as the unpolarized source of the exotic interaction. Due to the symmetric geometry, the spatially averaged $^{199}$Hg atoms experience a negligible net shift from the exotic interaction. In contrast, UCNs, being extremely low in energy, are strongly affected by gravity and develop a vertically varying density profile. This gravitational sag leads to an incomplete cancellation of the effective interaction-induced field across the neutron ensemble.
As a result, the neutron Larmor frequency acquires a shift that depends on the orientation of the applied magnetic field. Notably, the frequency ratio between UCNs and $^{199}$Hg atoms changes sign upon reversing the magnetic field, providing a sensitive probe of exotic spin-dependent interactions.

Despite suppression of magnetic field fluctuations in clock-comparison co-magnetometers, systematic errors persist. Ref.~\cite{Sheng2014PRL} identified two key sources: (1) higher-order magnetic field gradients, causing frequency shifts proportional to the cube of the gradient strength, and (2) thermal diffusion, where temperature gradients create gas concentration gradients, introducing sensitivity to first-order magnetic field gradients.
Dipolar and scalar (J-coupling) couplings between polarized $^3$He and $^{129}$Xe also shift the frequency~\cite{Terrano2019aPRA, Limes2019PRA}. In alkali-metal/noble-gas co-magnetometers, isotope-dependent spin-exchange interactions between alkali atoms and noble gases (e.g., $^{129}$Xe vs. $^{131}$Xe) further affect the frequency ratio~\cite{Bulatowicz2013PRL}. Accounting for these effects is crucial for maximizing sensitivity to new physics.\\

\noindent\textit{\S~1.1.5~Spin Relaxation}\\

Exotic spin-dependent interactions can affect spin relaxation in NMR measurements. In the ultracold chamber, interactions between neutrons and nucleons in the wall can induce neutron spin relaxation during collisions, leading to depolarization during storage. By comparing measured depolarization rates with theoretical models, constraints on the interaction $\V_{9+10}$ were obtained in Refs.~\cite{Ignatovich2009EPJC, Serebrov2009PLB}.

Another case involves the depolarization of polarized $^3$He gas, which, due to its long relaxation time (up to thousands of hours~\cite{Parnell2009NIM, Gemmel2010EPJD}), is sensitive to exotic spin-dependent interactions. The main relaxation sources include dipole-dipole interactions during atomic collisions, wall interactions, and magnetic field gradients~\cite{Gentile2017RMP}. Gradient-induced relaxation results from phase decoherence caused by the fluctuating magnetic field experienced during atomic motion. This has been analyzed by Cates \textit{et al.} using perturbation theory and by McGregor within Redfield theory~\cite{Mcgregor1990PRA}, both yielding consistent results. Similarly, relaxation from gradients in the effective magnetic field due to exotic interactions can be modeled, and comparing with measured
rates yields upper bounds on interaction strength. Considering $\V_{9+10}$ between nucleons in the wall and polarized neutrons in $^3$He, Petukhov \textit{et al.}~\cite{Petukhov2010PRL} and Guigue \textit{et al.}~\cite{Guigue2015PRD} derived constraints in the submillimeter range. In contrast, Fu \textit{et al.}~\cite{Fu2011PRD} tested this interaction by measuring changes in the transverse relaxation time of $^3$He when a mass block was moved near the cell.

Furthermore, spin relaxation of polarized $^3$He was used to search for exotic spin-velocity-dependent interactions in Ref.~\cite{Yan2015PRL}. They studied $\V_{12+13} \propto \vec{\sigma}\cdot\vec{v}$ between $^3$He and unpolarized nucleons in the Earth, where the relative velocity $\vec{v}$ arises from thermal motion of the $^3$He atoms. Although $\langle \vec{v} \rangle = 0$, fluctuations $\langle v^2 \rangle \neq 0$ produce a fluctuating effective magnetic field. By attributing transverse relaxation to this field, they set limit on the interaction at $\lambda\sim 10^8$~m. Notably, this method requires no bulk motion of either polarized or unpolarized masses.

In addition to decoherence, a fluctuating field can shift the Larmor frequency. This was explored by Xiao \textit{et al.}~\cite{xiao2024arxiv} via noise spectral density in an Rb magnetometer.\\

\noindent\textit{1.2~Atomic Spectroscopy}\\

The remarkable agreement between experimental measurements and theoretical calculations in atomic and molecular spectroscopy enables sensitive tests for exotic interactions at the atomic scales. Historically, atomic spectroscopy has played a pivotal role in discovering new physics: the Lamb shift in hydrogen~\cite{Lamb1947PR} and the electron’s anomalous magnetic moment~\cite{Foley1948PR} were key to the development of quantum electrodynamics---one of the most precise theories to date. In the modern era, advances in controlling matter and light have continuously improved spectroscopic precision and its utility in probing new physics~\cite{Safronova2018RMP}. For example, the accuracy of atomic clocks has increased by over three orders of magnitude in the past decade, reaching a frequency uncertainty of one part in $10^{18}$~\cite{Ushijima2015NP}.
To constrain exotic interactions, atomic spectroscopy typically compares experimental results with theoretical predictions and sets bounds based on the discrepancy. 
If theoretical precision exceeds experimental accuracy, the constraint is instead derived by attributing the experimental uncertainty to potential exotic effects.

Current searches for exotic interactions in atomic systems mainly focus on simple atoms~\cite{Karshenboim2010PRD, Karshenboim2011PRA, Kotler2015PRL, Ficek2017PRA, Ficek2018PRL, Cong2024arXiv1, Cong2025arXiv2}. In Ref.~\cite{Karshenboim2010PRD}, Karshenboim constrained the exotic spin-spin interaction $\V_2$ by comparing the ground-state hyperfine splitting $E_\mathrm{hfs}(1s)$ between experiment and QED predictions in six simple atoms: muonium, positronium, H, D, T, and $^3$He ion.
To suppress nuclear structure effects, he later considered the interval difference $D_{12} = 8E_\mathrm{hfs}(2s) - E_\mathrm{hfs}(1s)$~\cite{Karshenboim2011PRA}, improving the constraints by about an order of magnitude.
Beyond atomic bound states, equivalent two-particle systems can be realized in ion traps. 
Kotler \textit{et al.}~\cite{Kotler2015PRL} used a linear radiofrequency (RF) Paul trap to confine two $^{88}\mathrm{Sr}^+$ ions and probed $\V_2$ or $\V_{3}$ between electron spins at micrometer separations.

Helium is the simplest multielectron atom. While its electron energies cannot be solved analytically as in hydrogen, numerical calculations have reached an accuracy of order $m_e\alpha^7$, where $m_e$ is the electron mass and $\alpha$ the fine-structure constant. Combined with high-precision spectroscopy, helium thus provides a platform for testing QED and probing exotic interactions.
Ficek \textit{et al.} studied exotic spin-dependent interactions between electrons in $^4$He~\cite{Ficek2017PRA}, constraining dipole-dipole interactions similar to those in Refs.~\cite{Karshenboim2010PRD, Karshenboim2011PRA, Kotler2015PRL} by comparing theory and experiment for the $2^3P_2 - 2^3P_1$ and $2^3S_1 - 2^3P$ transitions. They also placed bounds on two types of spin-velocity-dependent dipole-dipole interactions. 
To explore matter-antimatter interactions, Ficek \textit{et al.}~\cite{Ficek2018PRL} extended this analysis to antiprotonic helium, considering interactions between the antiproton in the $(n,l) = (37,35)$ state and the electron in the $(1,0)$ state. Using the experimental-theoretical agreement at the $0.01\%$ level, they constrained exotic interactions at angstrom-scale ranges.\\

\subsubsection{Off-Resonance Measurement Method}\label{expinv_31}
\quad\\
\noindent\textit{2.1~Torsion Pendulum}\\

The torsion pendulum measures weak forces---such as gravity and electric forces---by detecting the torque on a suspended pendulum. A typical setup consists of a horizontal pendulum suspended by a fine fiber, with two symmetrically placed mass blocks serving as source masses. Their interaction generates torque, balanced by the fiber’s twist; measuring the twist angle reveals the magnitude of the force.
A key feature of the torsion pendulum is its sensitivity to the angular direction of applied forces: if the forces are parallel, no torque arises regardless of their magnitudes. Acting as a highly sensitive null instrument, the torsion pendulum compares the angles rather than the magnitudes of force vectors. As a result, torsion balances can compare test-body accelerations with a precision of parts in $10^{13}$, even though the dimensions and masses of the pendulum are known only to parts in $10^4$.
This extraordinary precision has made torsion pendulums essential tools in testing new physics, including the gravitational inverse-square law at short distances, the equivalence principle, and exotic spin-dependent interactions~\cite{Adelberger2009PPNP}. In measuring $G$, the relative uncertainty has improved from $10^4$ ppm in the 19th century to several $10$ ppm today, thanks to major technological advancements~\cite{Xue2020NCR}.

Detecting exotic spin-dependent interactions with torsion pendulums requires replacing the source masses, the pendulum, or both with spin-polarized materials, while carefully suppressing gravitational and electromagnetic torques. The first such experiment was performed by Ritter \textit{et al.}~\cite{Ritter1990PRD} in 1990 using spin-polarized $\mathrm{Dy_6Fe_{23}}$ as the source, chosen for its near-cancellation of spin and orbital magnetic moments, which minimizes magnetic field backgrounds.
The E\"{o}t-Wash group at the University of Washington later conducted a series of landmark experiments, developing torsion pendulum designs optimized for various interaction ranges~\cite{Heckel2006PRL, Heckel2008PRD, Hoedl2011PRL, Heckel2013PRL, Terrano2015PRL}. In Ref.~\cite{Heckel2008PRD}, they used unpolarized nucleons in the Earth, Moon, and Sun as sources, and polarized electrons in the pendulum as detectors. Their spin pendulum, containing $\sim 10^{23}$ polarized electrons, was designed to produce negligible external magnetic fields and maintain a uniform mass distribution. It consists of four stacked octagonal ``pucks,'' each composed of half Alnico and half SmCo$_5$. The SmCo$_5$ provides spin polarization with minimal external field, as its magnetization arises almost entirely from Co, with the orbital and spin moments of Sm$^{3+}$ nearly canceling. The Alnico, a soft ferromagnet, is magnetized to match the Co contribution in SmCo$_5$, allowing the net magnetic flux to close internally. To balance mass density, thin plates are added to the Alnico side.

This design produces a net spin $\vec{\sigma}$ in the plane perpendicular to the fiber while minimizing gravitational torque. The entire apparatus is inside a vacuum vessel and magnetic shielding. An effective magnetic field $\vec{B}^\prime$ from exotic interactions generates a torque $\vec{\tau} = \vec{\sigma} \times \vec{B}^\prime$, but only its vertical component---twisting the fiber---is detectable. Thus, the pendulum is sensitive only to $\vec{B}^\prime$ perpendicular to the fiber.
Using the Earth and Sun as sources, they constrained the $V_{\rm SP}$ interaction at astronomical-unit (AU) ranges and, exploiting the Earth's and Moon's motions, also constrained spin-velocity interactions $V_{\rm VA}$ and $V_{\rm AA}$. A variety of other spin-dependent scenarios were also explored.
In a later study~\cite{Heckel2013PRL}, the same apparatus was adapted to constrain exotic dipole-dipole interactions between electrons at a $\sim 1$ m range by surrounding the pendulum with spin-polarized SmCo$_5$ sources. To access shorter interaction ranges ($\sim1$ mm), Terrano \textit{et al.}~\cite{Terrano2015PRL} developed a new spin pendulum design enabling millimeter-scale separation between source and probe. The pendulum, composed of alternating SmCo$_5$ and Alnico segments arranged in an icosagonal puck, is paired with a matching rotating spin source placed 4.12 mm below. Both are enclosed in $\mu$-metal to suppress magnetic fields. The torque signal is detected at 10 times the spin source’s rotation frequency, helping isolate it from base and low-frequency noise.\\

\noindent\textit{2.2~Alkali-metal Magnetometers}\\

Alkali-metal magnetometers use the valence electrons of alkali atoms---such as K, Rb, and Cs---as magnetic-field sensors. Optical pumping transfers atomic vapor (natural or enriched isotopes) into specific ground-state sublevels, creating net spin polarization. A magnetic field induces spin precession, which is usually detected via Faraday rotation.
The fundamental sensitivity of alkali-metal magnetometers is limited by spin-projection noise~\cite{Allred2002PRL, Kominis2003N, Budlker22013PM}:
\begin{equation}
\label{deltaB}
\delta B=\frac{1}{\gamma \sqrt{nT_2Vt}},
\end{equation}
where $\gamma$ is the gyromagnetic ratio of the electron, $n$ the atomic density, $T_2$ the transverse relaxation time, $V$ the effective cell volume, and $t$ the integration time. Improving sensitivity involves increasing $n$, $V$, and $T_2$. Atomic density is determined by temperature; $V$ can be enlarged by using bigger vapor cells or multipass designs~\cite{sheng2013PRL}. The transverse relaxation time is mainly limited by spin-exchange collisions (SECs) between atoms~\cite{Budlker22013PM}.
The effect of SECs arises from the two hyperfine levels of the alkali ground state, which have opposite $g_F$ factors. SECs induce random transitions between these levels, effectively altering the Larmor precession axis and reducing coherence. In 1973, Happer \textit{et al.}~\cite{Happer1973PRL} discovered that this relaxation can be suppressed when operating in the so-called spin-exchange relaxation-free (SERF) regime, where $\omega \ll \Gamma_{SE}$, with $\omega$ the Larmor frequency and $\Gamma_{SE}$ the SEC rate. This condition requires high atomic density and low magnetic fields.

In 2002, the first SERF magnetometer achieved a sensitivity of $10~\mathrm{fT/\sqrt{Hz}}$~\cite{Allred2002PRL}, which was later improved to $0.54~\mathrm{fT/\sqrt{Hz}}$ using a gradiometer arrangement~\cite{Kominis2003N}.
The best sensitivity to date has reached $0.16~\mathrm{fT/\sqrt{Hz}}$ using ferrite shielding and a gradiometer configuration~\cite{Dang2010APL}.
Even commercial SERF magnetometers---with lower sensitivity but compact form factors---are widely used in array configurations for such searches~\cite{Kim2018PRL, Kim2019NC, Wu2022PRL}.
Moreover, by applying a magnetic bias field and increasing atomic density, SERF magnetometers can be extended to probe exotic spin-dependent interactions at frequencies up to $\sim$10 kHz~\cite{Bloch2023NC}.\\

\noindent\textit{2.3~Atomic Co-magnetometers}\\

Atomic co-magnetometers operate by simultaneously interrogating multiple spin species---typically one electron and one nucleus---within the same atomic cell. By exploiting their differing responses to magnetic and exotic fields, these systems suppress sensitivity to magnetic noise while retaining sensitivity to exotic interactions.
A typical co-magnetometer consists of an alkali-metal electron spin and a noble-gas nuclear spin. Magnetic or exotic fields rotate the polarized noble-gas nuclei, and this rotation is sensed by the co-located alkali spin, which serves as a magnetometer.
State-of-the-art atomic co-magnetometers have achieved energy sensitivities as low as $10^{-23}~\mathrm{eV/\sqrt{Hz}}$ in certain implementations~\cite{Wei2023PRL}.
Atomic co-magnetometers typically operate in two distinct regimes---self-compensation and strong-coupling---depending on the intensity of the holding magnetic field~\cite{Kornack2002PRL}.

In the self-compensation regime, magnetic noise is suppressed by canceling the fluctuating field sensed by the electron spin with the field generated by nuclear magnetization, which adiabatically follows the external field. This cancellation is effective for fields with frequencies much lower than the nuclear Larmor frequency and relaxation rate. As a result, self-compensation co-magnetometers are optimized for detecting exotic interactions modulated at low frequencies (typically $\sim$Hz), where their sensitivity peaks.
In Ref.~\cite{Vasilakis2009PRL}, a K-$^3$He co-magnetometer achieved a sensitivity of $0.75~\mathrm{fT/\sqrt{Hz}}$ at 0.18 Hz. Using this setup, Vasilakis \textit{et al.} searched for a dipole-dipole interaction $\V_3$ by modulating the spin polarization of a nearby $^3$He cell via adiabatic fast passage. In a separate study~\cite{Lee2018PRL}, a similar K-$^3$He system was used to search for the $\V_{9+10}$ interaction by modulating two Pb brick stacks placed symmetrically along the y-axis.
Smiciklas \textit{et al.}~\cite{Smiciklas2011PRL} achieved a sensitivity of $2~\mathrm{fT/\sqrt{Hz}}$ at 0.023 Hz by periodically rotating the entire co-magnetometer to probe Lorentz-violation backgrounds. Since $^{21}$Ne has a gyromagnetic ratio an order of magnitude smaller than $^3$He, co-magnetometers based on $^{21}$Ne offer improved energy resolution. For example, a Rb-$^{21}$Ne setup was used in Ref.~\cite{Almasi2020PRL} to search for $V_{\rm PP}$ interactions between polarized SmCo$_5$ electrons and $^{21}$Ne neutrons. A K-Rb-$^{21}$Ne system was later employed to search for spin-velocity interactions mediated by the {\color{black}$Z^\prime$} boson via a rotating tungsten ring~\cite{Wei2022NC}.
The $I=3/2$ spin of $^{21}$Ne also enables searches for tensor components of Lorentz-violation fields~\cite{Smiciklas2011PRL}.

The strong-coupling regime arises when the resonance frequencies of alkali-metal electrons and noble-gas nucleons match, satisfying:
\begin{equation}
\begin{aligned}
\gamma_e(B_z + B^0_n) &= \gamma_n(B_z + B^0_e),
\end{aligned}
\end{equation}
where $\gamma_e$ and $\gamma_n$ are the gyromagnetic ratios, and $B^0_e$, $B^0_n$ are the magnetizations from the polarized electron and nucleon, respectively. Given that $B^0_e \sim 10~\mu\mathrm{G}$ and $B^0_n \sim 1~\mathrm{mG}$, the required bias field $B_z$ is typically on the milligauss scale.
In this regime, electron and nuclear spins are strongly coupled, forming a hybrid oscillation mode with enhanced bandwidth. The nuclear response resonantly drives the electron spin, amplifying the signal. This feature is particularly beneficial for broadband searches. In Ref.~\cite{Wei2025ropp}, a K-Rb-$^{21}$Ne co-magnetometer operating in the strong-coupling regime was used to search for axion-like fields across a broad frequency range [$10^{-2}$, $10^3$] Hz. Enclosed in a five-layer $\mu$-metal shield lined with Mn-Zn ferrite, the device reached a sensitivity of $0.78~\mathrm{fT/\sqrt{Hz}}$ at $\sim$30 Hz, limited by magnetic shielding noise. A similar setup was used to search for the exotic interaction $V_{\rm VA}$, generated by rotating cubic lead blocks, as described in Ref.~\cite{heng2025PNAS}. 
In this configuration, the broadband nature of the comagnetometer enhances the contribution from higher harmonics of the exotic signal.

Furthermore, spin-exchange interactions between alkali metals and noble gases in co-magnetometers enable rich dynamical behavior when the external magnetic field is properly tuned, enhancing their potential for precision measurements. Such tuning induces constructive or destructive interference between the spin species, allowing for signal amplification or noise suppression.
In the frequency domain, this interference produces a characteristic Fano lineshape~\cite{miroshnichenko2010RMP, huang2025PRL}, marked by a peak at the nuclear resonance and a dip nearby---arising from out-of-phase responses under equal field amplitudes. Exploiting this effect, a novel self-compensation scheme has been proposed to suppress magnetic noise at higher frequencies (up to 160 Hz), significantly extending the bandwidth of self-compensated co-magnetometers~\cite{Qin2024PRL}.\\

\noindent\textit{2.4~Beam Methods}\\

\noindent\textit{\S~2.4.1~Neutron Beam Methods}\\

Because neutrons are electrically neutral, they can penetrate deeply into materials without Coulomb repulsion. As a spin-$1/2$ particle, it can be polarized and analyzed, making it an excellent probe of spin-dependent new physics~\cite{Dubbers2011RMP}. Its neutrality also allows subtle phase shifts from exotic interactions to accumulate over long distances, making them measurable through techniques such as interferometry.
As a result, slow neutrons are particularly well-suited for detecting exotic interactions at mesoscopic to macroscopic scales~\cite{Piegsa2012PRL, Baessler2007PRD, Yan2013PRL, Afach2015PLB, Haddock2018PLB, Lehnert2014PLB, Lehnert2017PLB, Voronin2009JETP, Voronin2020JETP}. A wide range of experimental techniques---neutron interferometry, scattering, optics, {\color{black}gravity resonance spectroscopy}, and spin rotation---have been developed to test fundamental symmetries and search for exotic spin-dependent interactions.
A comprehensive review of these neutron-based methods is provided in Ref.~\cite{Sponar2021NRP}. In the following, we focus on the search for exotic spin-dependent interactions using the neutron spin rotation technique in slow-neutron beams.

In neutron beam experiments, spin rotation induced by exotic spin-dependent interactions can be detected using Ramsey interferometry. A representative example was demonstrated by Piegsa \textit{et al.}~\cite{Piegsa2012PRL}. In this setup, polarized neutrons pass through a $\pi/2$ spin flipper, a magnetic field region with the leading field perpendicular to the neutron momentum, a second $\pi/2$ flipper, and finally a polarization analyzer. Scanning the spin flipper frequency produces a Ramsey fringe pattern in the analyzer signal. When a mass plate is placed beneath the neutron beam, the spin-velocity interaction $V_{\rm AA}$ induces an additional phase shift, resulting in a measurable shift in the Ramsey fringe frequency. To reduce common-mode noise, a second, co-propagating neutron beam is used as a reference.
Recently, Ref.~\cite{Fierlinger2024PRL} proposed a neutron beam experiment at the European Spallation Source (ESS) employing Ramsey interferometry to search for ultralight axion dark matter. The setup uses the coupling between axions and neutron spins, the high-intensity ESS HIBEAM neutron line, and the Ramsey-separated oscillating-field method. By comparing the neutron spin precession frequency to an external reference, the experiment aims to detect axion-induced frequency shifts. With a one-year integration time, the projected sensitivity improves by 2-3 orders of magnitude, covering the axion mass range $10^{-22}$-$10^{-16}$ eV.

The neutron spin rotation effect can also be measured directly by analyzing the neutron polarization before and after interaction. Since neutron beams can penetrate directly through the source mass, this method enables sensitivity to shorter interaction ranges. A notable example is the measurement of neutron spin rotation in liquid $^4$He to search for the $\V_{12+13}$ type interaction~\cite{Yan2013PRL}.
Experimental results in liquid $^4$He have shown no significant spin rotation~\cite{Snow2011PRC, Swanson2019PRC}, thereby placing upper limits on exotic parity-odd interactions. In 2013, Yan \textit{et al.}~\cite{Yan2013PRL} used these measurements to constrain $g_Vg_A$ by probing the spin-velocity potential $\V_{12+13}$. Although initially intended to detect the parity-violating weak nucleon-nucleon interaction, the experiment remains the most sensitive probe of spin-velocity interactions (Eq.~(\ref{va1_v})) at short distances. The resulting constraint exceeds previous laboratory limits in the micrometer-to-meter range by over $\sim 10^7$ and, more than a decade later, still holds the record for precision in this regime.
However, further improvements are limited by SM weak-interaction backgrounds, unless those contributions can be calculated with high theoretical accuracy. These experimental results have also been used to constrain spacetime nonmetricity~\cite{Lehnert2017PLB} and in-matter torsion~\cite{Lehnert2014PLB}.
Using similar techniques, the $\V_{4+5}$-type interaction was investigated in Ref.~\cite{Haddock2018PLB}. In that experiment, two orthogonally oriented polarizers block the neutron beam. If exotic interactions cause a small spin rotation, a fraction of neutrons can pass through the second polarizer, producing a detectable signal.

{\color{black}In the Earth’s gravitational field, UCNs occupy non-equidistant quantum gravitational states~\cite{nesvizhevsky2002N}. When UCNs are specularly reflected by a horizontal mirror, the combined effect of the mirror’s potential barrier and the Earth’s linear gravitational potential forms an effective potential well. Within this well, the vertical motion of UCNs is quantized into discrete energy levels rather than a continuous spectrum. Since the characteristic spatial extent of the UCN wavefunction in this potential is on the order of micrometers, transitions between these eigenstates provide a sensitive probe of exotic spin-dependent interactions~\cite{Baessler2007PRD, Ivanov2021PLB}.}

{\color{black}The development of Gravity Resonance Spectroscopy (GRS)~\cite{jenke2011NP} has further advanced this approach by enabling high-precision measurements of transition frequencies between quantum states, making it a versatile tool for testing a wide class of exotic potentials~\cite{Jenke2014PRL, Ivanov2016PRD, cronenberg2018NP}. As in magnetic resonance techniques, the use of Ramsey-type separated oscillating fields can substantially enhance the sensitivity of GRS. For confined UCNs, a Ramsey-type measurement with an integration time of $T = 130$~s is predicted to constrain $|g^N_s g^n_p|$ at the level of $\sim 10^{-23}$ for $\lambda \sim 10^{-4}$~m~\cite{Abele2010PRD}.}\\

\noindent\textit{\S~2.4.2~Muon Beams}\\

\begin{figure}[htbp]
    \centering
    \includegraphics[width=0.95\linewidth]{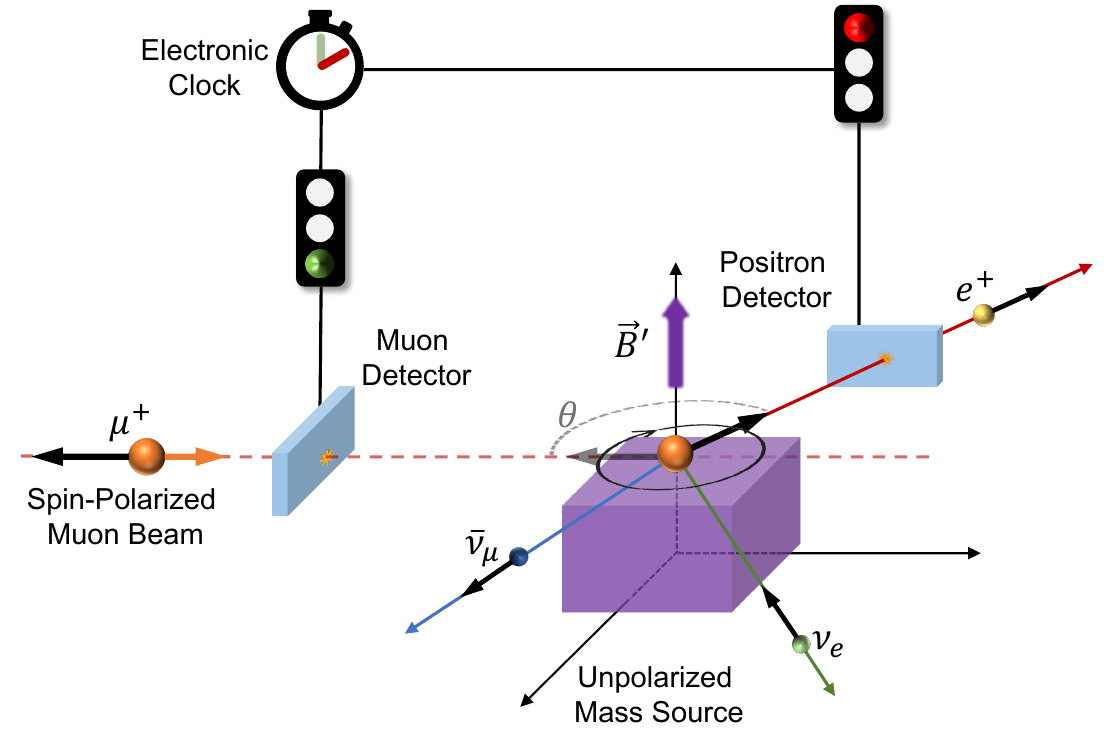}
    \caption{\justifying Conceptual illustration of a search for the monopole-dipole interaction using time-differential $\mu{\rm SR}$ measurement. Black arrows denote particle spins. A spin-polarized muon (shown as an orange sphere) enters from the left and is initially detected by the muon detector, triggering the electronic clock at time $t_0$. The muon then travels toward the unpolarized mass source, where an effective magnetic field $\vec{B}^\prime$---generated by an exotic spin-dependent interaction between the muon spin and nucleons in the mass---induces spin precession. Upon decay, the muon emits a positron preferentially along its spin direction. The positron is detected by a positron detector, which defines the emission direction and stops the clock at time $t$. By analyzing the time-dependent anisotropy in the positron emission distribution, the strength of the exotic interaction can be determined. Adapted from Ref.~\cite{Amato2024}.}
    \label{musr}
\end{figure}

Given the long-term development of muon beam techniques and their successful application in magnetic material characterization---particularly through the muon spin rotation, relaxation, or resonance (${\rm\mu SR}$) method~\cite{Amato2024}---this technique also offers a powerful tool for testing new physics effects. The muon produced from pion decay is naturally $100\%$ polarized, making it an ideal probe for spin-dependent interactions.
A schematic illustration of searching for exotic spin-dependent interactions using the ${\rm\mu SR}$ method is shown in Fig.~\ref{musr}. The basic idea is as follows: a beam of spin-polarized muons passes over the surface of an unpolarized mass source. An {\color{black}electronic clock} records the start timestamp $t_0$ as the muon passes through the muon detector. When the muon reaches the region above the mass source, the effective magnetic field $\vec{B}^\prime$---induced by exotic interactions---rotates the muon spin.
The muon subsequently decays via the weak interaction into a positron and two neutrinos. The emitted positron is detected by a positron detector, which records an arrival timestamp $t_1$. Due to the correlation between the positron emission direction and the muon spin orientation at the time of decay, the spin rotation angle can be inferred by analyzing the angular distribution of the detected positrons. By combining this angular information with the arrival time $t_1$, the strength of the effective field $\vec{B}^\prime$ can be determined.\\

\noindent\textit{\S~2.4.3~$^3$He Beams}\\

Precision measurements of nuclear, molecular, and atomic properties have a long history, dating back to the resonance method pioneered by Rabi \textit{et al.} in 1938~\cite{Rabi1938PR}. With the advent of laser cooling techniques, atomic beam methods have found increasing applications in fundamental physics, such as the search for the electron EDM. However, they have not yet been experimentally applied to probe exotic spin-dependent interactions. Compared with neutron or muon beams, atomic beams offer distinct advantages, including higher particle flux and lower operating costs.

Despite the high sensitivity of the aforementioned neutron spin rotation experiment, further improvements face substantial challenges. The precision is limited by statistical uncertainty, i.e., the total neutron count. Increasing the neutron flux is technically demanding, and the existing experiment already requires nearly a year of integration time. Moreover, any future observation of a nonzero rotation would be difficult to interpret due to the poorly known contribution from SM parity violation. The expected parity-odd spin rotation angle in liquid helium is on the order of $10^{-6}$ to $10^{-7}$ rad/m~\cite{Stodolsky1974PLB}, which lies near the experimental sensitivity limit, while theoretical predictions remain imprecise.
Alternative approaches using polarized noble gases allow for large numbers of probing particles, but also present limitations. Magnetic storage introduces systematic uncertainties, and sealed glass cells constrain the minimum probe-to-source distance, making it difficult to generate significant relative velocity, which is crucial for detecting velocity-dependent interactions. While neutron beams permit nearly zero probe-to-source separation and operate in well-shielded environments~\cite{Yan2013PRL, Snow2011PRC}, neutron flux remains a key limiting factor.

To overcome these issues, Yan \textit{et al.}~\cite{Yan2014EPJC} proposed using nuclear spin-polarized $^3$He atomic beams to search for spin-dependent short-range interactions. While other spin-1/2 species such as $^{129}$Xe could be used, the $^3$He atomic beam technique is particularly mature and has been extensively applied in surface dynamics studies in condensed matter physics~\cite{DeKieviet1995PRL, Jardine2001RSI}. A comprehensive review of polarized $^3$He spin-echo techniques is provided in Ref.~\cite{Jardine2009PSS}.
The $^3$He beam is produced using the standard atomic beam method~\cite{miller1988}, where compressed gas is expanded through a nozzle into a vacuum. Beam velocity can be tuned via the nozzle temperature. Polarization is achieved using hexapole magnet-based beam polarizers~\cite{Jardine2001RSI}. This setup can generate beams with high intensity ($1.5\times10^{14}$ atoms/s~\cite{Fouquet2005RSI}), narrow collimation (2 mm diameter~\cite{Fouquet2005RSI}), and high polarization (above $90\%$~\cite{DeKieviet1995PRL}).
The polarized $^3$He beam is directed over a dense target, such as a lead plate. If exotic spin-dependent interactions exist, they would induce a measurable rotation in the spin polarization of the beam. In contrast to neutron-based experiments, which accumulate $\sim10^{14}$ neutrons over a year, a comparable number of $^3$He atoms can be delivered in just one second using this method.\\

\subsection{Force-based  Detection}\label{expinv_2}

In this subsection, we review several experimental techniques that search for exotic spin-dependent interactions via force measurements. A representative approach involves mechanical oscillators, whose motion is described by the harmonic oscillator model (Eq.\ref{ho}). A high quality factor ($Q$) is essential for enhancing the response to forces near resonance ($\omega\sim \omega_0$) while suppressing off-resonance noise ($|\omega-\omega_0| \gg \omega_0/Q$). Oscillators used in these searches typically achieve $Q$ values in the thousands, while nanoscale devices at cryogenic temperatures can reach up to $10^6$~\cite{Bachtold2022RMP}. Both the natural frequency and $Q$ generally increase as device dimensions decrease. Higher $Q$ values are also attainable in vacuum and cryogenic environments, where dissipation is minimized.

Mechanical oscillators play a central role in ultrasensitive force detection, including nanoscale magnetometry~\cite{Rossel1996JAP} and atomic force microscopy (AFM)~\cite{Grütter1991JAP}. Their force sensitivity is fundamentally limited by thermal noise and can be improved by lowering the temperature. For example, Chiaverini \textit{et al.}~\cite{Chiaverini2003PRL} achieved a sensitivity of $10^{-16}\mathrm{N/\sqrt{Hz}}$ at 10 K. Advances in microfabrication have further enabled microscale oscillators, ideal for probing short-range forces. These attributes make mechanical oscillators powerful tools for testing deviations from Newtonian gravity at sub-millimeter scales~\cite{Chiaverini2003PRL, Long2003CRP, Wang2016PRD, Manley2024PRD}, and more recently, for exploring exotic spin-dependent interactions~\cite{Leslie2014PRD, Ding2020PRL, Ren2021PRD, Wang2023PRD}.

Mechanical oscillators generally operate in two modes: cantilever and torsional. In the cantilever mode, the structure vibrates transversely about its clamped base, whereas torsional oscillators rotate about a central axis. Torsional modes offer greater immunity to ambient vibrations and can operate in resonance without extensive vibration isolation. In contrast, cantilever oscillators---though capable of resonance-based amplification---are more sensitive to external disturbances and are often operated off-resonance.

\subsubsection{On-Resonance Measurement Method}\label{expinv_22}
\quad\\
\noindent\textit{1.1~Torsional Oscillator}\\

The torsion oscillator was employed in the experiment proposed by Leslie \textit{et al.}~\cite{Leslie2014PRD} to search for all 15 possible exotic spin-dependent interactions. The setup was adapted from a previous apparatus used to constrain Yukawa-type spin-independent interactions such as $\V_1$~\cite{Long2003N}, by covering half of the small rectangular area of the detector with a spin-polarized material, DyIG.
The oscillator consists of two coplanar rectangular plates joined by a central segment. Two symmetrically attached transducers on the larger plates measure the rotation angle. A polarized or unpolarized source mass is mounted on a cantilever located beneath the torsion oscillator and is driven at its resonance frequency (nominally 1 kHz) using a piezoelectric bimorph.
The planar geometry is optimized to suppress Newtonian background forces while maximizing the spatial overlap between the source and detector within the relevant interaction range. The first antisymmetric torsion mode is selected for its high $Q$ factor, which enhances sensitivity to weak spin-dependent forces.\\

\subsubsection{Off-Resonance Measurement Method}\label{expinv_21}
\quad\\
\noindent\textit{2.1 Cantilever Oscillator}\\

\begin{figure}
    \centering
    \includegraphics[width=0.9\linewidth]{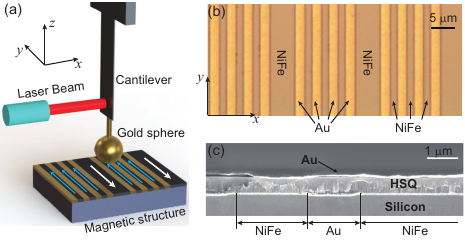}
    \caption{\justifying (a) Schematic of the cantilever-based search for exotic spin-dependent interactions. The interaction between an unpolarized gold sphere and a magnetic structure induces a minute deflection of the cantilever to which the sphere is attached. This deflection is measured using an optical fiber interferometer. The magnetic structure consists of periodic magnetic stripes, producing an alternating exotic force on the gold sphere as the structure is translated along the $x$-axis. By driving the magnetic structure with sinusoidal motion, the resulting force is modulated into higher harmonics of the driving frequency, facilitating the separation of the exotic signal from spurious mechanical noise. The signal is extracted using a lock-in detection method referenced to the relevant harmonics. (b) and (c): Images of the magnetic structure. Reprinted from Ref.~\cite{Ding2020PRL}.}
    \label{fig:hust_cant}
\end{figure}

A cantilever oscillator behaves like a spring, obeying Hooke’s law: $F = kx$, where $F$ is the applied force, $x$ is the cantilever displacement, and $k$ is the spring constant. Since the displacement is inversely proportional to $k$, a soft cantilever with a low spring constant is more sensitive to weak forces. To detect the tiny motion (on the order of ${\rm\sim p m}$), fiber optical interferometry is commonly employed.
In cantilever oscillator experiments, the source mass is modulated near the cantilever at a fixed frequency, and the resulting response of the oscillator is observed. If exotic interactions exist, the coupling between the source mass and the test mass mounted on the cantilever will induce oscillations at the modulation frequency or its harmonics.
In the search for exotic spin-dependent interactions, a cantilever oscillator was employed in the work of Ding \textit{et al.}~\cite{Ding2020PRL}, whose operating principle is illustrated in Fig.~\ref{fig:hust_cant}. In their setup, a cantilever with a gold sphere attached to its free end is paired with a magnetic structure designed to probe the spin-velocity-dependent interaction $\V_{4+5}$. The magnetic structure consists of periodic magnetic strips that are translated back and forth in a direction perpendicular to their magnetization. This modulation shifts the exotic signal to higher harmonics, enabling effective discrimination from spurious signals associated with mechanical motion.

\subsubsection{Systematic Effects in Short-Range Force Measurements}\label{expinv_23}

As discussed earlier, mechanical oscillators are effective tools for detecting weak forces at short distances. However, two main challenges must be addressed: the reduction in signal strength as the source and probe are miniaturized, and the rapid increase in background noise at short distances---both of which also affect the detection of exotic interactions.
Notably, background forces such as electromagnetic interactions become increasingly significant at short ranges. The electrostatic force from surface potentials scales as $r^{-2}$, while the Casimir force scales as $r^{-4}$~\cite{Long2003CRP, Ren2021PRD}. Without mitigation, these backgrounds can easily overwhelm the signal.
To address this, existing experiments have implemented several strategies. A well-grounded, rigid metal shield is commonly placed between the source and oscillator to suppress direct electromagnetic coupling~\cite{Long2003CRP, Chiaverini2003PRL, Leslie2014PRD}. The Casimir force, which depends solely on electronic properties, can be canceled by a common metallic coating thicker than the plasma wavelength, rendering it independent of underlying material differences. This allows differential measurement techniques to subtract the Casimir contribution~\cite{Decca2005PRL}.
Alternatively, suppose the source is composed of alternating high- and low-density materials. In that case, the oscillator experiences a static Casimir force from the uniform coating, while the exotic interaction induces a time-varying signal with a known frequency. This modulated signal can be extracted using lock-in detection methods~\cite{Wang2016PRD, Chen2016PRL, Ding2020PRL, Ren2021PRD, Wang2023PRD}.\\

\section{Existing Experimental Constraints}\label{expcon}

In this section, we compile the existing best-known experimental constraints on exotic spin-dependent interactions summarized in Sec.~\ref{theo}. We primarily focus on the Yukawa-type interactions mediated by ALPs, i.e., Eqs.~(\ref{va1_v})--(\ref{sp2_vgg}), as these have been the central targets of experimental searches over the past two decades and have been explored across various physical systems. Meanwhile, new experiments and notable results continue to emerge.
As discussed earlier, the interactions in Eqs.~(\ref{va1_v})--(\ref{sp2_vgg}) can be derived solely from the assumption of rotational invariance, independent of any specific theoretical model. As a result, the corresponding experimental limits can be reinterpreted in the context of various new physics scenarios, and vice versa.
Some of the most stringent bounds on these interactions have been obtained by reinterpreting data from experiments originally intended to probe other phenomena, such as Lorentz and CPT violation or atomic and molecular EDMs.
In addition to searches for ALP-mediated interactions, increasing efforts have been devoted to probing couplings between axion fields and spin as discussed in Sec.~\ref{theo_3}. 
These efforts have led to meaningful results, and we include a summary of the current constraints on the axion-nucleon coupling $g_{\rm aNN}$ and axion-electron coupling $g_{aee}$, which collectively cover the axion mass range $10^{-24}<m<10^{-10}$~eV.

Constraints on these Yukawa-type interactions are organized by coupling types: scalar-scalar (SS) $g_Sg_S$, scalar-pseudoscalar (SP) $g_Sg_P$, pseudoscalar-pseudoscalar (PP) $g_Pg_P$, vector-vector (VV) $g_Vg_V$, vector-axial-vector (VA) $g_Vg_A$, and axial-vector-axial-vector (AA) $g_Ag_A$. For each coupling, constraints are further grouped by the fermion pairs involved---lepton-lepton, lepton-nucleon, and nucleon-nucleon---with specific particle pairs indicated in the plot legends.
Most searches focus on interactions between electrons ($e$), protons ($p$), and neutrons ($n$). When an unpolarized nucleon acts as the source, it is denoted by $N$, without distinguishing between $p$ and $n$. Atoms involving---such as muons ($\mu$), positrons ($e^+$), and antimatter species---have also been investigated.
Each coupling type may correspond to multiple spin-dependent potentials, as summarized in Table~\ref{tab1}. 
Thus, experiments constraining the same coupling might probe different potentials.
The specific potential associated with each experimental constraint is detailed in the figure captions.
Note that the confidence level (CL) for each constraint varies across experiments and is presented in its original form.
{\color{black}To ensure a rigorous comparison, the experimental constraints from various original works have been converted into the unified $g_ig_j$ convention used throughout this manuscript concerning the different formalization and velocity definition of these interactions.}
Since a given potential can arise from different couplings, constraints on a particular configuration may apply to multiple coupling types. 
For example, the experiment by Ficek \textit{et al.}~\cite{Ficek2017PRA} constrained the $\V_{4+5}$ potential, which, according to Table~\ref{tab1}, corresponds to both $g_Sg_S$ and $g_Vg_V$ couplings.

The Yukawa interaction’s range dependence implies that each experimental setup loses sensitivity below a certain scale due to exponential suppression. Atomic spectroscopy provides the strongest bounds at short distances, while for $\lambda \gtrsim 10^{7}~\mathrm{m}$, astronomical sources yield tighter constraints. This trend, seen across all constraint plots, underscores the need for complementary experimental approaches. Constraints generally tighten with increasing $\lambda$ (or decreasing ALP mass) as more particles contribute, and level off once the interaction range becomes sufficiently large.
{\color{black} We primarily focus on exotic spin-dependent interactions mediated by ALPs with masses below the MeV scale, which are accessible through low-energy precision measurements. For ALPs heavier than $\sim$1~MeV, their implications in particle physics---such as contributions to the muon anomalous magnetic moment~\cite{Marciano2016PRD} and their role as mediators between the Standard Model and dark matter---have attracted substantial attention in collider and fixed-target experiments (see Refs.~\cite{Joerg2016PLB, dolan2017JHEP, Lucian2019PLB, dobrich2019JHEP, Beacham2020, Brdar2021PRL}). These high-energy searches are essential for probing this mass regime. In contrast, low-energy tabletop experiments, while compact, cost-effective, and rapidly deployable, provide highly sensitive probes of ultralight ALPs and therefore serve as an efficient and powerful complement to collider-based approaches.}\\

\subsection{Constraints on \texorpdfstring{$g_Sg_S$}{gsgs}}\label{expcon_1}

\begin{figure}[htbp]
\centering
\subfigure{
\includegraphics[width=0.75\textwidth]{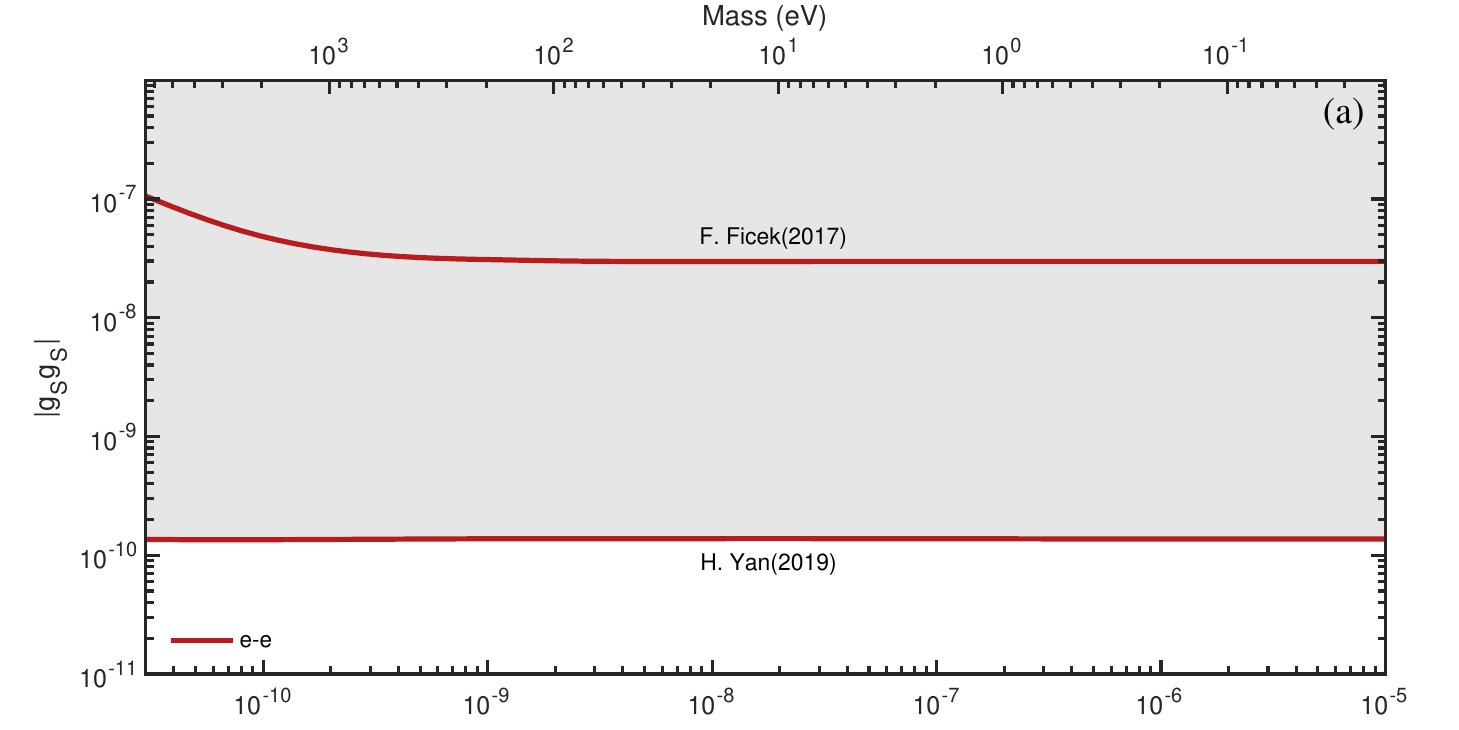}}
\subfigure{
\includegraphics[width=0.75\textwidth]{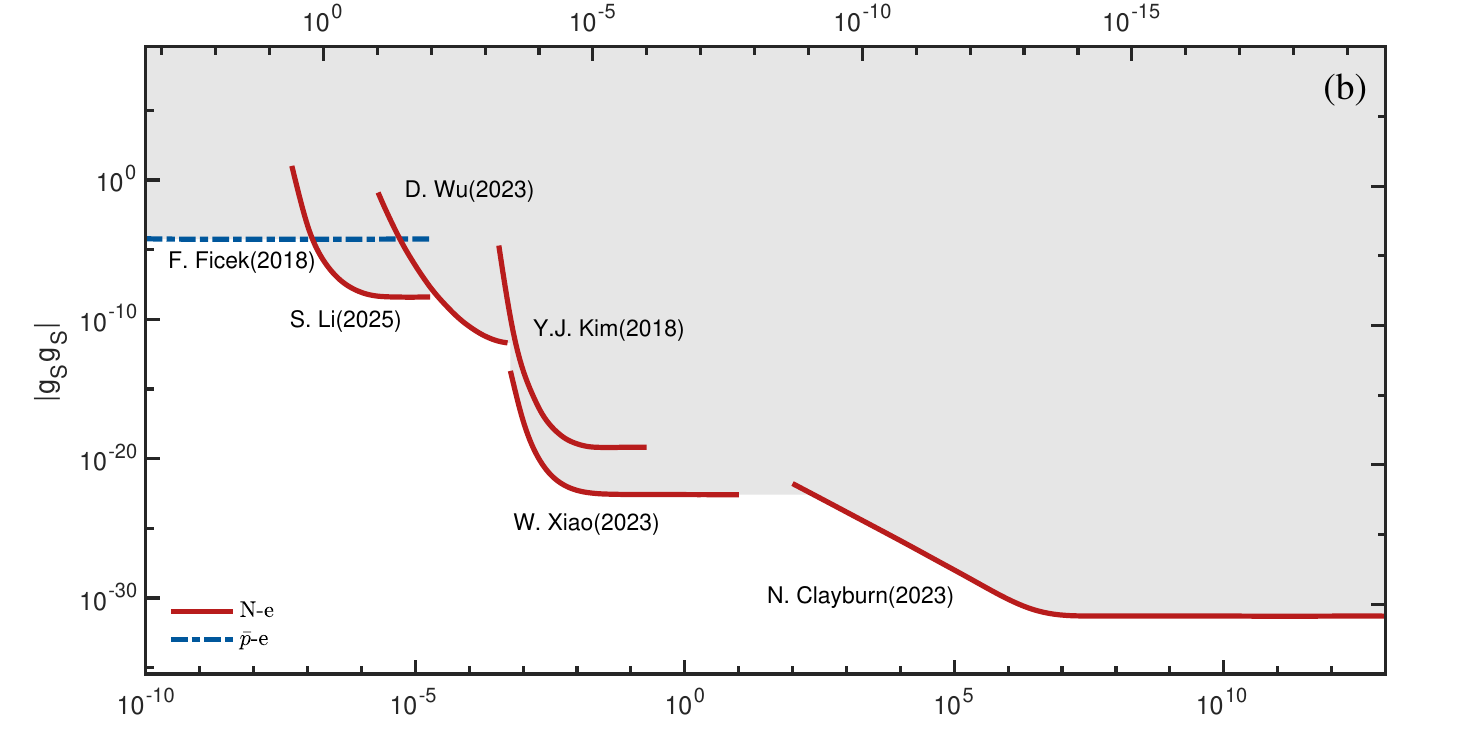}}
\subfigure{
\includegraphics[width=0.75\textwidth]{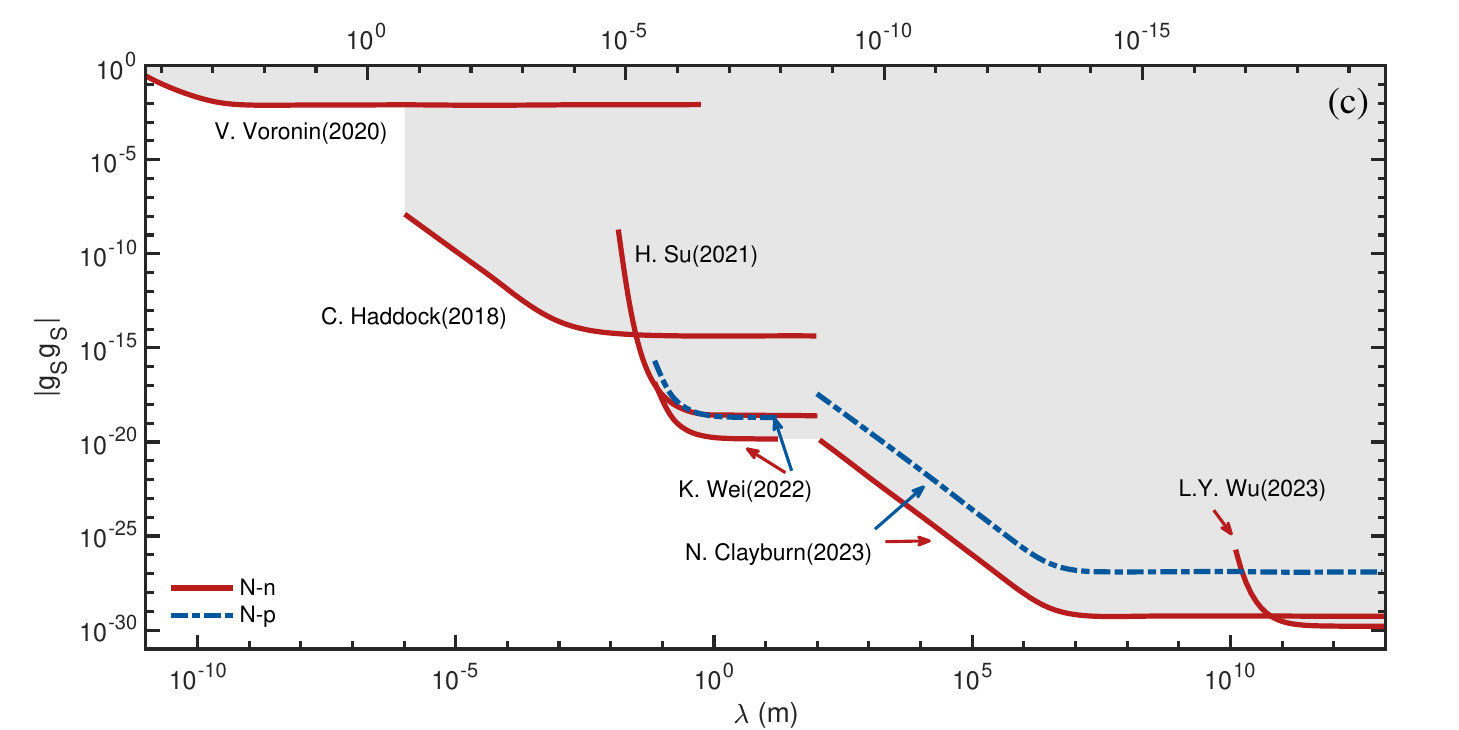}}
\caption{\justifying Experimental constraints on the scalar-scalar coupling $g_Sg_S$ as a function of interaction range or ALP mass. The curves represent upper bounds on the coupling strength, with shaded regions indicating excluded parameter space. In the legend, labels such as $N$-$n$ denote particle pairs, where the nucleon ($N$) couples to the left vertex and the neutron ($n$) to the right vertex in the Feynman diagram shown in Fig.~\ref{fig3}.
The constraints shown correspond to the $\V_{4+5}$ potential: Refs.~\cite{Ficek2017PRA, Yan2019EPJC, Ficek2018PRL, Li2025PRL, Wud2023PRL, Kim2018PRL, Xiao2023PRL, Clayburn2023PRD, Voronin2020JETP, Haddock2018PLB, Su2021SA, Wei2022NC, Wu2023PRL}.}
\label{fgsgs}
\end{figure}

The exotic interaction derived from the SS coupling involves two potentials: $\V_1$ and $\V_{4+5}$. The spin-independent $\V_1$ also appears in VV coupling; since this work focuses on spin-dependent effects, readers are referred to Refs.~\cite{Long2003CRP, adelberger2003ARNPS, Snow2022Sy} for details on spin-independent interactions.
The spin-velocity-dependent $\V_{4+5}$ term from SS coupling is constrained by various experiments, as shown in Fig.~\ref{fgsgs}: (a) for lepton-lepton, (b) for lepton-nucleon, and (c) for nucleon-nucleon pairs. 

\subsubsection{lepton-lepton pairs}
In the lepton-lepton sector, constraints arise from atomic-scale measurements: helium fine-structure spectroscopy~\cite{Ficek2017PRA} and the anomalous magnetic moment of the electron~\cite{Yan2019EPJC}.

\subsubsection{lepton-nucleon pairs}
For lepton-nucleon pairs, constraints on $e$-$\bar{p}$ arise from antiprotonic helium spectroscopy~\cite{Ficek2018PRL}. In the range $10^{-7} \lesssim \lambda \lesssim 10^{-3}$~m, compact experimental setups employing nitrogen-vacancy (NV) centers and cantilevers have produced strong bounds~\cite{Wud2023PRL, Ding2020PRL}, benefiting from reduced probe-source separation and high-frequency modulation.
The ensemble-NV-diamond magnetometer was employed by Wu \textit{et al.}~\cite{Wud2023PRL} to search for the $\V_{4+5}$ interaction, which was generated by a motion-modulated lead sphere placed in close proximity.
Li \textit{et al.}~\cite{Li2025PRL} employed a similar cantilever setup as described in Sec.~\ref{expinv_21}, with notable improvements including an increased modulation frequency and a higher spin-density source material compared to that used by Ding \textit{et al.} in Ref.~\cite{Ding2020PRL}, to place constraints on the potential $\V_{4+5}$.

For $\mathrm{10^{-3}\lesssim\lambda\lesssim 10}$~m, atomic magnetometers dominate.
BGO crystals serve as unpolarized sources due to their high nucleon density and low susceptibility~\cite{Yamamoto2003TNS}. 
Kim \textit{et al.}~\cite{Kim2018PRL} used commercial SERF magnetometers, whereas Xiao \textit{et al.}~\cite{Xiao2023PRL} developed a custom SERF magnetometer using spatially separated pump/probe beams and a gradiometer, reaching a sensitivity of $\mathrm{1.4~fT/\sqrt{Hz}}$.
For $\lambda\gtrsim 10^2$~m, using Earth as the source, Clayburn \textit{et al.}~\cite{Clayburn2023PRD} reanalyzed data from the spin torsion pendulum experiment~\cite{Heckel2008PRD}, setting the most stringent constraint to date.

\subsubsection{nucleon-nucleon pairs}
For nucleon-nucleon pairs, constraints in the range of $10^{-11} \lesssim \lambda \lesssim 10^{-2}$~m come from neutron spin rotation~\cite{Haddock2018PLB, Voronin2020JETP}. 
Su \textit{et al.}~\cite{Su2021SA} used a spin amplifier and rotating BGO source to set limits at $\lambda\sim\mathrm{m}$ scale. Wei \textit{et al.}~\cite{Wei2022NC} improved these with a K-Rb-$^{21}$Ne co-magnetometer and tungsten source.
In the range of $10^2\lesssim\lambda\lesssim10^{10}$~m, Clayburn \textit{et al.}~\cite{Clayburn2023PRD} adopted the experimental results from Refs.~\cite{Peck2012PRA, Zhang2023PRL} to derive constraints on the $N$-$n$ and $N$-$p$ pairs using the same strategy previously applied to the $N$-$e$ pair.
For the $N$-$n$ interaction, their analysis yielded the most stringent bounds within this distance range. At larger distances ($\lambda\gtrsim 10^{11}$m), Wu \textit{et al.}~\cite{Wu2023PRL} presented improved constraints based on a reinterpretation of existing data from the Lorentz and CPT violation test conducted by Allmendinger \textit{et al.}~\cite{Allmendinger2014PRL}.

\subsection{Constraints on \texorpdfstring{$g_Sg_P$}{gsgp}}\label{expcon_5}

\begin{figure}[htbp]
\centering
\subfigure{
\includegraphics[width=0.75\textwidth]{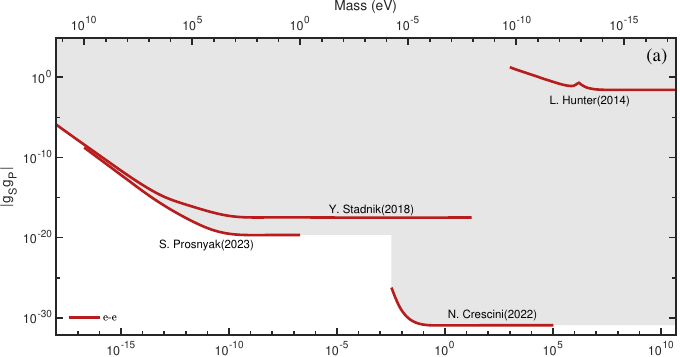}}
\subfigure{
\includegraphics[width=0.75\textwidth]{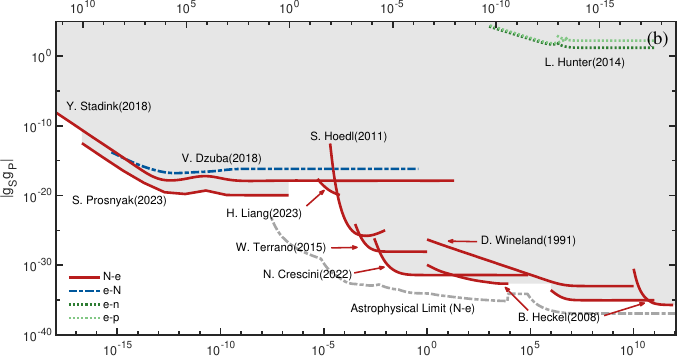}}
\subfigure{
\includegraphics[width=0.75\textwidth]{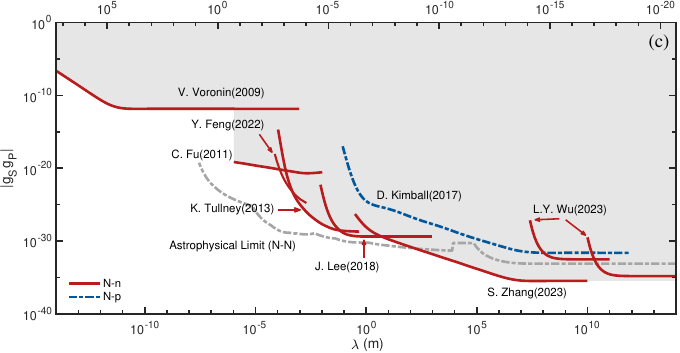}}
\caption{\justifying Experimental constraints on the scalar-pseudoscalar coupling $g_Sg_P$ as a function of interaction range or ALP mass. The curves indicate upper bounds on the coupling strength, with shaded regions representing excluded parameter space. In the legend, labels such as $N$-$n$ denote particle pairs, where the nucleon ($N$) couples to the left vertex and the neutron ($n$) to the right vertex in the Feynman diagram shown in Fig.~\ref{fig3}, corresponding to constraints on $g_S^Ng_P^n$. Curves with different line styles and colors represent results for different particle-pair combinations.
The constraints are primarily derived from experiments probing the $\V_{9+10}$ potential: Refs.~\cite{Voronin2009JETP, Stadnik2018PRL, Prosnyak2023S, Crescini2022PRD, Dzuba2018PRD, Liang2022NCR, Lee2018PRL, Hoedl2011PRL, Heckel2008PRD, Wineland1991PRL, Fu2011PRD, Feng2022PRL, Tullney2013PRL, Kimball2017PRD, Wu2023PRL, Zhang2023PRL, Terrano2015PRL}, while a few results correspond to the $\V_{15}$ potential: Ref.~\cite{Hunter2014PRL}. The astrophysical limit, shown as a gray dash-dotted line, is adapted from Ref.~\cite{OHare2020PRD}.}
\label{fgsgp}
\end{figure}

The exotic interaction arising from the SP coupling includes two types of potentials: the monopole-dipole type $\V_{9+10}$ and the dipole-dipole type $\V_{15}$. 
The $\V_{15}$ potential is of spin order 2, involves second-order derivatives ($\nabla$ power 2), and is velocity dependent. These characteristics make it particularly challenging to probe experimentally, resulting in weaker constraints on the coupling constant $g_Sg_P$.
As a result, experiments targeting the $\V_{9+10}$ potential are generally expected to place more stringent bounds on the SP coupling than those focusing on $\V_{15}$. 
Accordingly, most experimental efforts have concentrated on detecting the $\V_{9+10}$ potential. 
The compiled experimental constraints on the SP coupling are presented in Fig.~\ref{fgsgp}.

\subsubsection{lepton-lepton pairs}

With regard to the SP coupling between lepton-lepton pairs, existing experimental constraints have primarily focused on the $e$-$e$ and $\mu$-$\mu$ channels. 
In this section, we summarize the $e$-$e$ results, while the relatively sparse but intriguing constraints involving muons will be discussed separately in a later section.
Since the $\mathcal{V}_{9+10}$ interaction violates both $P$ and $T$ symmetries, EDMs of atomic systems serve as sensitive probes. 
Stadnik \textit{et al.}~\cite{Stadnik2018PRL} computed the EDMs of Hg, ThO, and HfF$^+$ induced by $\mathcal{V}_{9+10}$ using the relativistic Hartree-Fock-Dirac method. 
A similar calculation in HfF$^+$ was performed by Prosnyak \textit{et al.}~\cite{Prosnyak2023S}. 
By comparing theoretical predictions with experimental EDM results, they derived constraints on the SP interaction. 
Incorporating the latest EDM measurement~\cite{Tanya2023S}, Prosnyak \textit{et al.} improved the constraint by Stadnik \textit{et al.} for $ 10^{-16}\lesssim \lambda \lesssim 10^{-7}$~m.

In the range $10^{-2}\lesssim \lambda\lesssim 10^{5}$~m, the most stringent constraint on the SP coupling is set by Crescini \textit{et al.}~\cite{Crescini2017PLB, Crescini2022PRD}, who investigated the $\V_{9+10}$ interaction between a lead cylinder (source mass) and a paramagnetic $\mathrm{Gd_2SiO_5}$ (GSO) crystal (spin ensemble).
The effective magnetic field generated by $\V_{9+10}$ aligns spins in the GSO crystal, inducing a net magnetization detected by a SQUID. 
To isolate the signal, lead cylinders were uniformly mounted on a rotating disk driven by a brushless motor, periodically modulating the source-probe separation.
Both the GSO crystal and SQUID were enclosed in a superconducting shield and immersed in a 4.2 K liquid helium cryostat to suppress ambient magnetic noise and thermal fluctuations. 
The achieved magnetic field sensitivity at the signal frequency was $53~\mathrm{aT/Hz}$. 
After 11 hours of integration, they reported a constraint of $|g^e_S g^e_P| \leq 1.6 \times 10^{-31}$ (95\%-CL) for $\lambda \sim 1$~m.
This solid-state design offers two key advantages for probing spin-dependent interactions: (i) high spin number density, and (ii) reduced source-probe distance owing to the use of unpolarized materials. 
The use of solid paramagnets in such searches has a long history. Vorobyov and Gitarts~\cite{Vorobyov1988PLB} first employed induced ferromagnetism with SQUIDs to search for a long-range quasi-magnetic interaction mediated by a hypothetical Goldstone boson (``arion''). Later, Ni \textit{et al.}~\cite{Ni1999PRL} used a paramagnetic salt ($\mathrm{TbF}_3$) and a DC SQUID to constrain SP-type interactions mediated by ALPs and spin-gravity couplings, setting the leading bounds for over a decade.
Beyond exotic interactions, paramagnetic materials have also been widely used in precision measurements of the electron EDM~\cite{Eckel2012PRL, Kim2015PRD}.

The $\mathcal{V}_{15}$ interaction between $e$-$e$ pairs has been examined in the work of Hunter \textit{et al.}~\cite{Hunter2014PRL}.
Hunter \textit{et al.} derived constraints based on the orientation-dependent energy shifts of a spin torsion pendulum reported in Ref.~\cite{Heckel2008PRD}. 
Their analysis evaluated the interaction between polarized electrons within the Earth and the spin pendulum, using the geoelectron spin-density profile developed in Ref.~\cite{Hunter2013S}.

\subsubsection{lepton-nucleon pairs}

For lepton-nucleon pairs, existing experimental bounds primarily involve the $N$-$e$, $e$-$n$, and $e$-$p$ combinations, as shown in Fig.~\ref{fgsgp}(b). 
Among these, the strongest constraints across a wide range of interaction lengths consistently stem from experiments targeting the $\V_{9+10}$ potential, highlighting its greater sensitivity compared to $\V_{15}$ in probing macroscopic spin-dependent interactions.
Analogous to the $e$-$e$ case, Hunter \textit{et al.}~\cite{Hunter2014PRL} placed limits on the effective field generated by polarized electrons in the Earth through measuring the spin-precession frequencies of $^{199}$Hg-$^{133}$Cs~\cite{Peck2012PRA} and $^{199}$Hg-$^{201}$Hg~\cite{Venema1992PRL} co-magnetometers. Additionally, the constraints from Crescini \textit{et al.}~\cite{Crescini2022PRD}-originally derived for $e$-$e$ interactions using a lead mass and GSO crystal-can be reinterpreted as stringent bounds on $N$-$e$ couplings by considering the unpolarized nucleons in the lead as the source.

The upper bounds on the SP-type interaction between the $N$-$e$ pair, derived by Stadnik \textit{et al.}~\cite{Stadnik2018PRL} and Prosnyak \textit{et al.}~\cite{Prosnyak2023S}, follow a similar strategy to that previously employed for the $e$-$e$ case. 
The key difference lies in the order of the interaction: while the electron-electron coupling contributes to the electron EDM at the loop level, the electron-nucleon interaction contributes at the tree level, inducing an EDM for the entire atom.
In the work of Stadnik \textit{et al.}~\cite{Stadnik2018PRL}, the ALPs are assumed to couple to electrons through a pseudoscalar vertex and to nucleons through a scalar vertex. By contrast, Dzuba \textit{et al.}~\cite{Dzuba2018PRD} explored the opposite scenario, where ALPs couple to electrons via a scalar vertex and to nucleons via a pseudoscalar one. 
Utilizing the experimental upper limit on $^{199}$Hg EDM, they placed a constraint on the product $|g_S^e g_P^N|$.

In the range of $5 \lesssim \lambda \lesssim 50~\mathrm{\mu m}$, the most stringent constraint to date was set by Liang \textit{et al.}~\cite{Liang2022NCR}. Their experiment tested the $\V_{9+10}$ interaction between nucleons in a lead sphere and polarized spins in NV center ensembles in diamond. 
The lead sphere, mounted on a piezoelectric bender oscillating at 1.953 kHz, modulated the distance to the NV centers, which served as magnetometers detecting the resulting time-varying effective magnetic field via fluorescence. To suppress low-frequency noise, a frequency-modulated resonant microwave field was applied to the NV centers. While the magnetic sensitivity of NV magnetometers ($\sim\mathrm{pT/\sqrt{Hz}}$) is modest compared to SERF-based probes, their high spatial resolution allows closer placement of the mass source. 

For interaction ranges $\lambda \gtrsim 50~\mathrm{\mu m}$, the most stringent constraints predominantly come from spin-pendulum experiments~\cite{Hoedl2011PRL, Terrano2015PRL, Heckel2008PRD}, except within $10^{-2}\lesssim \lambda \lesssim 10$~m, where the result from Crescini \textit{et al.}~\cite{Crescini2022PRD} provides a tighter bound. 
The pendulums used by Heckel \textit{et al.}~\cite{Heckel2008PRD} and Terrano \textit{et al.}~\cite{Terrano2015PRL} shared similar designs: two spin-polarized materials arranged to produce net spin while suppressing magnetic leakage. The pendulum used by Terrano \textit{et al.}~\cite{Terrano2015PRL} incorporated design elements from short-range gravity experiments~\cite{Kapner2007PRL}, making it particularly suitable for probing spin-dependent forces at short distances. In their experiment, a copper attractor forming a mass 20-pole was rotated to modulate the signal at the tenth harmonic, allowing them to search for $\V_{9+10}$-induced torques.
Heckel’s experiment instead used celestial bodies (Earth, Moon, Sun) as attractors, relying on sidereal modulation caused by Earth’s rotation.
Hoedl \textit{et al.}~\cite{Hoedl2011PRL} employed a distinct scheme using an unpolarized pendulum and a polarized source. 
The pendulum, built from high-purity single-crystal silicon and titanium, was coated with a thin layer of paramagnetic terbium to minimize magnetic contamination. 
It was suspended within a split-toroidal electromagnet, where alternating current periodically reversed the source’s spin polarization. 
This configuration enabled the search for spin-dependent forces at millimeter-scale separations, free from magnetic interference. 
In the interaction range $10^5\lesssim \lambda \lesssim 10^6$~m, a notable constraint was reported by Wineland \textit{et al.}~\cite{Wineland1991PRL} in 1991. 
They measured the hyperfine transition frequency shift of trapped $^9\mathrm{Be}^+$ ions as the magnetic field was reversed relative to the local gravitational field, observing a shift below $13.4~\mathrm{\mu Hz}$. 
Interpreting this shift as arising from the potential $\V_{9+10}$ between the Earth and the $^9\mathrm{Be}^+$ ions, they derived a constraint on the coupling product $|g_S^N g_P^e|$. 
To eliminate spurious shifts from magnetic field variations, the field strength $B_0$ was tuned such that $\partial \nu_0 / \partial B_0 = 0$.

The gray dash-dotted line in Fig.~\ref{fgsgp}(b) represents a combined constraint derived from stringent laboratory limits on $g^N_S$-obtained from measurements of long-range forces-and astrophysical bounds on $g^e_P$~\cite{OHare2020PRD}.
The constraint on $g^N_S$ is inferred by reinterpreting experimental tests of Newton’s inverse-square law and potential violations of the weak equivalence principle (WEP) in terms of a spin-independent interaction of the form $\mathcal{V}_1$. This interpretation relies on results from well-established experiments, such as those conducted by the E\"{o}t-Wash group~\cite{Kapner2007PRL, Smith1999PRD, Lee2020PRL}; see also the comprehensive review by Adelberger \textit{et al.}~\cite{Adelberger2009PPNP}.

\subsubsection{nucleon-nucleon pairs}

For interaction range $\lambda \lesssim 10^{-6}$~m, the most stringent constraint is set by {\color{black} Voronin} \textit{et al.}~\cite{Voronin2009JETP} by measuring the neutron spin rotation in a perfect non-centrosymmetric crystal caused by potential $\V_{9+10}$.
In the range $10^{-6}\lesssim \lambda \lesssim 10^{-4}$~m, the constraint cast by Fu \textit{et al.}~\cite{Fu2011PRD} through studying the change of $^3$He spin relaxation time upon the motion of an unpolarized test mass near it remains the most strict to date.
In the range $10^{-4}\lesssim \lambda \lesssim 10^{-1}$~m, the strongest constraints are set by two clock-comparison experiments that measure Larmor precession frequencies of dual-species noble gases~\cite{Tullney2013PRL, Feng2022PRL}.
In the work of Feng \textit{et al.}~\cite{Feng2022PRL}, they constrained the $\V_{9+10}$ potential by comparing the frequency ratio of $^{129}$Xe and $^{131}$Xe with a BGO crystal placed near and far from the vapor cell.
Tullney \textit{et al.}~\cite{Tullney2013PRL} measured the frequency shift of collocated $^3$He and $^{129}$Xe in a ultrasensitive low-field NMR co-magnetometer caused by a BGO crystal.
At an interaction range of $\lambda\sim 1$~m,
the constraint on $|g^N_Sg^n_P|$ was derived by Lee \textit{et al.}~\cite{Lee2018PRL} using the K-$^3$He comagnetometer through considering the $^3$He instead of the K atom in the $N$-$e$ case as a sensitivity probe for the potential $\V_{9+10}$.

For interaction ranges $\lambda\gtrsim 10$~m, terrestrial experiments leverage unpolarized nucleons in large-scale natural sources such as the Earth~\cite{Zhang2023PRL, Kimball2017PRD}, the Sun, and the Moon~\cite{Wu2022PRL}.
Zhang \textit{et al.}~\cite{Zhang2023PRL} employed a co-magnetometer similar in design to that of Feng \textit{et al.}~\cite{Feng2022PRL}, with the key enhancement being the use of a rotating platform to enable precise gyroscopic measurements of Earth’s rotation. From a measured rotation rate accuracy of $\pm 2.6~\mathrm{nHz}$, they derived a stringent upper bound of $|g^N_S g^n_P| \leq 3.6 \times 10^{-36}$ (95$\%$-CL) for $\lambda \sim 10^7$~m.
At a comparable range, Kimball \textit{et al.}~\cite{Kimball2017PRD} utilized a dual-isotope Rb co-magnetometer to probe exotic spin-dependent interactions between Earth’s unpolarized nucleons and the nuclear spins of $^{87}$Rb and $^{85}$Rb. 
Since both isotopes have valence protons, their nuclear spin is predominantly protonic~\cite{Klinkenberg1952RMP, Kimball2015NJP}, making this setup especially sensitive to SP coupling in the $N$-$p$ channel.
In their experiment, synchronous optical pumping~\cite{Korver2015PRL} was employed to induce transverse spin polarization in both isotopes, followed by light-off detection of spin precession. 
Through detailed analysis and mitigation of systematic effects---including magnetic field gradients, vector and tensor light shifts, and the gyrocompass effect---they established an upper bound of $\Delta\mathcal{R}(g^N_S g^p_P) \leq 1.5 \times 10^{-8}$ (90\%-CL), corresponding to a constraint of $|g^N_S g^p_P| \leq 2.5 \times 10^{-32}$.

At astrophysical distance $\lambda\gtrsim 10^{11}$~m, Wu \textit{et al.}~\cite{Wu2023PRL} derived the constraint through reinterpreting the experimental limits on CPT and Lorentz violation reported in Ref.~\cite{ Allmendinger2014PRL} in the context of $\V_{9+10}$ potential sourced by the unpolarized nucleons in the Sun.

The astrophysical limit $g^N_Sg^N_P$, represented by the gray dash-dotted line, combines astrophysical constraints on $g^N_P$ and laboratory constraints on $g^N_S$ derived from inverse-square law or WEP tests~\cite{OHare2020PRD}.

\subsection{Constraints on \texorpdfstring{$g_Pg_P$}{gpgp}}\label{expcon_2}

\begin{figure}[htbp]
\centering
\subfigure{
\includegraphics[width=0.75\textwidth]{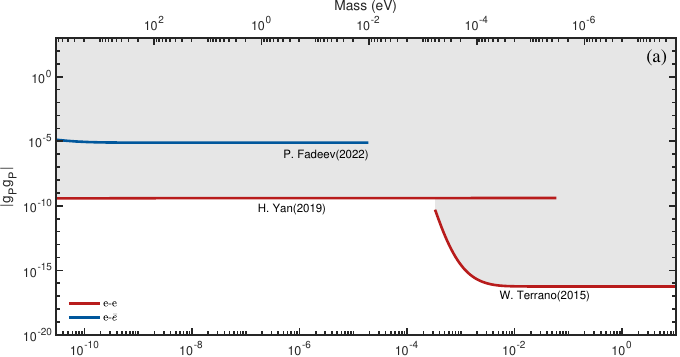}}
\subfigure{
\includegraphics[width=0.75\textwidth]{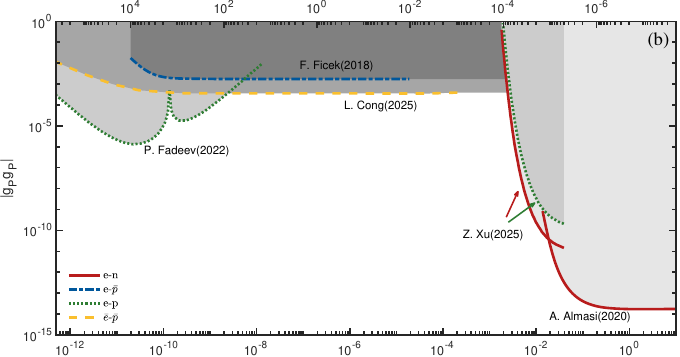}}
\subfigure{
\includegraphics[width=0.75\textwidth]{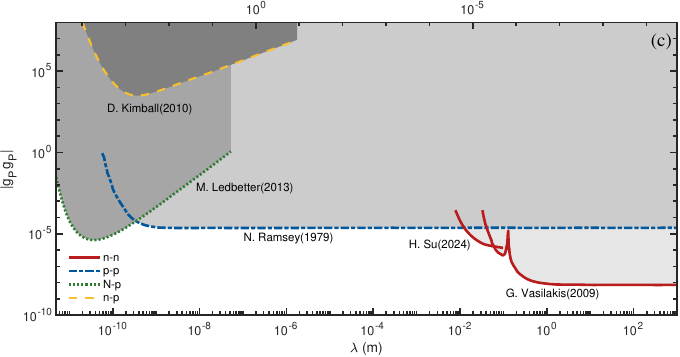}}
\caption{\justifying Experimental constraints on the pseudoscalar-pseudoscalar coupling $g_Pg_P$ for different interaction ranges or ALP mass. The curves represent upper bounds on the coupling strength, and the shaded regions correspond to the excluded parameter space.
In the legend, labels such as $e$-$n$ denote particle pairs, where the electron ($e$) couples to the left vertex and the neutron ($n$) to the right vertex in the Feynman diagram shown in Fig.~\ref{fig3}, corresponding to constraints on $g^e_Pg^n_P$.
The curves with different line types and colors are experimental constraints between different particle pairs.
The constraints are derived from experiments probing the $\V_3$ potential: Refs.~\cite{Terrano2015PRL, Ficek2018PRL, Fadeev2022PRA, Yan2019EPJC, Almasi2020PRL, Xu2025PRL, Cong2025arXiv2, ledbetter2013, Ramsey1979PA, Su2024PRL, Vasilakis2009PRL, Kimball2010PRA}.}
\label{fgpgp}
\end{figure}

{\color{black}The PP coupling arises from a single type of potential, $\V_3$, which shares the same mathematical form as the magnetic dipole interaction.
As a result, techniques that distinguish the exotic field from magnetic interactions based on discrete symmetry differentiation are ineffective in this case.}
Experimental searches for potential $\V_3$ at mesoscopic and macroscopic scales require a spin-polarized source, which inevitably produces a magnetic dipole field.
For solid-state spin sources with high spin density, this necessitates advanced magnetic shielding and careful optimization of the trade-off between reducing the source-probe distance and minimizing magnetic leakage~\cite{Terrano2015PRL, Xu2025PRL}.
At shorter distances, low spin-density gaseous sources---whose magnetic fields are more easily shielded---are employed to suppress this effect~\cite{Vasilakis2009PRL, Su2024PRL}.
Given the high precision of atomic spectroscopy, such interactions can be tightly constrained at the atomic scale~\cite{ledbetter2013, Fadeev2022PRA}.
Moreover, the strong agreement between theoretical predictions and experimental results in antimatter spectroscopy extends these constraints to exotic interactions involving antimatter systems~\cite{Ficek2018PRL, Cong2025arXiv2}.
Figure~\ref{fgpgp} summarizes the current experimental constraints on the PP coupling.

\subsubsection{lepton-lepton pairs}

In the range $\lambda\sim 10^{-9}$~m, experimental constraints on the PP coupling between $e$-$e$ pairs have been established via atomic spectroscopy~\cite{Fadeev2022PRA} and measurements of the electron anomalous magnetic moment~\cite{Yan2019EPJC}.
Fadeev \textit{et al.}~\cite{Fadeev2022PRA} compared experimental spectra and QED-based calculation of positronium ($e$-$\bar{e}$) to place constraints on the PP coupling.
The most stringent constraint in the range comes from Yan \textit{et al.}~\cite{Yan2019EPJC}, who analyzed the one-loop correction to the electromagnetic vertex arising from the PP coupling between electrons and ALPs.
By exploiting the remarkable agreement between theoretical and experimental values of the electron anomalous magnetic moment, they set a bound of $|g^e_P g^e_P| \leq 4 \times 10^{-10}$ (95\%-CL).

For longer ranges $\lambda \sim 10^{-2}$~m, Terrano \textit{et al.}~\cite{Terrano2015PRL} derived a constraint using the same spin-pendulum setup employed in their SP coupling study.
In this case, the unpolarized copper source was replaced with a polarized spin ring consisting of 20 equally magnetized segments made from alternating high-spin-density (Alnico) and low-spin-density (SmCo$_5$) materials.
This configuration yielded the most stringent constraint to date in this range: $|g^e_P g^e_P| \leq 5.6 \times 10^{-17}$ (95$\%$-CL).

\subsubsection{lepton-nucleon pairs}

For the PP coupling between leptons and nucleons, existing experimental constraints have been established for $e$-$n$, $e$-$p$, $e$-$\bar{p}$, and $\bar{e}$-$\bar{p}$ pairs.
The exotic interaction between $e$ and $p$ was investigated by Fadeev \textit{et al.}~\cite{Fadeev2022PRA}, who analyzed the energy difference $D_{21} = 8E_\mathrm{hfs}(2s) - E_\mathrm{hfs}(1s)$ between hyperfine transitions in hydrogen.
This observable is highly sensitive to spin-dependent interactions and enables probing the $\V_3$ potential between the electron and proton.
For exotic matter-antimatter interactions, Ficek \textit{et al.}~\cite{Ficek2018PRL} derived constraints on the PP coupling between $e$-$\bar{p}$ pairs using antiprotonic helium, a three-body system in which an antiproton replaces one electron of a helium atom.
By constructing an appropriate wavefunction for this exotic system, they theoretically evaluated the energy shift in its hyperfine structure induced by the $\V_3$ potential and compared it with experimental measurements, arriving at a constraint of $|g^e_P g_P^{\bar{p}}| \leq 2 \times 10^{-2}$ (90\%-CL) for $\lambda \sim 10^{-10}$~m.
In a pure antimatter system, Cong \textit{et al.}~\cite{Cong2025arXiv2} recently analyzed the contribution of the $\V_3$ potential to the hyperfine splitting interval $D_{21}$ in antihydrogen and obtained a constraint on the PP coupling in $\bar{e}$-$\bar{p}$ channel.

For interaction ranges $\lambda \gtrsim 10^{-3}$~m, experimental constraints on the PP coupling have been established using co-magnetometers as probes and bulk spin-polarized sources~\cite{Almasi2020PRL, Xu2025PRL}.
In the experiment by Wang \textit{et al.}~\cite{Wang2022PRL}, a $^{87}$Rb-$^{129}$Xe co-magnetometer serves as the probe, while a nearby cell containing optically pumped $^{87}$Rb atoms functions as the spin-polarized source.
The potential $\V_3$, sourced by the polarized $^{87}$Rb atoms, couples to the $^{129}$Xe nuclear spins, inducing their precession. This precession in turn generates a magnetic field detectable by the co-located Rb magnetometer.
Xu \textit{et al.}~\cite{Xu2025PRL} and Almasi \textit{et al.}~\cite{Almasi2020PRL} employed solid-state spin sources, which inherently generate stronger magnetic fields. To suppress these fields, both experiments utilized Rb-$^{21}$Ne co-magnetometers operated at the self-compensation point, where sensitivity to magnetic fields is minimized. In both setups, the spin source consisted of a SmCo$_5$ permanent magnet shielded by iron and enclosed in a $\mu$-metal shell. Additionally, Almasi \textit{et al.} implemented a cosine coil around the magnet to produce a compensating field and further reduce magnetic leakage.
With a source-probe separation of approximately 25 cm, they constrained the PP coupling to $|g^e_P g^n_P| \leq 1.7 \times 10^{-14}$ (95\%-CL) at $\lambda \sim 10^{-1}$~m.
Xu \textit{et al.}~\cite{Xu2025PRL} designed a multi-layer magnetic shielding structure that suppressed ambient magnetic fields by up to 11 orders of magnitude, allowing the source to be placed just 5 cm from the co-magnetometer cell.
This setup yielded the most stringent limit on $|g^e_P g^n_P|$ for $\lambda \sim 10^{-2}$~m.
Moreover, since the nuclear spin of $^{21}$Ne arises from both the proton and the neutron, their result also constrains the proton coupling $|g^e_P g^p_P|$ under the Schmidt model~\cite{Kimball2015NJP}.

\subsubsection{nucleon-nucleon pairs}

Regarding the PP coupling between nucleons, Ramsey~\cite{Ramsey1979PA} was the first to discuss a possible non-magnetic tensor force between protons, such as the $\V_3$ interaction.
By comparing high-precision experimental measurements and theoretical predictions of the rank-2 tensor interaction in the H$_2$ molecule, he derived an upper bound of $|g^p_P g^p_P| < 2.3 \times 10^{-5}$ (90\%-CL) for interaction ranges $\lambda \sim 10^{-9}~\mathrm{m}$.

At comparable angstrom-scale ranges, Ledbetter \textit{et al.}~\cite{ledbetter2013} obtained a constraint by analyzing the J-coupling strength in the HD molecule.
Also within this range, Kimball \textit{et al.}~\cite{Kimball2010PRA} investigated the possible contribution of the exotic dipole-dipole interaction $\V_3$ to spin-exchange collisions between Na and $^3$He atoms.
Based on limited experimental data and the theoretical framework for spin-exchange cross sections outlined in Refs.~\cite{Walker1989RPA, Walker1997RMP}, they placed a constraint on the coupling constant $|g^n_P g^p_P|$.

For longer interaction ranges $\lambda \gtrsim 10^{-2}\mathrm{m}$, Vasilakis \textit{et al.}~\cite{Vasilakis2009PRL} employed a K-$^3$He co-magnetometer operating at the self-compensation point, with an optically polarized $^3$He cell serving as the spin source, to search for the $\V_3$ interaction between polarized neutrons in $^3$He.
This experiment yielded the most stringent constraint to date on the neutron-neutron coupling: {\color{black}$|g^n_P g^n_P| < 8.1 \times 10^{-9}$ (68\% CL) at $\lambda \sim 1~\mathrm{m}$}.

More recently, Su \textit{et al.}~\cite{Su2024PRL} reported a new constraint on $|g^n_P g^n_P|$ at $\lambda \sim 10^{-2}\mathrm{m}$.
In addition to enhancing the exotic signal via Rb-Xe spin exchange, they implemented a template-filtering technique~\cite{Abbott2004PRD} to extract the signal with optimal SNR.
This method can be extended to enhance sensitivity in searches for axion fields or other exotic phenomena, provided the signal's functional form is known.

\begin{figure}[htbp]
\centering
\subfigure{
\includegraphics[width=0.75\textwidth]{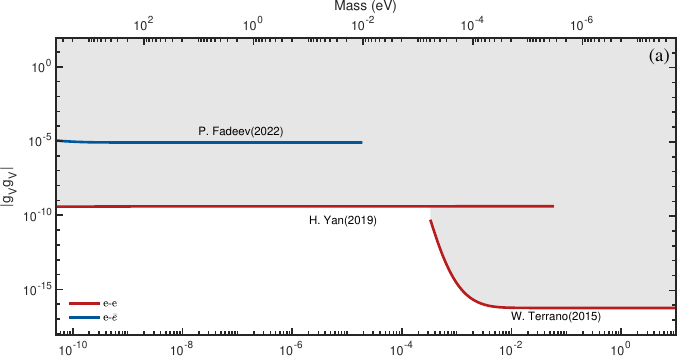}}
\subfigure{
\includegraphics[width=0.75\textwidth]{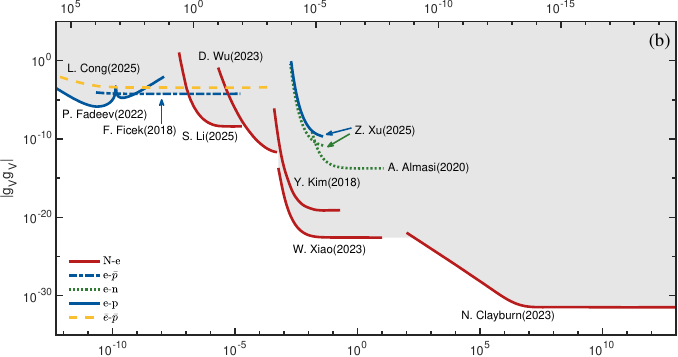}}
\subfigure{
\includegraphics[width=0.75\textwidth]{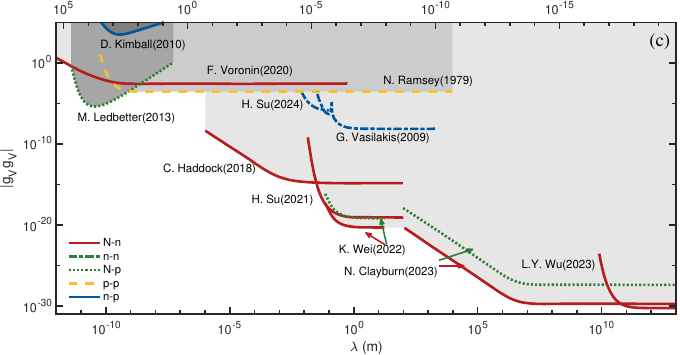}}
\caption{\justifying Experimental constraints on the vector-vector coupling $g_Vg_V$ for different interaction ranges or ALP mass. The curves represent upper bounds on the coupling strength, and the shaded regions correspond to the excluded parameter space.
In the legend, labels such as $e$-$n$ denote particle pairs, where the electron ($e$) couples to the left vertex and the neutron ($n$) to the right vertex in the Feynman diagram shown in Fig.~\ref{fig3}.
The curves with different line types and colors are experimental constraints between different particle pairs.
The potentials involved in various experiments are summarized as follows:
$\V_3$: Refs.~\cite{Terrano2015PRL, Fadeev2022PRA, Yan2019EPJC, Almasi2020PRL, Xu2025PRL, Cong2025arXiv2, ledbetter2013, Ramsey1979PA, Su2024PRL, Vasilakis2009PRL, Kimball2010PRA};
$\V_{4+5}$: Refs.~\cite{Ficek2018PRL, Li2025PRL, Wud2023PRL, Kim2018PRL, Xiao2023PRL, Clayburn2023PRD, Voronin2020JETP, Haddock2018PLB, Su2021SA, Wei2022NC, Wu2023PRL}.}
\label{fgvgv}
\end{figure}

\subsection{Constraints on \texorpdfstring{$g_Vg_V$}{gvgv}}\label{expcon_3}

The VV coupling can be constrained via the potentials $\V_1$, $\V_3$, and $\V_{4+5}$, where $\V_1$ is spin-independent. Since $\V_3$ and $\V_{4+5}$ arise from PP and SS couplings, respectively, existing experimental limits on these interactions can be translated into constraints on the VV coupling by relating their coefficients. These rescaled constraints are shown in Fig.~\ref{fgvgv}, and further discussed in Secs.~\ref{expcon_1} and \ref{expcon_2}.

\subsection{Constraints on \texorpdfstring{$g_Vg_A$}{gvga}}\label{expcon_6}

The VA coupling can be constrained via the potentials $\V_{12+13}$, $\V_{11}$, and $\V_{16}$. Among these, $\V_{12+13}$ is a spin-dependent potential with spin order 1 and $\nabla$ power 0, making it the most accessible for constraining $g_V g_A$. In contrast, both $\V_{11}$ and $\V_{16}$ involve spin order 2 and $\nabla$ power 1. However, since $\V_{11}$ is velocity-independent, it is somewhat easier to constrain than the velocity-dependent $\V_{16}$.

\subsubsection{lepton-lepton pairs}
Constraints on the VA coupling between electrons have been derived by Heckel \textit{et al.}~\cite{Heckel2013PRL} and Hunter \textit{et al.}~\cite{Hunter2013S}, based on the potential $\V_{11}$. Heckel \textit{et al.}~\cite{Heckel2013PRL} employed a spin torsion pendulum---described in Ref.~\cite{Heckel2008PRD}---surrounded by four spin sources arranged in a specific configuration to enhance sensitivity to the $\V_{11}$ interaction. Hunter \textit{et al.}~\cite{Hunter2013S} obtained a constraint for interaction ranges $\lambda \gtrsim 10^{3}$~m by reinterpreting experimental data originally intended to test Lorentz invariance.

\subsubsection{lepton-nucleon pairs}
For the VA coupling between lepton-nucleon pairs, the potentials $\V_{11}$ and $\V_{16}$ have been investigated for both electron-neutron and electron-proton interactions, where the electron couples to ALPs via either an axial-vector or vector vertex. Since both interactions are symmetric under particle exchange, constraints on the $e$-$n$ ($e$-$p$) pair can be translated to the $n$-$e$ ($p$-$e$) pair by applying a proper factor.
The constraint reported by Hunter \textit{et al.}~\cite{Hunter2013S} was based on the velocity-independent potential $\V_{11}$. Building upon their model of the Earth's polarized spin density and incorporating the relative motion between geoelectrons and laboratory spins due to Earth's rotation, they later derived a constraint on the velocity-dependent potential $\V_{16}$ in a follow-up study~\cite{Hunter2014PRL}.
More recently, Wang \textit{et al.}~\cite{Wang2023SA} constrained the VA coupling for the electron-neutron pair using an experimental setup with spin-polarized $^{87}$Rb as the source and a $^{87}$Rb-$^{129}$Xe spin amplifier as the detector, similar to their apparatus for probing the $\V_3$ interaction~\cite{Wang2022PRL}. By alternatively considering the polarized valence protons in $^{87}$Rb as the spin source, they also derived a constraint on the VA coupling for the neutron-proton pair.

In addition, most experimental efforts have focused on searching for the monopole-dipole potential $\V_{12+13}$ between nucleons and electrons ($N$-$e$ pairs). Because $\V_{12+13}$ violates parity inversion symmetry, it mixes atomic states of opposite parity, leading to observable parity-violation effects. This P-odd interaction induces energy shifts in atomic transitions. Dzuba \textit{et al.}~\cite{Dzuba2017PRL} evaluated such shifts in atoms including $^{133}\mathrm{Cs}$, $^{174}\mathrm{Yb}$, and $^{205}\mathrm{Tl}$, and by comparing experimental measurements with SM predictions, set the most stringent constraint on the VA coupling between $e$-$N$ pairs at interaction ranges $\lambda \lesssim 10^{-6}\mathrm{m}$.

When $\V_{12+13}$ acts between nucleons, it contributes to nuclear anapole moments~\cite{Safronova2018RMP}. Dzuba \textit{et al.} further analyzed this effect in the $^{133}$Cs nucleus, placing bounds on the VA coupling between $N$-$p$ pairs.
At larger interaction ranges around $\lambda\sim 10^{-5}\mathrm{m}$, the most stringent constraints have been obtained by Jiao \textit{et al.}~\cite{Jiao2021PRL} and Liang \textit{et al.}~\cite{Liang2022NCR}, using a NV center diamond magnetometer as a sensitive detector. These experiments achieved the required relative motion between the source and probe by periodically modulating the position of a low-mass source, enabling modulation frequencies in the kilohertz range.
At the millimeter scale, Tian \textit{et al.}~\cite{Tian2025PRL} recently employed a diamagnetically levitated force sensor to detect the force generated by the $\V_{12+13}$ interaction between rotating, modulated magnets and a sapphire crystal attached to the probe. By measuring the displacement induced by the distance-dependent interaction, they derived constraints on the strength of the VA coupling.
Levitated force sensors are emerging as a promising approach for detecting dark matter and exotic spin-dependent interactions, with sensitivity that may be comparable to or potentially exceed that of atomic magnetometers~\cite{Kalia2024PRD, Higgins2024PRD, Ahrens2025PRL}. In particular, theoretical studies suggest that levitated ferromagnetic probes could offer greater sensitivity than existing spin-based magnetometers~\cite{Kimball2023PRA}, indicating their potential to probe effective magnetic fields associated with new physics.
Additional constraints have been reported by Kim \textit{et al.}~\cite{Kim2019NC} and Wu \textit{et al.}~\cite{Wu2022PRL}, both of whom used a commercial SERF magnetometer as the probe and a BGO crystal as the source mass. Details of these experiments are provided in earlier sections.

In the force range $\lambda\gtrsim 10^{7}\mathrm{m}$, the most stringent constraints have been derived from spin torsion pendulum experiments using the Moon and Sun as moving sources, as conducted by Heckel \textit{et al.}~\cite{Heckel2008PRD}. As noted by Hunter \textit{et al.} in Ref.~\cite{Hunter2014PRL}, Earth's rotation induces non-negligible relative velocities between typical Earth-bound atoms and laboratory spins. By utilizing this effect, and based on the electron spin orientation-related energy parameter $\beta$ measured in Ref.~\cite{Heckel2008PRD}, Clayburn \textit{et al.}~\cite{Clayburn2023PRD} placed improved constraints on the VA coupling in the range $\lambda \lesssim 10^{10}\mathrm{m}$.
Similarly, using nuclear spin orientation-related energy extracted from Ref.~\cite{Hunter2013S}, Clayburn \textit{et al.}~\cite{Clayburn2023PRD} also derived corresponding bounds on the VA coupling between $N$-$n$ and $N$-$p$ pairs.

\begin{figure}[htbp]
\centering
\subfigure{
\includegraphics[width=0.75\textwidth]{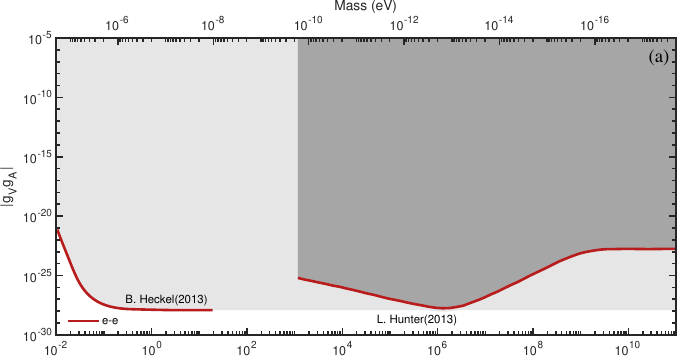}}
\subfigure{
\includegraphics[width=0.75\textwidth]{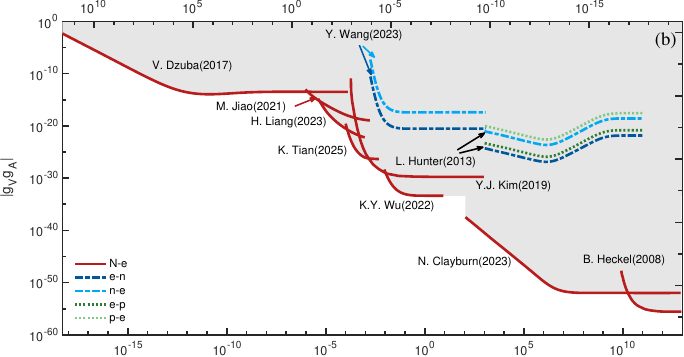}}
\subfigure{
\includegraphics[width=0.75\textwidth]{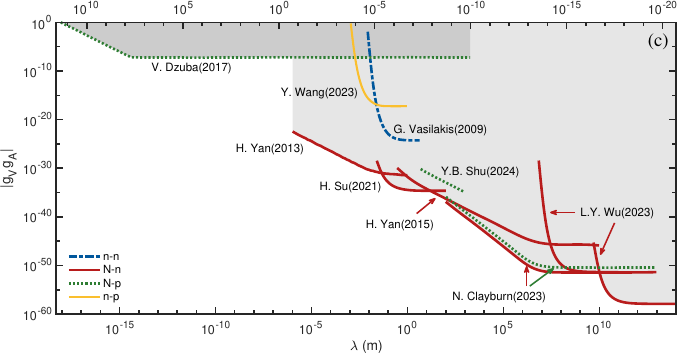}}
\caption{\justifying Experimental constraints on the vector-axial-vector coupling $g_Vg_A$ for different interaction ranges or ALPs mass. The curves represent upper bounds on the coupling strength, and the shaded regions correspond to the excluded parameter space.
In the legend, labels such as $e$-$n$ denote particle pairs, where the electron ($e$) couples to the left vertex and the neutron ($n$) to the right vertex in the Feynman diagram shown in Fig.~\ref{fig3}.
The curves with different line types and colors are experimental constraints between different particle pairs.
The potentials involved in various experiments are summarized as follows: 
$\V_{11}$: Refs.~\cite{Heckel2013PRL, Hunter2013S, Wang2023SA, Vasilakis2009PRL};
$\V_{12+13}$: Refs.~\cite{Dzuba2017PRL, Clayburn2023PRD, Jiao2021PRL, Liang2022NCR, Tian2025PRL, Wu2022PRL, Kim2019NC, Heckel2008PRD, Su2021SA, Shu2024PRL, Wu2023PRL, Yan2013PRL, Yan2015PRL}.}
\label{fgvga}
\end{figure}

\subsubsection{nucleon-nucleon pairs}

For the VA coupling between nucleon pairs, the constraint on the $n$-$n$ interaction obtained by Vasilakis \textit{et al.}~\cite{Vasilakis2009PRL} was derived through a reinterpretation of the upper limit using the $\V_{11}$ potential. In the interaction range $10^{-6} \lesssim \lambda \lesssim 10^{-1}$~m, Yan \textit{et al.}~\cite{Yan2013PRL} reported the most stringent constraint on the $N$-$n$ pair, based on measurements of parity-violating neutron spin rotation in liquid $^4$He. The underlying experimental principle is introduced in Sec.~\ref{expinv_31}.

Su \textit{et al.}~\cite{Su2021SA} presented a constraint on the $\V_{12+13}$ interaction via a reanalysis of spin-amplifier data originally acquired for probing the $\V_{4+5}$ potential. To explore interactions at ranges near $\lambda \sim 10^{7}$~m, Yan \textit{et al.}~\cite{Yan2015PRL} considered the contribution of a fluctuating $\V_{12+13}$ interaction-arising from Earth’s nucleons and thermally moving $^3$He atoms-to the spin relaxation rate of polarized $^3$He gas. Leveraging the long spin relaxation time and the large number of source nucleons, they placed a constraint on the VA coupling between $N$-$n$ pairs.

At even longer ranges ($\lambda \gtrsim 10^{10}\mathrm{m}$), Wu \textit{et al.}~\cite{Wu2023PRL} derived the most stringent constraint to date by reinterpreting Lorentz-violation test data from Ref.~\cite{Allmendinger2014PRL} in terms of a possible $\V_{12+13}$ interaction between solar nucleons and polarized terrestrial spins.

More recently, Shu \textit{et al.}~\cite{Shu2024PRL} employed an atom interferometer to search for the $\V_{12+13}$ interaction between the $N$-$p$ pair. In their experiment, spin-polarized $^{87}$Rb atoms were launched from a magneto-optical trap and interrogated using a Bragg atom interferometer to measure free-fall acceleration. The atoms were prepared in states $|F, m_F = \pm1\rangle$ so that spin-dependent acceleration induced by the $\V_{12+13}$ interaction could be detected. To suppress contributions from the electron spin in $^{87}$Rb, differential measurements were performed by comparing results between atoms in $|F, m_F = +1\rangle$ and $|F, m_F = -1\rangle$ states.

\subsection{Constraints on \texorpdfstring{$g_Ag_A$}{gaga}}\label{expcon_4}

\begin{figure}[htbp]
\centering
\subfigure{
\includegraphics[width=0.75\textwidth]{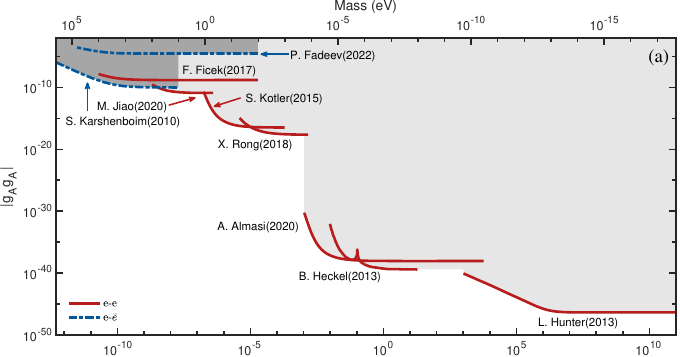}}
\subfigure{
\includegraphics[width=0.75\textwidth]{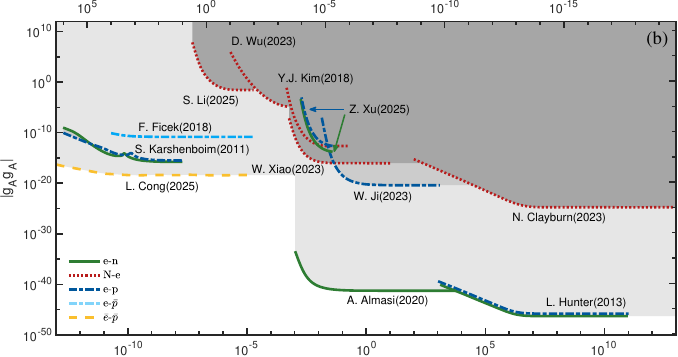}}
\subfigure{
\includegraphics[width=0.75\textwidth]{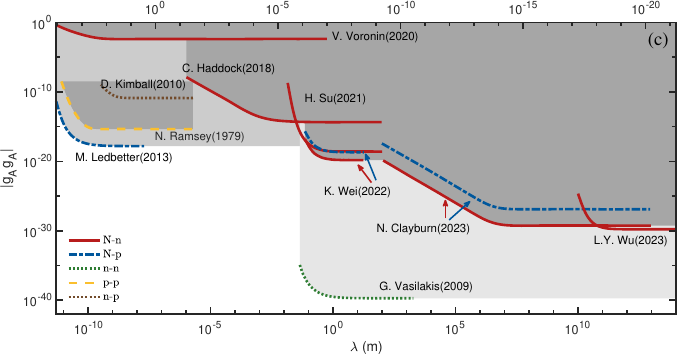}}
\caption{\justifying Experimental constraints on the axial-vector-axial-vector coupling $g_Ag_A$ for different interaction ranges or ALP mass. The curves represent upper bounds on the coupling strength, and the shaded regions correspond to the excluded parameter space.
In the legend, labels such as $e$-$n$ denote particle pairs, where the electron ($e$) couples to the left vertex and the neutron ($n$) to the right vertex in the Feynman diagram shown in Fig.~\ref{fig3}.
The curves with different line types and colors are experimental constraints between different particle pairs.
The potentials involved in various experiments are summarized as follows: 
$\V_2$: Refs.~\cite{Kimball2010PRA, ledbetter2013, Ficek2018PRL, Karshenboim2010PRD, Karshenboim2011PRA, Jiao2020PRD, Kotler2015PRL, Rong2018PRL, Almasi2020PRL, Hunter2013S, Heckel2013PRL};
$\V_3$: Refs.~\cite{Fadeev2022PRA, Almasi2020PRL, Xu2025PRL, Cong2025arXiv2, Ramsey1979PA, Vasilakis2009PRL};
$\V_{4+5}$: Refs.~\cite{Li2025PRL, Wud2023PRL, Kim2018PRL, Xiao2023PRL, Clayburn2023PRD, Voronin2020JETP, Haddock2018PLB, Su2021SA, Wei2022NC, Wu2023PRL};
$\V_8$: Refs.~\cite{Ji2023PRL, Ficek2017PRA}.}
\label{fgaga}
\end{figure}

The potentials $\V_2$, $\V_3$, $\V_{4+5}$, and $\V_8$ are associated with the AA coupling. As in the case of the VV coupling, constraints on the AA coupling-particularly those involving the $\V_3$ and $\V_{4+5}$ interactions-can be inferred from existing limits on the SS and PP couplings, as discussed in Secs.\ref{expcon_1} and \ref{expcon_2}, respectively. These constraints are summarized in Fig.\ref{fgaga}, where they have been rescaled using the appropriate coefficient factors to reflect the AA coupling.

\subsubsection{lepton-lepton pairs}

Constraints on the AA coupling arising from the $\V_2$ and $\V_8$ interactions are exclusive to this coupling type. At the atomic scale, the most stringent limit on the AA coupling between electrons is provided by Ficek \textit{et al.}~\cite{Ficek2017PRA}, based on the $\V_8$ interaction. They considered contributions from both $\V_2$ and $\V_8$ to transition energies in helium fine-structure spectroscopy. Specifically, the constraint from $\V_2$ was obtained by comparing theoretical and experimental values for the $2^3S_1$-$2^3P$ transition, while that from $\V_8$ was based on the $2^3P_1$-$2^3P_2$ fine-structure splitting. Since the discrepancy in the $2^3P_1$-$2^3P_2$ transition is much smaller, the resulting constraint on $|g^e_A g^e_A|$ from $\V_8$ is slightly stronger in practice.

Karshenboim \textit{et al.}~\cite{Karshenboim2011PRA} constrained the AA coupling between the electron and antimuon ($e$-$\bar{\mu}$) by analyzing the additional hyperfine splitting in the ground state of muonium induced by the $\V_2$ interaction.

To probe the AA interaction at nanometer scales, Jiao \textit{et al.}~\cite{Jiao2020PRD} developed a molecular ruler composed of two electron spins separated by a shape-persistent polymer chain, with tunable inter-spin distances. By measuring the magnetic dipole-dipole interaction between electrons and fitting it to a theoretical model, they derived an upper limit on the $\V_2$ interaction.

Kotler \textit{et al.}~\cite{Kotler2015PRL} also provided a constraint on the AA coupling using the same experimental system as in their search for the $\V_3$ interaction.

At distances near $\lambda\sim 100$~m, Rong \textit{et al.}~\cite{Rong2018PRL} used a single NV center in diamond to sense the $\V_2$ interaction generated by optically polarized electrons in pentacene molecules. The interaction strength was extracted from the phase shift of the NV center's electron spin under a dynamical decoupling sequence~\cite{Du2009Na}.

In the intermediate interaction range of $10^{-3} \lesssim \lambda \lesssim 10^{3}$~m, Almasi \textit{et al.}~\cite{Almasi2020PRL} constrained the AA coupling using the same apparatus previously employed in their search for the PP coupling, focusing instead on the $\V_2$ potential. Similarly, Heckel \textit{et al.}~\cite{Heckel2013PRL} surrounded their spin torsion pendulum (described in Ref.~\cite{Heckel2008PRD}) with four spin sources configured to enhance sensitivity to the $\V_2$ interaction.

\subsubsection{lepton-nucleon pairs}

In the context of AA coupling between leptons and nucleons, constraints on the $e$-$N$ pair primarily arise from searches targeting the $\V_{4+5}$ interaction, as discussed in the sections on SS and VV couplings. For the AA coupling involving $e$-$n$ and $e$-$p$ pairs, the most stringent limits across a broad range of interaction distances---from atomic to astronomical scales---have been established by Karshenboim \textit{et al.}~\cite{Karshenboim2011PRA}, Almasi \textit{et al.}~\cite{Almasi2020PRL}, and Hunter \textit{et al.}~\cite{Hunter2013S}, all based on investigations of the $\V_2$ potential.

For the $e$-$\bar{p}$ pair, Ficek \textit{et al.}~\cite{Ficek2018PRL} evaluated the $\V_2$ interaction using antiprotonic helium spectroscopy, following a similar methodology to their analysis of the $\V_3$ potential for constraining the PP coupling.

Ji \textit{et al.}~\cite{Ji2023PRL} probed the AA coupling via the $\V_8$ interaction using a setup consisting of two iron-shielded SmCo$_5$ electron-spin sources and two commercial QuSpin SERF magnetometers. To minimize magnetic interference, the magnetometers and spin sources were housed in separate $\mu$-metal shields. Following a strategy similar to that of Wu \textit{et al.}~\cite{Wu2022PRL}, they employed signal combination from two magnetometers to suppress common-mode noise and enhance sensitivity.

\subsubsection{nucleon-nucleon pairs}
For the AA coupling between nucleons, atomic-scale constraints reported by Kimball \textit{et al.}~\cite{Kimball2010PRA} and Ledbetter \textit{et al.}~\cite{ledbetter2013} were obtained by reinterpreting the same experimental data previously used to constrain the PP coupling, this time in terms of the $\V_2$ interaction. Similarly, the strong constraint on the $n$-$n$ pair at an interaction range of $\lambda \sim 1$~m, as derived by Vasilakis \textit{et al.}~\cite{Vasilakis2009PRL}, is based on the same dataset originally analyzed for PP coupling. Additional constraints for AA couplings between nucleons are obtained by rescaling results established for the SS coupling.

\subsection{Constraints on muons}\label{expcon_7}

\begin{figure}[t!]
    \centering
    \subfigure{\includegraphics[width=0.95\textwidth]{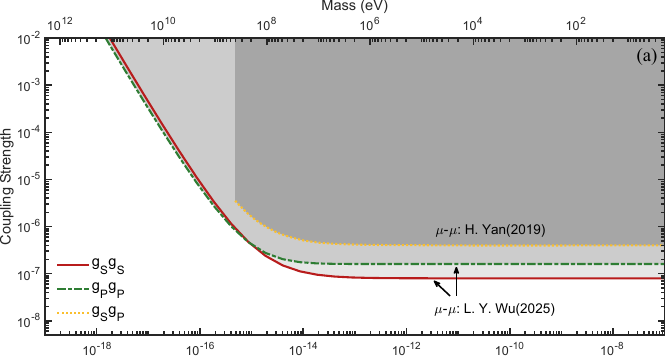}}
    \subfigure{\includegraphics[width=0.95\textwidth]{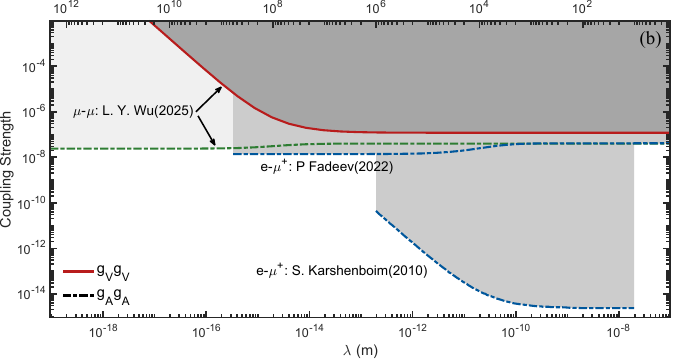}}
    \caption{\justifying Experimental constraints on exotic spin-dependent interactions involving muons, shown as a function of interaction range or ALP mass. The curves represent upper bounds on the coupling strength, while the shaded regions indicate excluded parameter space.
(a) Constraints on scalar-boson-mediated interactions. The green dotted line corresponds to the result of Ref.~\cite{Yan2019EPJC}.
(b) Constraints on vector-boson-mediated interactions. The blue dotted lines are derived from Refs.~\cite{Fadeev2022PRA, Karshenboim2010PRL}.
The constraint on the AA coupling strength obtained in this work is expressed as $|g^\mu_A|^2 m_\mu^2 / m_X^2$, where the rescaling factor $m_\mu^2 / m_X^2$ accounts for the contribution of the longitudinal polarization mode in axial couplings, which diverges as $m_X \to 0$~\cite{Fadeev2022PRA}. For consistency, the result of Fadeev \textit{et al.}~\cite{Fadeev2022PRA} is similarly rescaled and shown as $g^\mu_A g^e_A m_\mu m_e / m_X^2$.}
    \label{gamumu}
\end{figure}

Muons are arguably among the most intriguing fermions in the search for new physics. Their involvement in several persistent anomalies---such as the proton and deuteron charge radius puzzles observed in muonic hydrogen and deuterium systems~\cite{Pohl2016S,krauth2021na}, as well as the long-standing discrepancy in the muon anomalous magnetic moment~\cite{Abi2021PRL, Aoyama2020PR}---suggests that muons may participate in unknown interactions beyond the SM.
While parity-violating interactions that couple exclusively to muons (so-called ``muonic'' interactions) and are mediated by new massive gauge bosons in the MeV-GeV range have been proposed to address the proton radius puzzle~\cite{Carlson2015PPNP}, to the best of our knowledge, no experimental studies have explored such interactions at longer ranges (i.e., beyond the nanometer scale), which would correspond to mediator masses below approximately 100~eV.
By contrast, numerous experiments have investigated exotic spin-dependent interactions over length scales ranging from micrometers to astronomical distances---corresponding to energy scales below ~10~eV---but these searches have focused almost exclusively on electrons, protons, and neutrons. Experimental constraints on long-range interactions involving other fermions, such as muons, remain exceedingly scarce.
Despite the compelling theoretical motivations, including possible ALP-mediated muonic interactions, no dedicated experimental efforts have yet been undertaken to probe long-range spin-dependent forces involving muons. Given their prominent role in multiple unresolved puzzles in modern physics, muons represent a particularly promising target for future high-precision tests of exotic interactions at extended length scales.

Here, we present a focused overview of current experimental constraints involving muons. These constraints on exotic muonic interactions are summarized in Fig.~\ref{gamumu}.
The bounds on the AA coupling between $e$-$\mu$ are derived from precision measurements of the hyperfine splitting in muonium~\cite{Fadeev2022PRA, Karshenboim2010PRL}. For purely muonic systems, constraints arise from measurements of the muon’s EDM and anomalous magnetic moment ($g-2$).
Since SP interactions violate both P and T symmetries, radiative corrections involving new particles coupling to muons via scalar and pseudoscalar vertices can induce a nonzero EDM. The yellow dotted line~\cite{Yan2019EPJC} in Fig.~\ref{gamumu}(a) represents the constraint on SP coupling between muons, derived from the most stringent experimental upper limit on the muon EDM reported by the Muon $(g-2)$ Collaboration in 2009: $|d_\mu| < 1.8 \times 10^{-19}~e\cdot\mathrm{cm}$.
With ongoing progress at Fermilab and J-PARC in precision measurements of muon $g-2$, these constraints are expected to improve. 
In parallel, the frozen-spin technique currently under development at PSI aims to enhance muon EDM sensitivity by several orders of magnitude, with a projected reach of $6 \times 10^{-23}~e\cdot\mathrm{cm}$~\cite{Sakurai2022JPSP}.
If this four-order-of-magnitude improvement can be achieved experimentally, the corresponding constraint on the product $g_S g_P$ could be tightened by a similar factor.

Reference~\cite{Yan2019EPJC} discussed the possibility of explaining the discrepancy between theoretical predictions and experimental measurements of the muon anomalous magnetic moment at that time.
Similarly, if radiative corrections originate from S, P, V, or A couplings, the corresponding SS, PP, VV, or AA interactions can also be constrained through precision measurements of the muon $g-2$. 
Based on the recent White Paper~\cite{aliberti2025arxiV}, which reports consistency between experimental results and theoretical predictions of the muon $g-2$, updated constraints on these couplings have been provided in Ref.~\cite{Wu2025arXiv}.

\subsection{Constraints on the Axion Field}\label{expcon_8}

\begin{figure}[t!]
    \centering
    \subfigure{
    \includegraphics[width=0.85\linewidth]{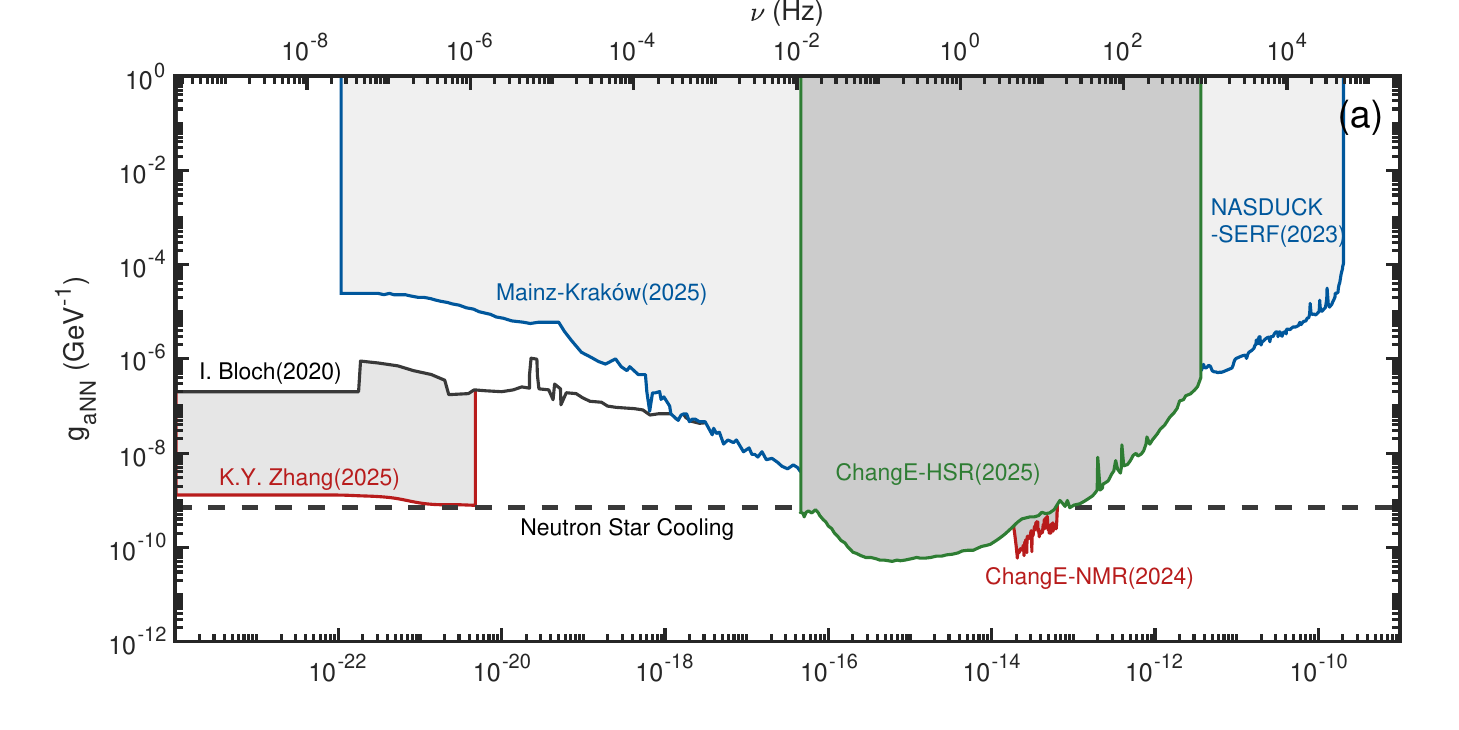}}
    \subfigure{
    \includegraphics[width=0.85\linewidth]{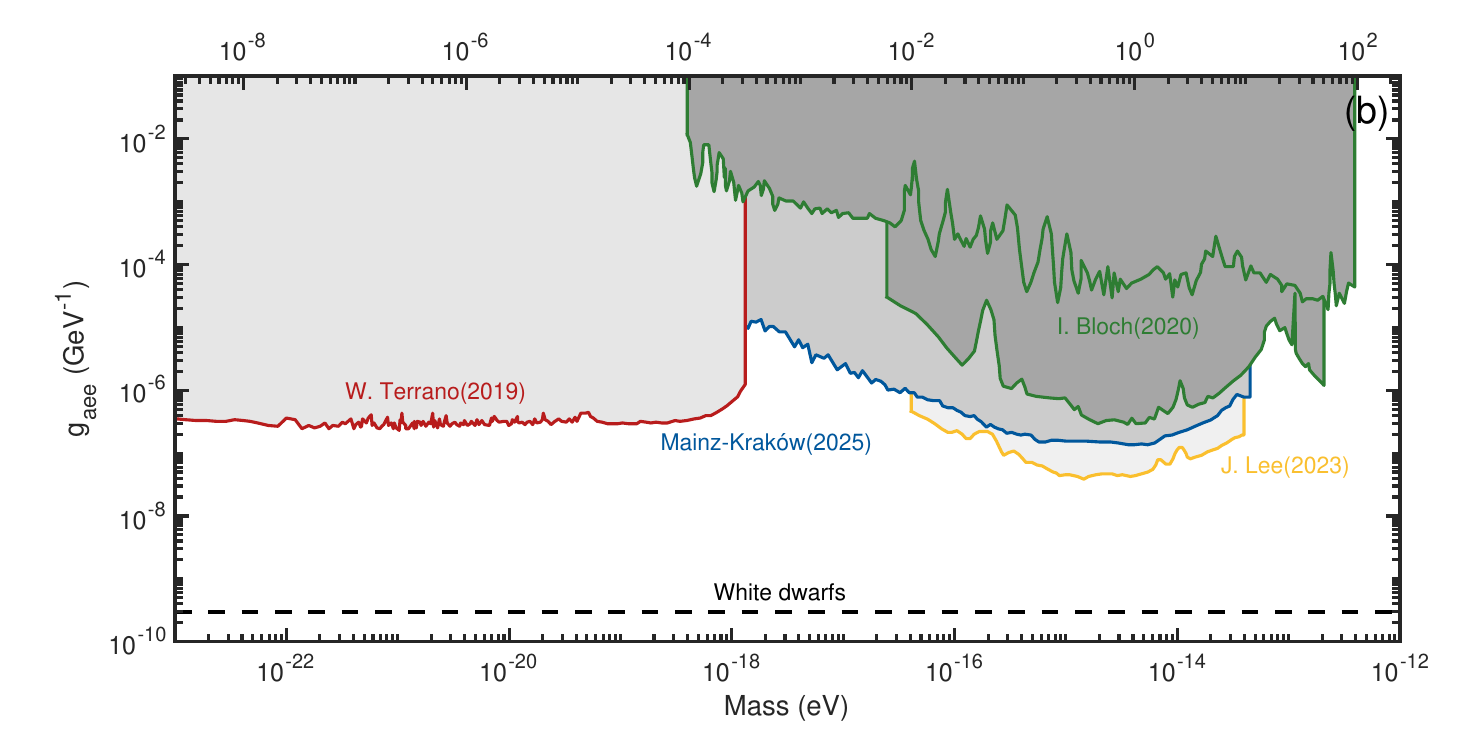}}
    \caption{\justifying Experimental constraints on axion-field couplings: axion-nucleon coupling $g_{\rm aNN}$ (upper panel) and axion-electron coupling $g_{aee}$ (lower panel), plotted as functions of ALP mass. Shaded regions are excluded by laboratory experiments~\cite{ Terrano2019PRL, bloch2020JHEP, Bloch2023NC, xu2024aCP, Wei2025ropp, zhang2025arXiv}. Dashed lines represent astrophysical bounds derived from neutron star and white dwarf cooling~\cite{Buschmann2022PRL, Bertolami2014JCAP}.}
    \label{gapsi}
\end{figure}

\subsubsection{Axion-Field Coupling to Nucleons}

The resonance methods involve detecting either modulations in the spin-precession frequency~\cite{zhang2025arXiv} or the excitation of transverse magnetization near the resonance condition~\cite{Bloch2023NC, xu2024aCP}, both induced by the axion field.
Because these methods are sensitive only to axion fields within a narrow resonance bandwidth, scanning the magnetic field is necessary to probe different axion frequencies.

At frequencies above 1 kHz, the ``NASDUCK-SERF'' region is constrained by a K-$^3$He magnetometer operating in the SERF regime~\cite{Bloch2023NC}, where $^3$He resonantly responds to the axion field, and the K atoms read out the response. 
The SERF condition is maintained by increasing the K atom density as the magnetic field is scanned to match the target axion frequency.
Xu \textit{et al.}~\cite{xu2024aCP} performed a similar resonant search using a K-Rb-$^{21}$Ne co-magnetometer. With an ultra-high sensitivity of $1.29~{\rm fT}/\sqrt{\rm Hz}$ at 5 Hz, they successfully extended the scanning range beyond the intrinsic bandwidth of the $^{21}$Ne resonance without significant sensitivity loss. This yielded the most stringent constraint to date on $g_{\rm aNN}$ in the 4.5-15.5 Hz range, labeled ``ChangE-NMR''.
In the ultralight axion mass region ($\sim 10^{-23}$~eV), the most recent constraint is provided by Zhang \textit{et al.}~\cite{zhang2025arXiv}, based on a reanalysis of Lorentz-violation data obtained from $^{129}$Xe-$^3$He NMR co-magnetometer measurements.

Off-resonance methods are dominated by atomic co-magnetometers operating in self-compensation or strong-coupling regimes~\cite{bloch2020JHEP, Lee2023PRX, Wei2025ropp}.
The regions labeled ``I. Bloch'' is derived from reanalyses of existing data originally collected to search for exotic spin-dependent interactions and Lorentz/CPT violation using a K-$^3$He magnetometer operating at the self-compensation point~\cite{bloch2020JHEP}.
The ``ChangE-HSR'' constraint, covering the $10^{-2}$-$10^3$ Hz range, was obtained by operating the K-Rb-$^{21}$Ne co-magnetometer in the self-compensation regime at low frequencies and the strong-coupling regime at high frequencies~\cite{Wei2025ropp}.

When the measurement duration is shorter than the axion coherence time, the axion field can be approximated as an oscillating field with constant amplitude and direction.
Using this approximation, Martin \textit{et al.} performed an interference measurement of the axion field with two K-$^3$He co-magnetometers-one located in Mainz and the other in Krak\'{o}w. The resulting constraint is shown as the blue curve labeled ``Mainz-Krak\'{o}w''.
The black dashed line in Fig.\ref{gapsi}(a) indicates the astrophysical bound derived from neutron star cooling~\cite{Buschmann2022PRL}.

\subsubsection{Axion-Field Coupling to electrons}

Figure~\ref{gapsi}(b) shows the constraints on the axion---electron coupling $g_{aee}$---the exclusion regions labeled ``I. Bloch'' and ``Mainz-Krak\'{o}w'' are obtained by scaling the corresponding constraints on $g_{\rm aNN}$ by the gyromagnetic ratio factor $\gamma_N / \gamma_e$, since the underlying measurements are sensitive to spin couplings and can be interpreted as coupling to either the nucleon or the electron~\cite{bloch2020JHEP, Daniel2025nc}. 
For axion masses down to $10^{-23}~{\rm eV}/c^2$, additional constraints are derived from the analysis of torque signals measured by a spin torsion pendulum over a 6.7-year period~\cite{Terrano2019PRL}, shown as the region labeled ``W. Terrano''.
The region above the black dashed line in Fig.\ref{gapsi}(b) is excluded by white dwarf cooling constraints~\cite{Bertolami2014JCAP}.

\section{Conclusions and Perspectives}
\label{concl}
R.P. Feynman once remarked~\cite{Gleick_feynman}, ``All mass is interaction.'' Indeed, not only do fundamental particles acquire mass through their coupling to the Higgs field, but the strong interaction---mediated by gluons---constitutes the dominant contribution to the mass of hadrons such as the proton~\cite{PhysRevLett.121.212001}. Interactions lie at the heart of the universe’s richness and complexity; without them, detection would be impossible. It is thus unsurprising that the introduction of new interactions could offer solutions to unresolved problems in modern physics. In this review, we have systematically organized and categorized relevant theoretical models, experimental strategies, and existing constraints to facilitate the ongoing search for exotic spin-dependent interactions.

ALPs may mediate such interactions, providing a promising pathway for their detection. Theoretically, these particles offer elegant solutions to persistent puzzles in the SM. The Peccei-Quinn mechanism, for instance, addresses the strong CP problem and establishes a compelling framework with broad implications for both cosmology and particle physics. Moreover, the potential role of ALPs as cold dark matter candidates further motivates experimental exploration.
Experimental searches for spin-dependent interactions are also driven by their deep connections to the SME, which parameterizes possible violations of fundamental symmetries. However, the diversity and complexity of these models present challenges in reconciling theoretical predictions with experimental results, highlighting the need for more refined theoretical tools and precise formulations.
A unifying feature across many beyond-the-Standard-Model scenarios is the spin-dependent character of the predicted interactions, particularly in the low-energy regime. Consequently, experimental constraints derived for one class of models often carry implications for others. Notably, limits on ALP-mediated spin-dependent couplings have been set not only by dedicated source-probe setups but also by well-established methods such as atomic EDM measurements, gravitational tests, and precision atomic spectroscopy.

We summarize the principal experimental approaches for detecting spin-dependent interactions, highlighting their underlying physical principles and respective advantages. These methods are complementary and, together, probe a wide range of parameter space. For six distinct types of couplings, we compile the most stringent constraints as functions of the interaction range or ALP mass. Notably, in certain long-range regimes, laboratory-based experiments that utilize celestial bodies as source masses now impose tighter limits on SP couplings than those derived from astrophysical observations. Additionally, we present the most stringent current bounds on axion couplings to both nucleons and electrons.

\textcolor{black}{A consolidated roadmap of the field can be organized according to the targeted interaction class, the corresponding mediator mass or force range, the relevant coupling product, and the experimental platform best suited to that region. Short-range monopole--dipole interactions, such as SP couplings proportional to $g_s g_p$, are naturally probed by laboratory source--probe experiments using dense unpolarized source masses placed close to polarized spins, including torsion balances, atomic magnetometers, NMR systems, ultracold neutrons, and related precision platforms. These experiments are most powerful when the interaction range is comparable to the source--probe separation, typically from micrometers to meters. Dipole--dipole and velocity-dependent interactions, such as AA, VA, and related couplings involving $g_p^2$, $g_A^2$, or $g_V g_A$, require either polarized spin sources, controlled relative motion, or sidereal/rotational modulation; they are therefore well matched to comagnetometers, spin-polarized torsion pendulum, atomic beam experiments, and precision spectroscopy. For macroscopic or astronomical force ranges, terrestrial source masses become less competitive, and Earth-, Sun-, Moon-, or astrophysical-source-based analyses can provide unique sensitivity to extremely light mediators. In the ultralight dark-matter regime, where the boson mass corresponds to an oscillation frequency rather than a static force range, resonant and broadband spin-precession searches based on atomic magnetometers, comagnetometers, and NMR  are especially important. Preferred-frame and Lorentz-violating spin couplings are instead characterized by sidereal, annual, or boost-dependent signatures and are most cleanly tested with clock-comparison experiments and comagnetometers. Thus, no single platform covers all parameter space: force and torque experiments are complementary for short- and intermediate-range static potentials, while coherent spin-precession platforms dominate oscillating-field and preferred-frame searches. }

\textcolor{black}{The complementarity between laboratory searches and astrophysical or cosmological constraints should also be emphasized. Astrophysical and cosmological bounds are extremely valuable because they can probe very weak couplings, ultralight mediators, and macroscopic or astronomical length scales that are difficult to access using ordinary laboratory-source masses.  However, these bounds often rely on additional model-dependent assumptions, such as the production mechanism and abundance of the new particle, its coupling to different Standard-Model sectors. Consequently, a constraint that is very strong in one specific model may become weaker or even inapplicable when these assumptions are modified.}

\textcolor{black}{Laboratory searches are therefore indispensable as controlled and reproducible tests of exotic spin-dependent interactions. They allow direct control over the spin probe, readout scheme, magnetic shielding, modulation frequency, detection bandwidth, and systematic-error analysis. Importantly, laboratory tests are not limited to artificial source masses. Several laboratory analyses have used macroscopic astronomical bodies, such as the Earth, the Sun, and the Moon, as unpolarized source masses for spin-dependent interactions~\cite{Heckel2008PRD,Wu2023PRL}. This strategy combines the advantages of laboratory precision measurement with the enormous source mass and long baseline provided by celestial bodies, making it especially powerful for macroscopic and astronomical interaction ranges.}

\textcolor{black}{Such laboratory-astronomical hybrid searches are particularly important for interactions that are spin-, velocity-, or direction-dependent, for cases in which astrophysical energy-loss arguments do not directly apply, and for scenarios involving local dark-matter coherence or preferred-frame signatures. Laboratory experiments can also test specific coupling products, such as $g_s g_p$, $g_p^2$, $g_A^2$, and $g_V g_A$, with minimal reliance on assumptions about stellar evolution or cosmological history. Thus, astrophysical, cosmological, and laboratory constraints should be regarded as complementary: astrophysical and cosmological observations often provide the broadest reach, while laboratory experiments provide controlled validation, direct access to the microscopic interaction structure, and, in some cases, sensitivity to astronomical-scale forces by using the Earth, the Sun, or the Moon as the source.}

While most existing studies target electrons, protons, neutrons, and other nucleons, searches involving muons remain surprisingly underexplored, despite their relevance to several unresolved anomalies. These include the muon anomalous magnetic moment, the proton radius puzzle, and discrepancies in the helium charge radius observed in muonic atoms. Such anomalies, divergent from electron-based measurements, hint at possible muon-specific new interactions, and some theoretical models predict exotic couplings exclusive to muons.
Muons are inherently polarized and benefit from well-established experimental techniques such as $\mu$SR spectroscopy, widely used in material science. Adapting $\mu$SR methods to search for exotic spin-dependent muonic interactions is both practical and timely. Including muons in such investigations could address a major gap in the experimental landscape and potentially uncover entirely new classes of physics.

Moving forward, interdisciplinary collaboration will be crucial. Integrating data from astrophysical observations and terrestrial experiments will help narrow the parameter space and refine viable models. Such efforts hold promise for deepening our understanding of fundamental interactions and constituents of the universe, with significant implications for both particle physics and cosmology.
In parallel, artificial intelligence (AI) has demonstrated significant potential for improving experimental precision. AI and machine learning techniques are increasingly applied to optimize complex experimental configurations, automate multiparameter tuning, and enhance signal extraction under low signal-to-noise conditions. For instance, deep learning and Bayesian optimization have been employed for real-time feedback control and adaptive data acquisition, improving exploration across high-dimensional parameter spaces.
For noise reduction, approaches such as neural network-based denoising, blind source separation, and anomaly detection have shown promise in isolating weak signals from environmental interference. These methods are particularly well-suited to processing the large datasets generated by atomic, neutron, and muon-based precision measurements. As experimental sensitivities approach limits set by quantum noise and environmental fluctuations, AI-assisted techniques will likely play an increasingly central role in accelerating discovery, guiding hypothesis development, and enhancing data interpretation.
Notably, emerging reports of AI-assisted magnetometer designs suggest that such tools may soon become standard in next-generation precision experiments~\cite{meng2023NC, duan2025NP}. Continued development in this direction may further expand the reach of spin-dependent interaction searches, bridging theoretical aspirations with experimental realizations.

\section*{Data availability statement}
The data and programs for producing the plots are openly available~\cite{wu_2026_20589502}.

\section*{Acknowledgments}
We acknowledge support from the National Natural Science Foundation of China under grant U2230207.
\appendix
\section{Complete form of \texorpdfstring{$\V_i$}{Vi}}
\label{appendix:a}
Including the tensor-type interactions, the total Lagrangians can be rewritten as
\begin{equation}
\label{aL}
    \begin{aligned}
        \mathcal{L}_{\rm SP}&=\bar{\psi}\phi(g_S+g_Pi \gamma_5)\psi, \\  \mathcal{L}_{\rm VA}&=X_\mu\bar{\psi}\gamma^\mu(g_V+\gamma_5 g_A)\psi,\\
        \mathcal{L}_{T}&=\frac{v_h}{M^2}P_{\mu\nu}\bar{\psi}\sigma^{\mu\nu}(\mathrm{Re}(C)+i\gamma_5\mathrm{Im}(C))\psi,
    \end{aligned}
\end{equation}
where the dimensionless coupling constants $g_S$, $g_P$, $g_V$, $g_A$, $\mathrm{Re}(C)$, and $\mathrm{Im}(C)$ parametrize the strengths of the scalar, pseudoscalar, vector, axial-vector, tensor, and axial-tensor interactions, respectively. $m_\psi$ denotes the fermion mass, and $\gamma_i$ represent the Dirac matrices. The third term, $\mathcal{L}{T}$, describes magnetic- and electric-like dipole couplings between the $X\mu$ field and fermions, where $v_h$ is the Higgs vacuum expectation value, $M$ is the ultraviolet cutoff scale, $\sigma^{\mu\nu} = \frac{i}{2}[\gamma^\mu, \gamma^\nu]$, and $P_{\mu\nu} = \partial_\mu X_\nu - \partial_\nu X_\mu$. 

Taking the scalar-pseudoscalar coupling between fermions and ALPs as an example, and following the Feynman diagram in Fig.~\ref{fig3}, the scattering amplitude for the interaction between ALPs and fermions is given by
\begin{equation}
\label{apM}
    i\mathcal{M}=[-i\bar{u}(p_{2f})g^1_Su(p_{2i})][-i\bar{u}(p_{1f})i\gamma^5g^2_Pu(p_{1i})]\frac{i}{q^2-m^2},
\end{equation}
where $u$ is a fermion spinor. In the non-relativistic limit, we have $q^2=q_0^2-\vec{q}^2\approx -\vec{q}^2$, and the spinor product in Eq.~(\ref{apM}) simplified to 
\begin{equation}
    \begin{aligned}
        \bar{u}(p_{2f})i\gamma^5u(p_{2i})&=-\vec{\sigma}_2\cdot\vec{q},\\
        \bar{u}(p_{1f})u(p_{1i})&=2m_1,\\
    \end{aligned}
\end{equation}
where $\vec{q} = \vec{p}_{2f} - \vec{p}_{2i}$. The coordinate-space potential is obtained from the scattering amplitude via a Fourier transform:
\begin{equation}
\begin{aligned}
        V(r)&=-\int\frac{d^3q}{(2\pi)^3}\frac{\mathcal{M}}{4m_1m_2}\\
        &=g^1_Sg^2_P\int\frac{d^3q}{(2\pi)^3}\frac{i\vec{\sigma}_1\cdot\vec{q}}{2m_2}\frac{1}{m^2+\vec{q}_1^2}e^{i\vec{q}\cdot\vec{r}},\\
        &=g^1_Sg^2_P\vec{\sigma}_2\cdot\nabla\frac{e^{-mr}}{8\pi m_2r},
\end{aligned}
\end{equation}
where $m=1/\lambda$.
Following a similar procedure, the remaining interactions can be derived. A complete list of the interactions generated by Lagrangian (\ref{aL}) is provided below.
\vspace{1.0 em}

\paragraph{Spin Order-0}
\begin{itemize}[left=0pt,itemsep=2pt,topsep=2pt]
    \item $\nabla^0$ Potentials
    \vspace{-0.5em}
    \begin{equation}
        \V_1(r)=\Big(g^1_Vg^2_V-g^1_Sg^2_S\Big)\frac{1}{4\pi r}e^{-r/\lambda},
    \end{equation}
    \vspace{-0.5em}
\end{itemize}
\paragraph{Spin Order-1}
\begin{itemize}[left=0pt,itemsep=2pt,topsep=2pt]
    \item $\nabla^0$ Potentials
    \vspace{-0.5 em}
    \begin{equation}
    \begin{aligned}
            \V_{12+13}(r)&=-\Big[(1+\frac{m_1}{m_2})\Big]g^1_Ag^2_V\vec{\sigma}_1\cdot\vec p\frac{e^{-r/\lambda}}{4\pi m_1r}+\Big[(1+\frac{m_2}{m_1})\Big]g^1_Vg^2_A\vec{\sigma}_2\cdot\vec p\frac{e^{-r/\lambda}}{4\pi m_2r},
        \end{aligned}
    \end{equation}
    \vspace{-0.5em}
    \item $\nabla^1$ Potentials
    \vspace{-0.5em}
    \begin{equation}
    \begin{aligned}
            \V_{9+10}(r)&=-(\frac{1}{2}g^1_Pg^2_S+\frac{2v_h m_1}{M^2}g^2_V{\rm Im}(C_1))\vec{\sigma}_1\cdot\nabla \frac{e^{-r/\lambda}}{4\pi m_1r}\\
            &+(\frac{1}{2}g^1_Sg^2_P+\frac{2v_h m_2}{M^2}g^1_V{\rm Im}(C_2))\vec{\sigma}_2\cdot\nabla \frac{e^{-r/\lambda}}{4\pi m_2r},\\
            \V_{4+5}(r)&=\Big[(\frac{1}{2}+\frac{m_1}{m_2})g^1_Vg^2_V+\frac{m^2_1}{2m^2_2}g^1_Ag^2_A+\frac{1}{2}g^1_Sg^2_S+4(1+\frac{m_1}{m_2})g^2_V\frac{\nu m_2}{M^2}\mathrm{Re}(C_1)\Big]\vec{\sigma}_1\cdot[\vec p\times\nabla]\frac{e^{-r/\lambda}}{8\pi m_1^2r}\\
    &-\Big[(\frac{1}{2}+\frac{m_2}{m_1})g^2_Vg^1_V+\frac{m^2_2}{2m^2_1}g^2_Ag^1_A+\frac{1}{2}g^2_Sg^1_S+4(1+\frac{m_2}{m_1})g^1_V\frac{\nu m_2}{M^2}\mathrm{Re}(C_2)\Big]\vec{\sigma}_2\cdot[\vec p\times \nabla]\frac{e^{-r/\lambda}}{8\pi m_2^2r},
        \end{aligned}
    \end{equation}
    \vspace{-0.5em}
\end{itemize}
\paragraph{Spin Order-2}
\begin{itemize}[left=0pt,itemsep=2pt,topsep=2pt]
    \item $\nabla^0$ Potentials
    \vspace{-0.5em}
    \begin{equation}
    \begin{aligned}
            \V_2(r)&=-g^1_Ag^2_A\frac{1}{4\pi r}\vec{\sigma}_1\cdot\vec{\sigma}_2e^{-r/\lambda},\\
            \V_8(r)&=-\frac{1}{2}(\frac{m_1+m_2}{m_1})^2g^1_Ag^2_A\frac{1}{4\pi m_2^2r}(\vec{\sigma}_1\cdot\vec p)(\vec{\sigma}_2\cdot\vec p)e^{-r/\lambda},
        \end{aligned}
    \end{equation}
    \vspace{-0.5em}
    \item $\nabla^1$ Potentials
    \vspace{-0.5em}
    \begin{equation}
    \begin{aligned}
            \V_{11}(r)&=\Big(\frac{1}{2}g^1_Ag^2_V+\frac{m_2}{m_1}g^1_Vg^2_A-\frac{2v_hm_2}{M^2}[g^2_A\mathrm{Re}(C_1)-g^1_A\mathrm{Re}(C_2)]\Big)(\vec{\sigma}_2\times\vec{\sigma}_1)\cdot\nabla\frac{e^{-r/\lambda}}{4\pi m_2r},\\
            \V_{16}(r)&=\Big[\frac{m_2}{4m_1}(1+\frac{m_2}{m_1})(g^1_Ag^2_V-g^1_Vg^2_A)\\
    &+\frac{v_h m_2}{M^2}[(1+2\frac{m_2}{m_1}+2\frac{m^2_2}{m^2_1})g^2_A\mathrm{Re}(C_1)+(2+2\frac{m_2}{m_1}+\frac{m^2_2}{m^2_1})g^1_A\mathrm{Re}(C_2)]\Big]\\
    &\{
    \vec{\sigma}_2\cdot(\vec p\times \nabla)(\vec{\sigma}_1\cdot \vec p)+(\vec{\sigma}_2\cdot \vec p)\vec{\sigma}_1\cdot(\vec p\times \nabla)
    \}\frac{e^{-r/\lambda}}{8\pi m_2^3r},\\
    \V_{6+7}(r)&=\frac{2v_hm_2}{M^2}[(1+
    \frac{m_2}{m_1})g^2_A\mathrm{Im}(C_1)](\vec{\sigma}_2\cdot\vec p)(\vec{\sigma}_1\cdot\nabla)\frac{e^{-r/\lambda}}{4\pi m_2^2r}\\
    &-\frac{2v_hm_1}{M^2}[(1+
    \frac{m_1}{m_2})g^1_A\mathrm{Im}(C_2)](\vec{\sigma}_1\cdot\vec p)(\vec{\sigma}_2\cdot\nabla)\frac{e^{-r/\lambda}}{4\pi m_1^2r}
    \end{aligned}
    \end{equation}
    \vspace{-0.5em}
    \item $\nabla^2$ Potentials
    \vspace{-0.5em}
    \begin{equation}
    \begin{aligned}
            \V_3(r)&=-\Big(\frac{m_2}{4m_1}g^1_Vg^2_V+(\frac{1}{8}+\frac{m^2_2}{8m^2_1})g^1_Ag^2_A-\frac{m_2}{4m_1}g^1_Pg^2_P\\
    &-\frac{v_hm_2}{M^2}(g^2_V\mathrm{Re}(C_1)-\frac{m_2}{m_1}g^1_V\mathrm{Re}(C_2))-\frac{4\nu^2_hm^2_2}{M^4}\mathrm{Re}(C^*_2C_1)\Big)(\vec{\sigma}_1\times\nabla)(\vec{\sigma}_2\times\nabla)\frac{e^{-r/\lambda}}{4\pi m_2^2r},\\
    \V_{15}(r)&=-\Big[\frac{2v_hm_2}{M^2}[(\frac{1}{2}+\frac{m_2}{m_1})g^2_V\mathrm{Im}(C_1)+\frac{m_2}{m_1}(1+\frac{m_2}{2m_1})g^1_V\mathrm{Im}(C_2)]\\
    &+\frac{8v_h^2m^2_2}{M^4}(1+\frac{m_2}{m_1})[\mathrm{Re}(C_2)\mathrm{Im}(C_1)-\mathrm{Im}(C_2)\mathrm{Re}(C_1)]+\frac{m_2}{4m_1}(g^1_Pg^2_S-\frac{m_2}{m_1}g^1_Sg^2_P)\Big]\\
    &\{
    [\vec{\sigma}_1\cdot(\vec p\times\nabla)](\vec{\sigma}_2\cdot\nabla)+(\vec{\sigma}_1\cdot\nabla)[\vec{\sigma}_2\cdot(\vec p\times\nabla)]
    \}\frac{e^{-r/\lambda}}{8\pi m_2^3r}.\\
        \end{aligned}
    \end{equation}
    \vspace{-0.5em}
\end{itemize}

Here, we organize these interactions using the notation adapted from Dobrescu \textit{et al.} in Ref.~\cite{Dobrescu2006JHEP}. The interactions are expressed in the two-body center-of-mass frame, where $\vec{p}$ denotes the relative momentum between the fermions. In the laboratory frame, this relative momentum $\vec{p}$ is related to the individual momenta $\vec{p}_1$ and $\vec{p}_2$ of the two fermions by:
\begin{equation}
    \vec p=\frac{m_1\vec p_2-m_2\vec p_1}{m_1+m_2}.
\end{equation}


\providecommand{\href}[2]{#2}\begingroup\raggedright\endgroup
\end{document}